\documentclass{aa}  
\usepackage{graphicx}
\usepackage{caption}
\usepackage{subcaption}
\usepackage{gensymb}   
\usepackage{txfonts}
\usepackage[utf8]{inputenc}

\begin{document}

\def\arcsec {\hbox{$^{\prime\prime}$}}
\def\arcmin {\hbox{$^{\prime}$}}
\def\ffam {\hbox{$\,.\!\!^{\prime}$}}
\def\ffas {\hbox{$\,.\!\!^{\prime\prime}$}}   
\def\ffcirc {\hbox{$\,.\!\!^{\circ}$}}
\def\ffM {\hbox{$\,.\!\!^{\rm M}$}}
\def\ffm {\hbox{$\,.\!\!^{\rm m}$}}
\def\ffs {\hbox{$\,.\!\!^{\rm s}$}}

   \title{Molecular gas in the northern nucleus of Mrk\,273:\\
   Physical and chemical properties of the  disc and its outflow}

   \author{R. Aladro \inst{1,2}
          \and             
          S. K\"{o}nig  \inst{2}
              \and
          S. Aalto\inst{2}
             \and  
             E. Gonz{\'a}lez-Alfonso  \inst{3}
             \and
         N. Falstad\inst{2}
             \and
             S. Mart{\'i}n  \inst{4,5}
            \and          
          S. Muller  \inst{2}
               \and           
            S. Garc{\'i}a-Burillo \inst{6}
           \and
                C. Henkel  \inst{1,7}
           \and
             P. van der Werf  \inst{8}  
           \and
           E. Mills  \inst{9}  
           \and
           J. Fischer\inst{10}    
           \and
             F. Costagliola  \inst{2}                 
           \and
                   M. Krips \inst{11}                
          }

   \institute{Max-Planck-Institut f{\"u}r Radioastronomie, Auf dem H\"ugel 69, 53121, Bonn, Germany\\
              \email{aladro@mpifr-bonn.mpg.de} 
        \and
        Chalmers University of Technology, Department of Space, Earth and Environment, Onsala Space Observatory, 43992 Onsala, Sweden
        \and  
        Universidad de Alcal{\'a}, Departamento de F{\'i}sica y Matem{\'a}ticas, Campus Universitario, 28871 Alcal{\'a} de Henares, Madrid, Spain     
         \and
             European Southern Observatory, Alonso de C{\'o}rdova 3107, Vitacura  763 0355, Santiago, Chile
        \and
             Joint ALMA Observatory, Alonso de C{\'o}rdova 3107, Vitacura 763 0355, Santiago, Chile
            \and
 Observatorio de Madrid, OAN-IGN, Alfonso XII, 3, E-28014-Madrid, Spain
        \and
                   Astron. Dept., King Abdulaziz University, P.O. Box 80203, 21589 Jeddah, Saudi Arabia
        \and
                        Leiden Observatory, Leiden University, P.O. Box 9513, NL-2300 RA Leiden, The Netherlands     
\and
                                San Jose State University, 1 Washington Square, San Jose, CA 95192, USA
                                           \and 
                                George Mason University, Department of Physics \& Astronomy, MS 3F3, 4400 University Drive, Fairfax, VA 22030, USA                             
        \and                                            
           Institut de Radio Astronomie Millim{\'e}trique, 300 rue de la Piscine, Dom. Univ., 38406 St Martin d’H{\`e}res, France
}
   \date{Received ; accepted }

 
  \abstract{Aiming to characterise the properties of the molecular gas in the ultra-luminous infrared galaxy Mrk\,273 and its outflow, we used the NOEMA interferometer to image the dense-gas molecular tracers HCN, HCO$^+$, HNC, HOC$^+$ and HC$_3$N at $\sim$86\,GHz and $\sim$256\,GHz with angular resolutions of 4$\ffas$9$\times$4$\ffas$5 ($\sim$3.7$\times$3.4\,kpc) and 0$\ffas$61$\times$0$\ffas$55 ($\sim$460$\times$420\,pc). We also modelled the flux of several H$_2$O lines observed with {\emph{Herschel}} using a radiative transfer code that includes excitation by collisions and far-infrared photons. The disc of the Mrk\,273 north nucleus has two components with decoupled kinematics. The gas in the outer parts (R$\sim$1.5\,kpc) rotates with a south-east to north-west direction, while in the inner disc (R$\sim$300\,pc)  follows a north-east to south-west rotation. The central 300\,pc, which hosts a compact starburst region, is filled with dense and warm gas, and contains a dynamical mass of $(4-5)\times10^9M_\odot$, a luminosity of $L'_{\rm HCN}=(3-4)\times10^8$\,K\,km\,s$^{-1}$pc$^2$, and a dust temperature of 55\,K. At the very centre, a compact core with R$\sim$50\,pc has a luminosity of $L_{\rm IR}=4\times10^{11}L_\odot$ (30\% of the total infrared luminosity), and a dust temperature  of 95\,K . The core is expanding at low velocities $\sim$50-100\,km\,s$^{-1}$, probably affected by the outflowing  gas. We detect the blue-shifted component of the outflow, while the red-shifted counterpart remains undetected in our data. Its cold and dense phase reaches fast velocities up to $\sim$1000\,km\,s$^{-1}$, while the warm outflowing gas has more moderate maximum velocities of $\sim$600\,km\,s$^{-1}$.  The outflow is compact, being detected as far as  460\,pc from the centre in the northern direction, and has a mass of dense gas $\le8\times10^8M_\odot$. 
        The difference between the position angles of the inner disc ($\sim$70$\degree$) and the outflow ($\sim$10$\degree$) indicates that the outflow is likely  powered by the  AGN, and not by the starburst. Regarding the chemistry in Mrk\,273, we measure an extremely low HCO$^+$/HOC$^+$ ratio of 10$\pm$5 in the inner disc of Mrk\,273.}
 

 \maketitle
%

\section{Introduction}
\label{intro}

Tracers of dense molecular gas are good probes of the central regions of active galaxies, where molecular regions are subjected to strong radiation fields (X-rays, cosmic rays, and ultraviolet (UV) fields) created by massive star formation and/or active galactic nuclei (AGNs). In particular, the rotational transitions of HCN and HCO$^+$  have bright emission and high dipole moments (hence large critical density) and are therefore convenient tracers of the dense gas component in galactic centres.

In the particular case of ultra-luminous infrared galaxies (ULIRGs), the nuclei  are  embedded in large quantities of gas and dust produced by merging processes and massive star formation. In these conditions,  the visual extinctions can be as high as Av$>$1000 mag when the H$_2$ column densities exceed $>10^{24}$cm$^{-2}$  in very compact regions of only a few pc (e.g. \citealp{Costagliola15,Aalto17}). In these extremely obscured environs, even the millimetre (mm)/sub-mm emission  can be significantly attenuated by self- or continuum absorption, therefore probing only the  gas emitted at distances $\ge$100\,pc from the central engines (e.g. \citealp{Aalto15,Martin16}). Complementing  mm/sub-mm observations with radiatively pumped molecular lines emitted in the infrared (IR), where most of the ULIRGs luminosity is emitted, allows us to probe  the regions of the dusty cores more deeply. In particular, H$_2$O lines are excited via absorption and re-emission of IR photons produced very close to the central engines, and therefore provide essential information about the physical conditions of the hidden power sources of ULIRGs.

Mrk\,273 (F13428+5608) is a ULIRG ($L_{\rm IR}=1.3\times10^{12}L_\odot$,  \citealp{Gao99}) located at a distance of 157\,Mpc ($V^{\rm optical}_{\rm LSR}$=11339\,km\,s$^{-1}$, z=0.037780, 1$''$=761\,pc\footnote{https://ned.ipac.caltech.edu}). Its complex morphology reflects a recent merger event between two or more galaxies.
Near-infrared (NIR), radio emission, and HI images  show two nuclei separated by $\sim$1$''$ in the northeast-southwest direction (hereafter the northern and southern nucleus) plus a weaker third source that might be another nucleus or a  starburst region triggered by the merger (Majewski et al. 1992, Cole et al. 1999, Condon et al. 1991). 
The nature of the progenitors has been the subject of several studies that claim either AGN or starburst activities.  Mrk\,273 is  classified as a Seyfert\,2 galaxy in the optical and NIR \citep{U,Rodriguez,Iwasawa17}, having an AGN bolometric luminosity of log$(L_{\rm AGN})$=44.73 erg\,s$^{-1}$, and a ratio between the bolometric luminosity of the AGN and the total bolometric luminosity of the galaxy of $L_{\rm AGN}/L_{\rm Bol}=0.08$ \citep{Nardini09}. Nevertheless, far infrared (FIR) and mm data point to a compact ultra-luminous starburst region in the northern nucleus (Condon et al. 1991, Majewski et al. 1992, Downes \& Solomon 1998). The total star formation rate (SFR) of the galaxy is 139\,$M_\odot$yr$^{-1}$ \citep{Cicone14}.

Molecular observations reveal a complex structure in the centre of Mrk\,273. Carbon monoxide maps show  extended gas streamers in the north-south direction, a nuclear disc (of $2''$ size) oriented east-west, and a very compact core ($0.35''\times0.2''$)  containing a powerful starburst  (Downes \& Solomon 1998). All these components  belong to the northern nucleus (which is the strongest radio source), while there is no evidence of significant amounts of molecular gas in the southern objects.  

A cool molecular outflow  has been detected in Mrk\,273 using H$_2$, CO, and OH observations \citep{U,Veilleux13,Cicone14,Gonzalez17}. From CO\,$(1-0)$ observations, the outflow appears clearly in both line wings at high velocities |v|$>$400\,km\,s$^{-1}$, and extends from the northern nucleus about 600\,pc to the north. The  CO channel-velocity maps also show a low velocity component (v$>$150\,km\,s$^{-1}$) of the red-shifted outflow expanding to the north. The total mass of the outflow estimated from CO is $\sim2\times10^8M_\odot$.

The multi-phase composition of the outflow has been revealed by infrared and optical observations.  Hydroxide  (OH) detections \citep{Veilleux13,Gonzalez17} probe a more compact and warmer phase of the outflow that  expands at  velocities of 100-700\,km\,s$^{-1}$. This  component travels shorter distances (160\,pc) before it presumably cools down, and its mass and mass-loss rate are 5$\times10^7M_\odot$ and 120$M_\odot$yr$^{-1}$. \citet{Colina99}  detected an ionised component of the outflow  by using the [OIII] $\lambda$=5007$\AA$  optical line. The hot, ionised  gas goes much further, as far as $\pm$5$''$ ($\sim$3.8\,kpc) along the north-south direction, and reaches velocities as high as $\pm$1200\,km\,s$^{-1}$. Other IR and optical lines, namely H$_2$, HeI,  Br$\gamma$, and [CII], also show the low-velocity component of the outflow ($\pm$200\,km\,s$^{-1}$), as well as a moderate velocity component (600$\pm$300)\,km\,s$^{-1}$  heading towards the north \citep{U,Janssen16}.

Despite being one of the most luminous ULIRGs in the local universe, molecular detections towards Mrk\,273 were still scarce and limited to  CO, OH, and H$_2$. In this paper we present observations of  molecules detected for the first time in this galaxy in the mm/sub-mm and FIR ranges. In Sect.\,\ref{observations} we present our observations with the NOEMA and \emph{Herschel} telescopes and the data reduction. The continuum and spectroscopic data analyses are presented in Sect.\,\ref{results}, where we also describe our modelling of H$_2$O.  The asymmetric, double-peaked line profiles of the inner disc are discussed in Sect.\,\ref{asymmetry}. The  detection of the Mrk\,273 outflow  and its properties  are presented in Sect.\,\ref{outflow}. A brief discussion of the expansion of the galaxy core can be found in Sect.\,\ref{expansion}. The non-detection of vibrationally excited HCN and HC$_3$N emission is addressed in Sect.\,\ref{vib}. We also discuss the HCN/HNC, HCN/HCO$^+$, and HCO$^+$/HOC$^+$   brightness temperature ratios  (Sect.\,\ref{lineratios}).  The different origin of HOC$^+$, the only species not peaking at the very centre,  is discussed  in Sect.\,\ref{hoc+}. Finally, our conclusions are summarised in Sect.\,\ref{conclusions}. %

\section{Observations and data reduction}
\label{observations}
\subsection{NOEMA}
\label{obs-NOEMA}
The HCN, HCO$^+$, HNC\,$(1-0)$ and HC$_3$N\,$(10-9)$ lines were observed simultaneously with the NOEMA interferometer on April 7, 2015 (with seven antennae), and June 12 and 15, 2017 (with eight antennae during a total time of 8.9 hours  (precipitable water vapour  (pwv)$\sim$5-9\,mm), while the HCN, HCO$^+$ and HOC$^+(3-2)$ transitions were observed, also simultaneously, for 5.6 hours  (pwv$\sim$1-2\,mm). The receivers were tuned to centre their lower side bands at 85.8\,GHz ($\lambda\sim$3\,mm) and 257.5\,GHz ($\lambda\sim$1\,mm), respectively.  The receivers were connected to the WideX correlator and provided a 3.6\,GHz bandwidth in two orthogonal polarisations (which were averaged). The data were calibrated and imaged with CLIC and MAPPING within the GILDAS$'$ package\footnote{http://www.iram.fr/IRAMFR/GILDAS}. Source 3C84 was used as a bandpass calibrator, and the phase and flux calibrations were done by observing 1418+546 and MWC349, respectively. 

The observations were centred on RA\,(J2000)\,=\,13$^{\rm h}$:44$^{\rm m}$:$42\fs1$
, DEC\,(J2000)\,=\,55$^\circ$:53$^{\rm '}$:$13\ffas5$. The original spectral resolutions of 6.8\,km\,s$^{-1}$ (3\,mm) and 2.3\,km\,s$^{-1}$ (1\,mm)  were smoothed to  68-70\,km\,s$^{-1}$. The final rms of the cubes, averaged across the spectral channels that do not contain line emission, are 0.3\,mJy\,beam$^{-1}$ (3\,mm) and 1.3\,mJy\,beam$^{-1}$ (1\,mm). We used a natural weighting mode and 0${''}$.8 (for the 3\,mm data) and 0${''}$.05 (for the 1.3\,mm data) pixel sizes to create the data cubes, and the Hogbom deconvolution method to clean them. The sizes of the primary beams were 58$\ffas$6 and 19$\ffas$6, and the angular resolutions achieved were ($4''.9\times4''.5$) and ($0''.61\times0''.55$) with position angles (P.A.) of -80$\degree$ and +34$\degree$  for the 3\,mm and 1\,mm transitions, respectively. At the adopted distance of Mrk\,273, these resolutions correspond to  ($3.7\times3.4$)\,kpc and  ($460\times420$)\,pc spatial scales. 

Given the NOEMA configurations used, the maximum recoverable scales for our 1\,mm and 3\,mm observations are 2$\ffas$5 and 17$''$ respectively, which are well above (about one order of magnitude) the emission sizes of the molecules (see Sect.\,\ref{intint} and Table\,\ref{table3}).  Furthermore, our HCN and HCO$^+$ fluxes are consistent with those obtained by \citet{Gracia08} with the IRAM 30\,m single-dish telescope. We can, therefore, safely claim that there was no flux filtered out in our observations.

\subsection{Herschel}
\label{obs-Herschel}
The H$_2$O data were taken with the Photodetector Array Camera and Spectrometer -PACS \citep{Ott10,Poglitsch10} and the Spectral and Photometric Imaging Receiver -SPIRE \citep{Griffin10} on  December 16, 2012 and  November 21, 2010 respectively.  The absorption water lines observed with Herschel/PACS are new detections in Mrk\,273, while the emission transitions detected with Herschel/SPIRE were already reported by \citet{Lu17}.  The PACS observations\footnote{OBSIDs:1342257290-1342257294} (PI: Gonz\'alez-Alfonso) were performed in high spectral sampling range spectroscopy mode in first and second orders of the grating, resulting in a velocity resolution of $\sim170-265$\,km\,s$^{-1}$.  The spectra were reduced with the standard product generation pipeline version 14.2.0. The nuclear FIR emission from Mrk~273 was unresolved with the PACS 5spaxel$\times$5spaxel IFU with 9.$''$4 x 9.$''$4 spaxels, so the spectra were extracted using the point source correction task available in the \emph{Herschel} interactive processing environment -HIPE \citep{Ott10} version 14.0.1. The spectra were then scaled to the integrated flux level of the central $3 \times 3$ PACS spaxels to compensate for pointing offsets and jitter which act to move flux out of the central spaxel. The H$_2$O spectra were  sampled in velocities of 20-40\,km\,s$^{-1}$ per channel width. Polynomial baselines of orders  lower than three were then removed, and the final root mean squares (rms) are 0.2-0.3\,Jy\,km\,s$^{-1}$. The lines were fitted with Gaussian line profiles using the GILDAS software CLASS (Fig.\,\ref{fig3}).

The SPIRE observation\footnote{OBSID:1342209850} (PI: P.~P.~van~der~Werf) was conducted using a single pointing centred on Mrk~273 in high-spectral-resolution sparse image sampling mode with a resolution of $1.2$~GHz  ($\sim$250-360\,km\,s$^{-1}$) in the two observing bands ($447-989$~GHz and $958-1545$~GHz). In total, 99 repetitions (198 FTS scans) were performed for a total on source time of $13187$~s. The data reduction was done with the standard single pointing pipeline available in HIPE version 14.0.1 and a bootstrap method was used to extract the line fluxes. The individual scans were averaged together and a polynomial baseline was extracted from each detector before all lines were fitted simultaneously using Gaussian profiles convolved with a sinc function (Fig.\,\ref{fig4}). After 1000 repetitions of this procedure, Gaussians were fitted to the resulting flux distribution of each line to get the mean line flux together with its standard deviation.

\begin{figure}[t!]
        \centering 
        \includegraphics[width=0.4\textwidth]{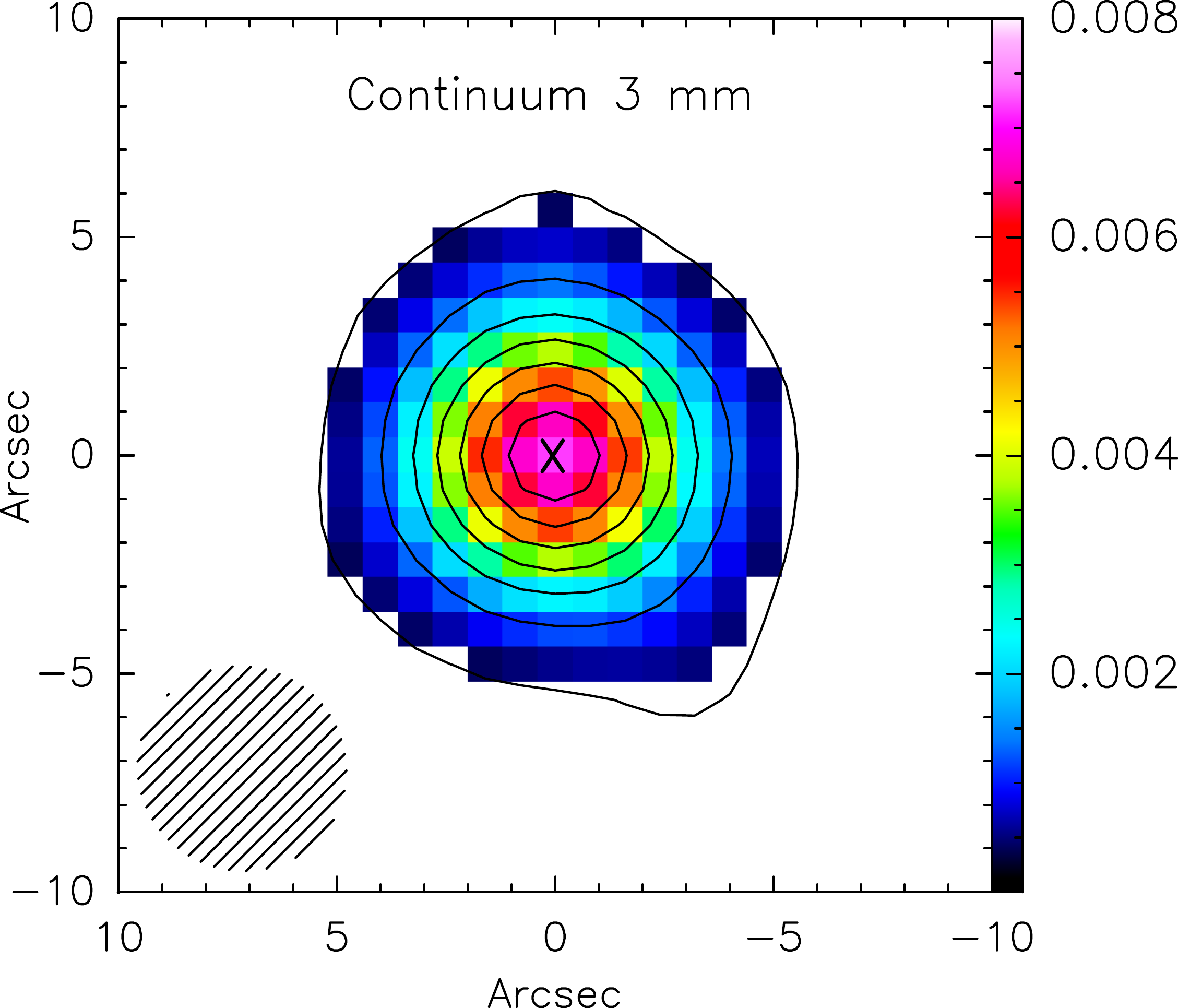}
        \includegraphics[width=0.4\textwidth]{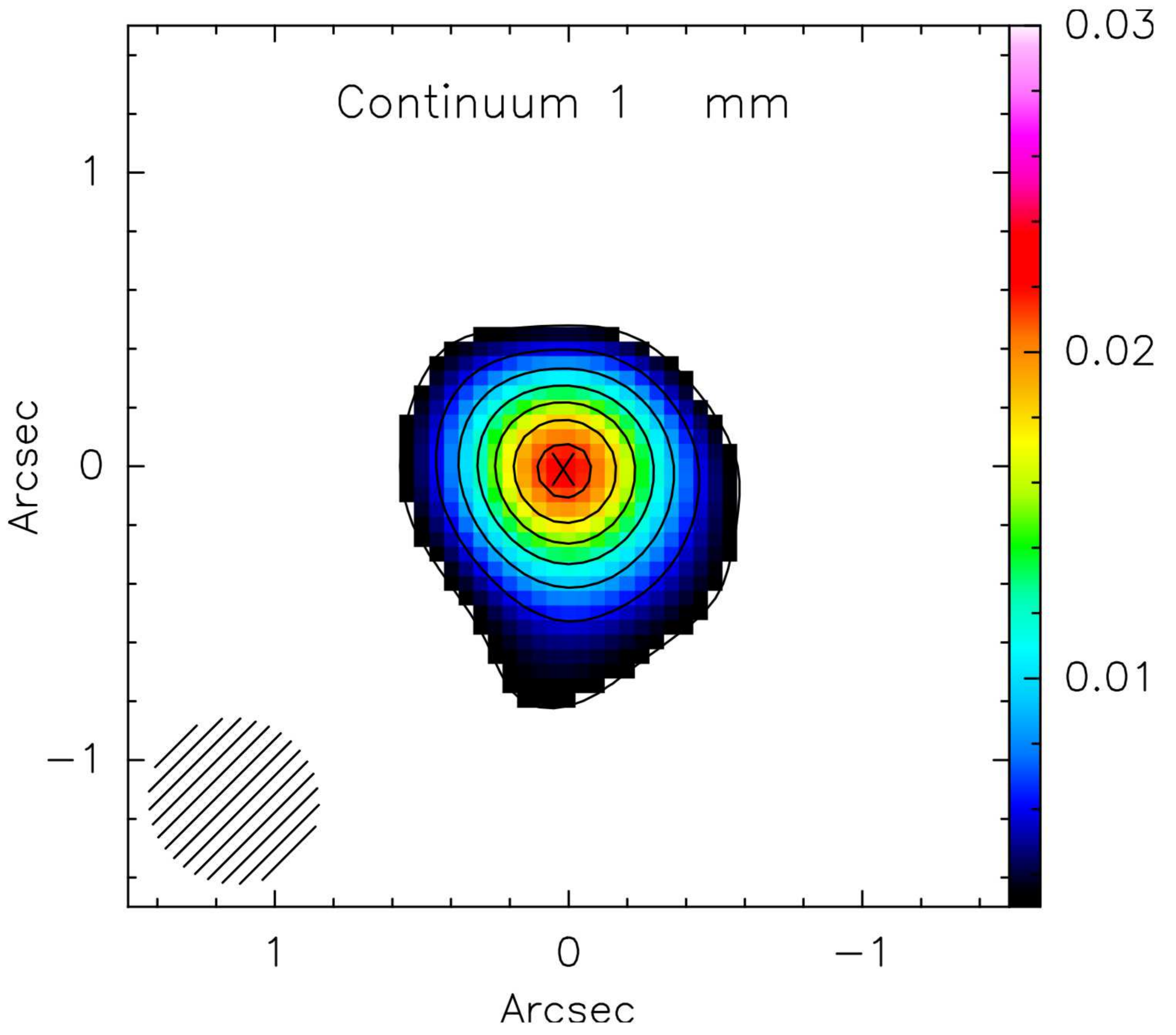}      
        \caption{Continuum maps at 3\,mm and 1\,mm. 
                Contour levels start at a significance of 5$\sigma$ with respect to the rms measured in both images (rms = 0.07\,mJy\,channel$^{-1}$ and 0.6\,mJy\,channel$^{-1}$ for the 3\,mm and 1\,mm maps respectively). The contour steps are 1 and 3\,mJy\,\,km\,s$^{-1}$\,beam$^{-1}$. The crosses at the centres mark the continuum peaks, which we take as the location of the northern nuclear source (see Sect.\,\ref{intro}). The synthesised beams are shown in the bottom-left corner. The colour flux  scales are in Jy\,km\,s$^{-1}$\,beam$^{-1}$.}
        \label{fig1}
\end{figure}

\section{Results}
\label{results}
\subsection{Continuum}
\label{continuum}

We first created the 3\,mm  ($\sim$89\,GHz) continuum visibilities  including only the channels free of line emission after smoothing to the final spectral resolution. These visibilities were then subtracted from the total emission in the uv plane. Using the task {\emph{uv\_fit}} within the MAPPING package, we measured the size and flux of the 3\,mm continuum. Our values were calculated  by fitting an elliptical Gaussian in the Fourier plane. Circular and elliptical fits gave consistent values within the errors. Given that the images of the continuum and line emissions are quite round, but not entirely, we opted for using elliptical fits in order to take into account small asymmetries in the emission. We measured a spatially integrated flux density of the 3\,mm continuum  of $8.24\pm0.07$\,mJy and a deconvolved size of $(1.92\pm0.06)'' \times (1.83\pm0.07)''$ with a P.A.=$(-90\pm30)\degree$. This is  similar to the 111\,GHz continuum flux density of 11$\pm$2\,mJy obtained by \citet{DS98}.

 Due to the very broad line widths at zero intensity (500-1100\,km\,s$^{-1}$) in the 1\,mm band ($\sim$265\,GHz) and the narrow bandwidth of the correlator,  almost all channels contain line emission. Therefore, the continuum visibilities were created using only nine line-emission-free channels (15\% of the total number of channels) at both edges of the spectrum. Including more channels could potentially lead to an overestimation of the continuum flux. A fit of an elliptical Gaussian in the uv plane gives an integrated flux of $28.6\pm0.9$\,mJy, with a deconvolved size (FWHM -full with at half maximum- of the Gaussian) of $(0.36\pm0.03)''\times(0.27\pm0.03)''$  $\sim$(270$\times$200\,pc) with a position angle of $(24\pm12)\degree$. The integrated intensities of the  continuum at 1\,mm and 3\,mm are plotted in Fig.\,\ref{fig1}.

\subsection{Line profiles}
  \begin{figure*}[ht!]
        \centering
        \begin{subfigure}{0.3\textwidth}
        \includegraphics[width=\textwidth]{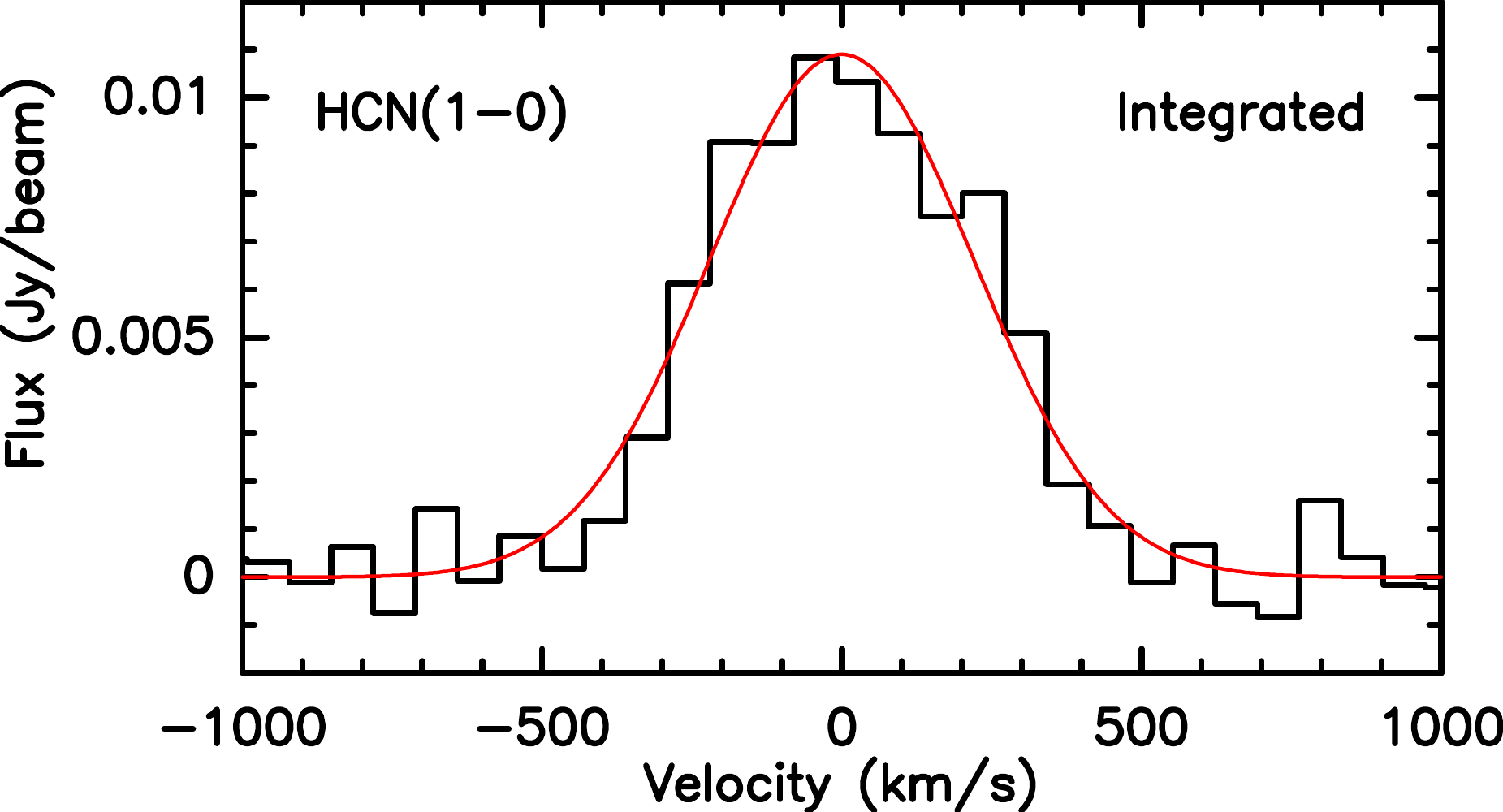}
\end{subfigure}\hspace*{5mm}
\vspace*{2mm}
        \begin{subfigure}{0.3\textwidth}
        \includegraphics[width=\textwidth]{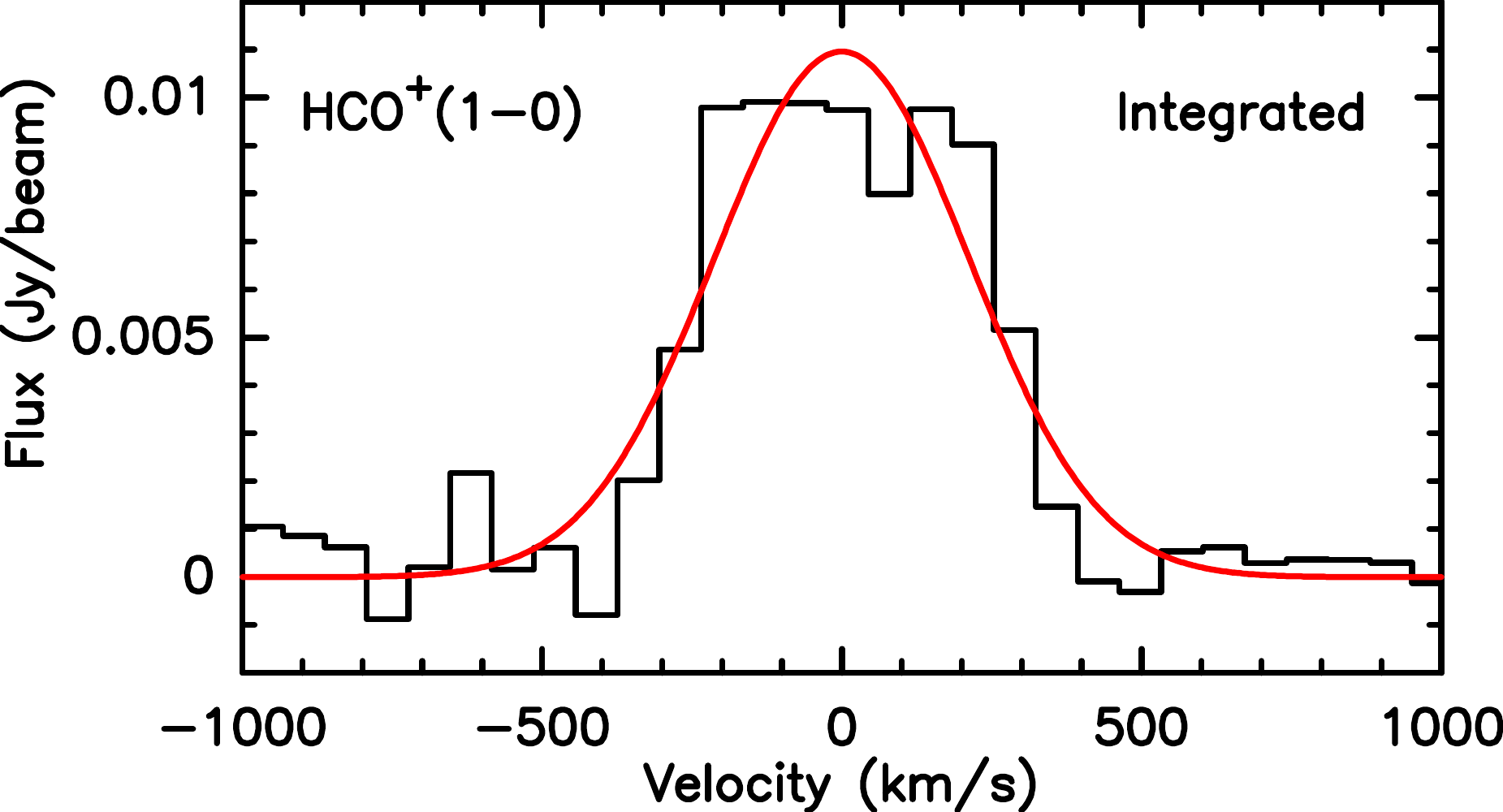}  
\end{subfigure}\hspace*{\fill}
        \begin{subfigure}{0.3\textwidth}
        \includegraphics[width=\textwidth]{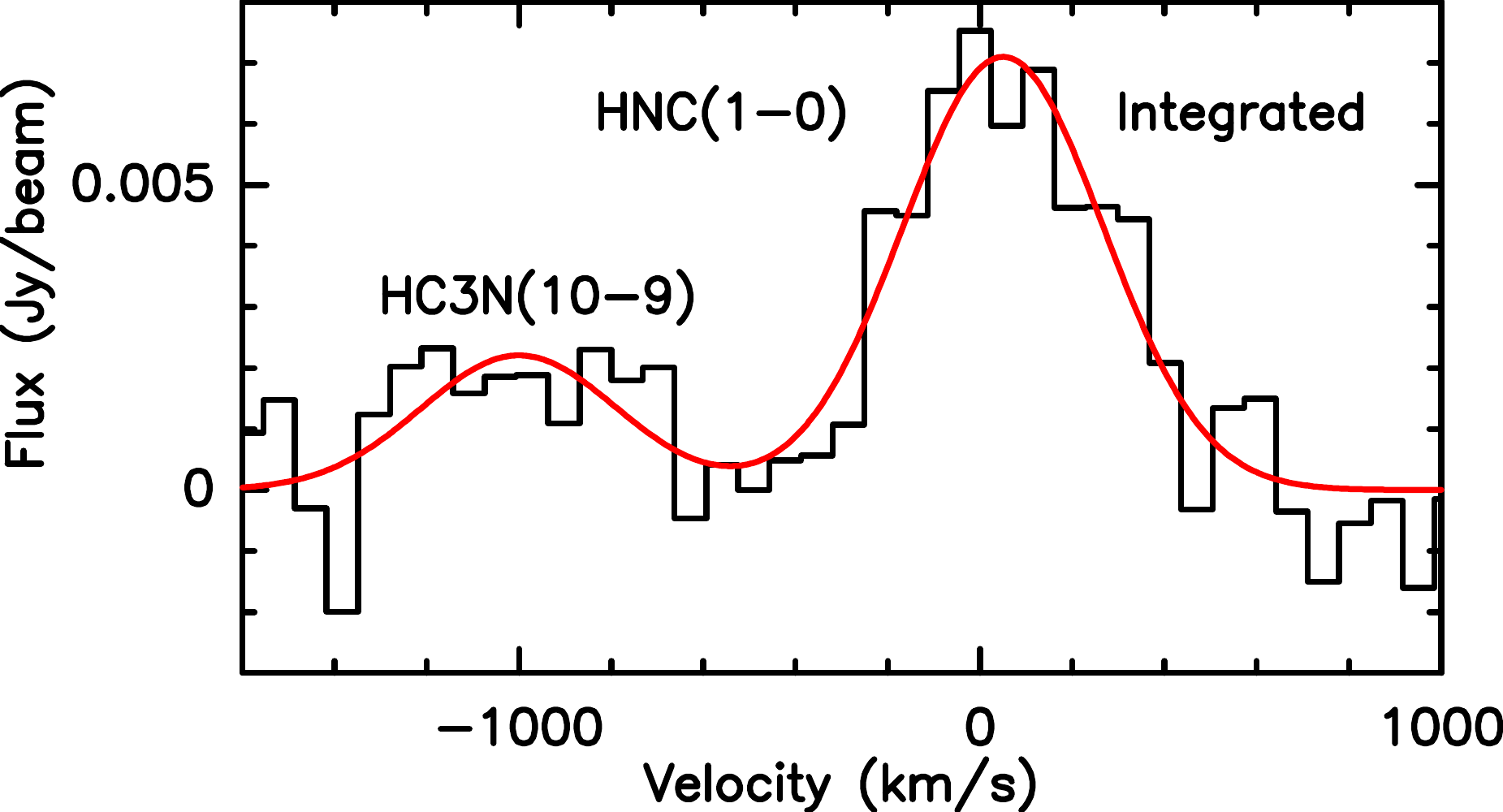}  
\end{subfigure}\hspace*{\fill}                  
        \vfill{}        
        \begin{subfigure}{0.3\textwidth}
        \includegraphics[width=\textwidth]{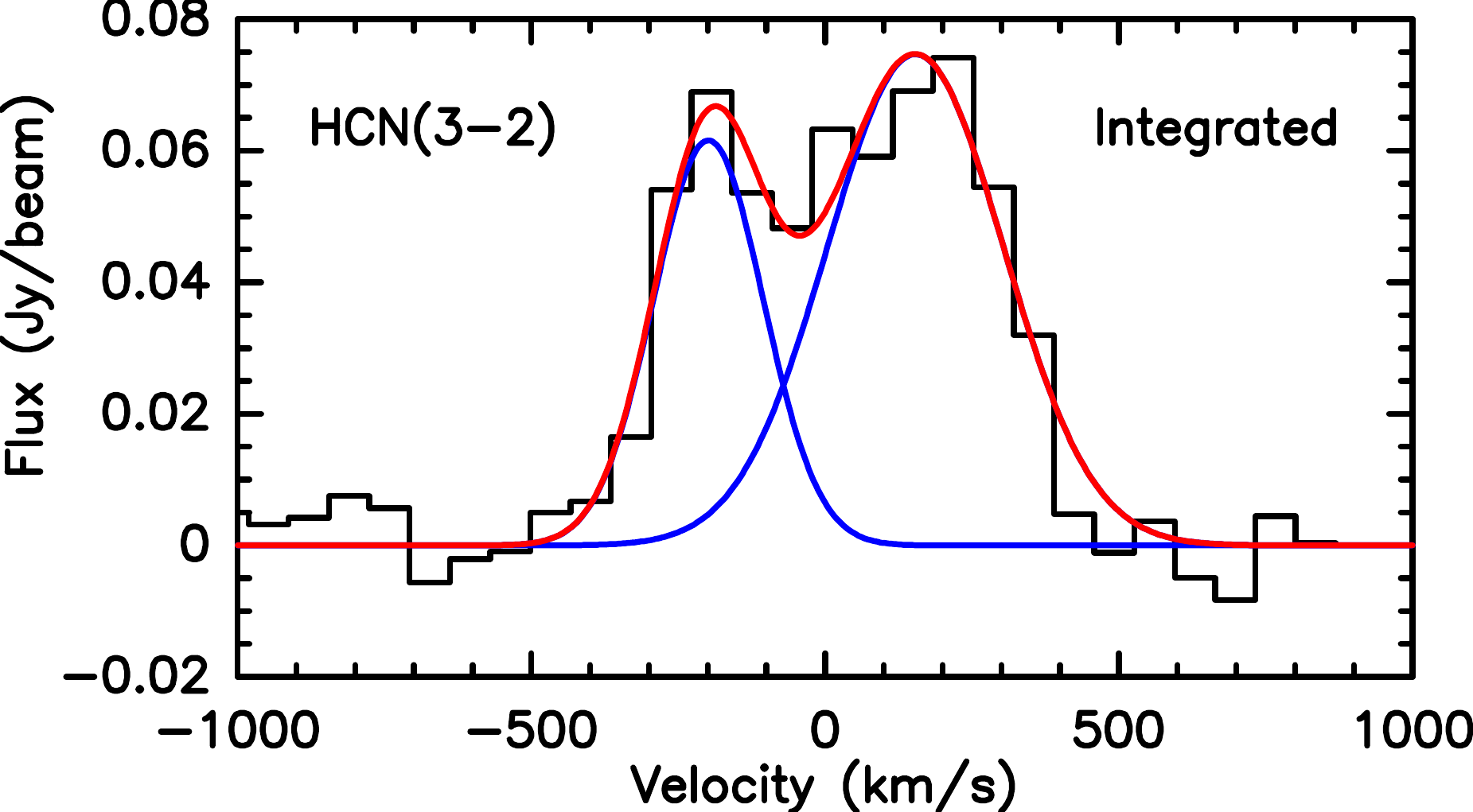}                 
\end{subfigure}\hspace*{\fill}
        \begin{subfigure}{0.3\textwidth}
        \includegraphics[width=\textwidth]{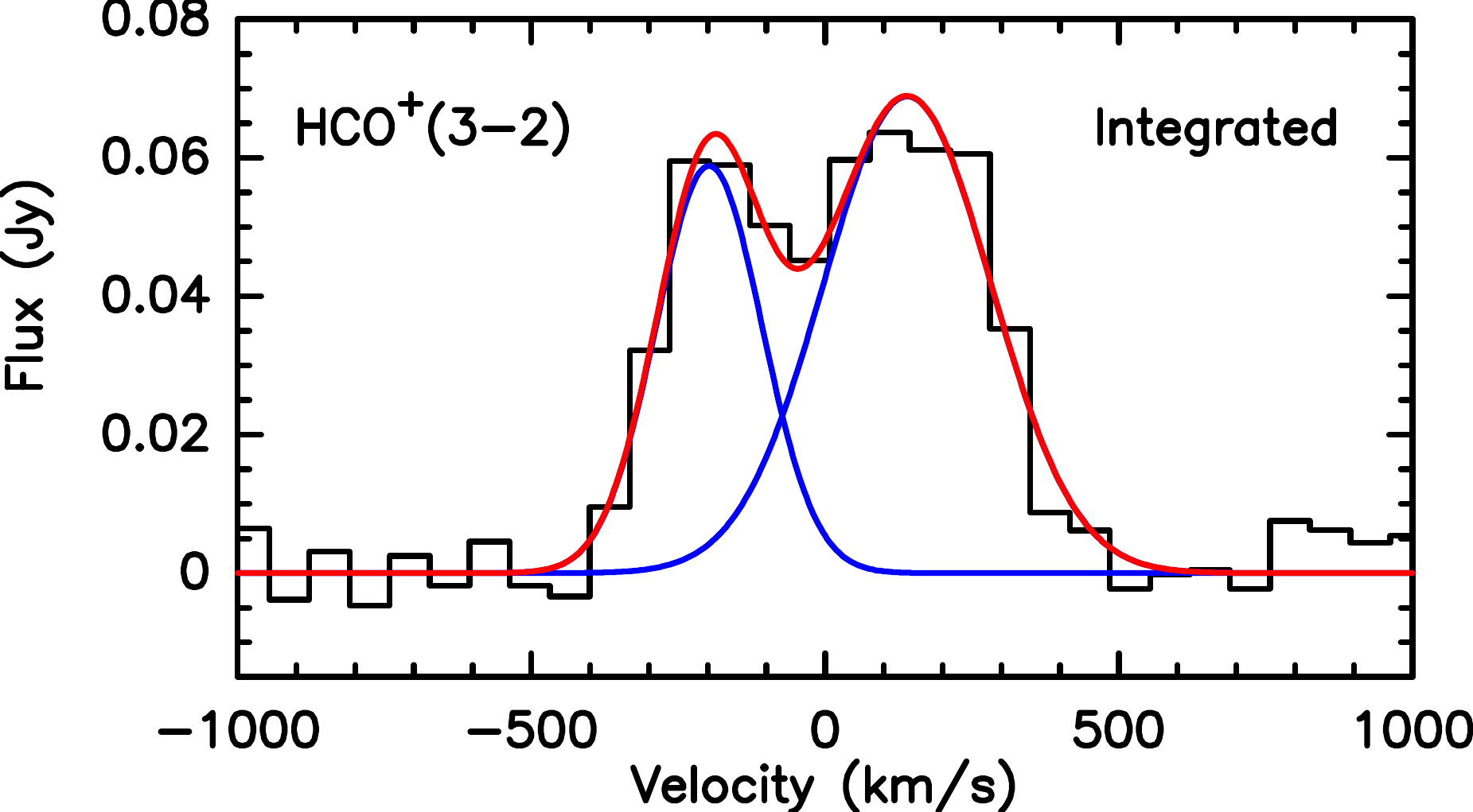}                 
\end{subfigure}\hspace*{\fill}
        \begin{subfigure}{0.3\textwidth}
        \includegraphics[width=\textwidth]{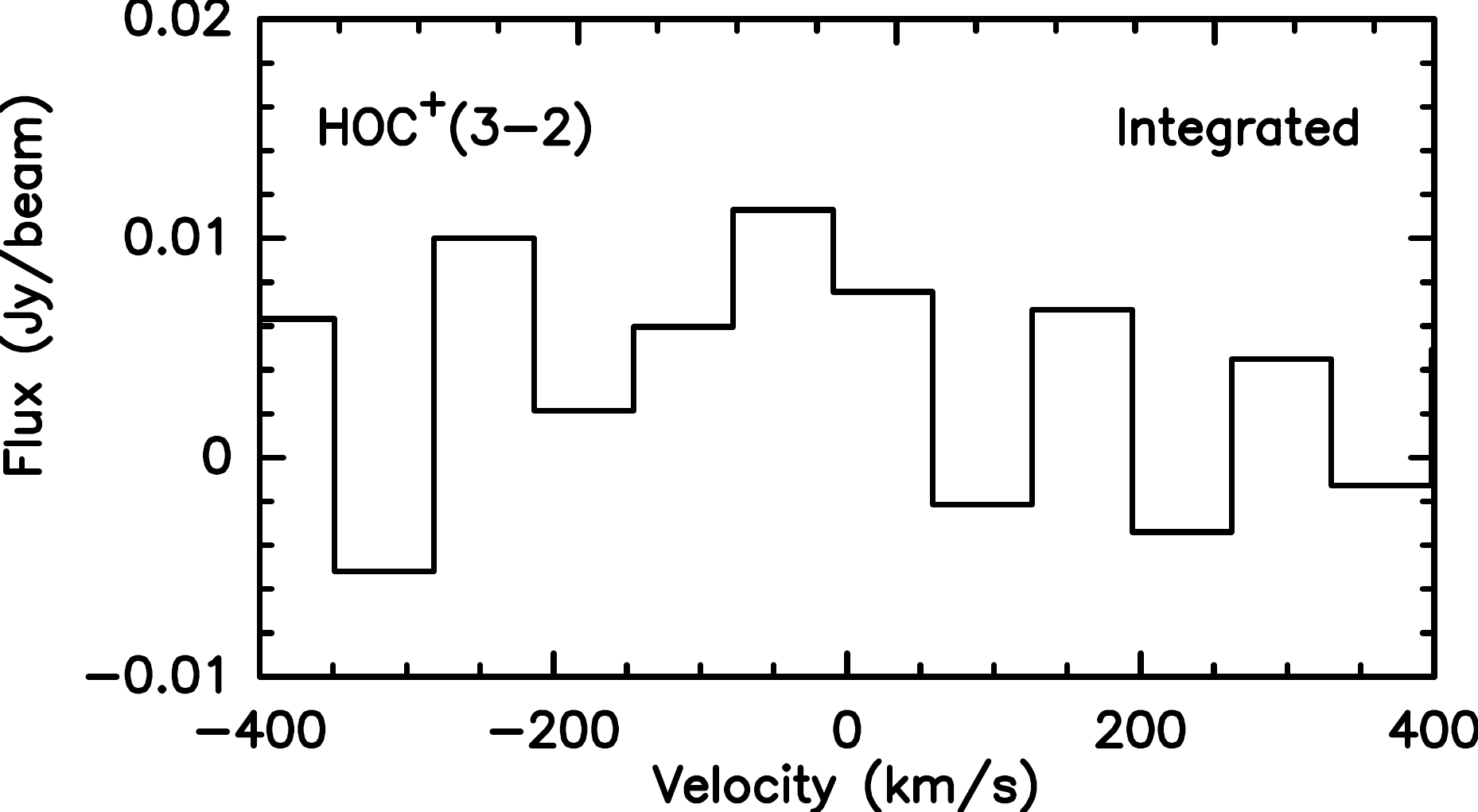}                 
\end{subfigure}\hspace*{\fill}
        \vfill{}        
\vspace{8mm}
        \begin{subfigure}{0.3\textwidth}
                \includegraphics[width=\textwidth]{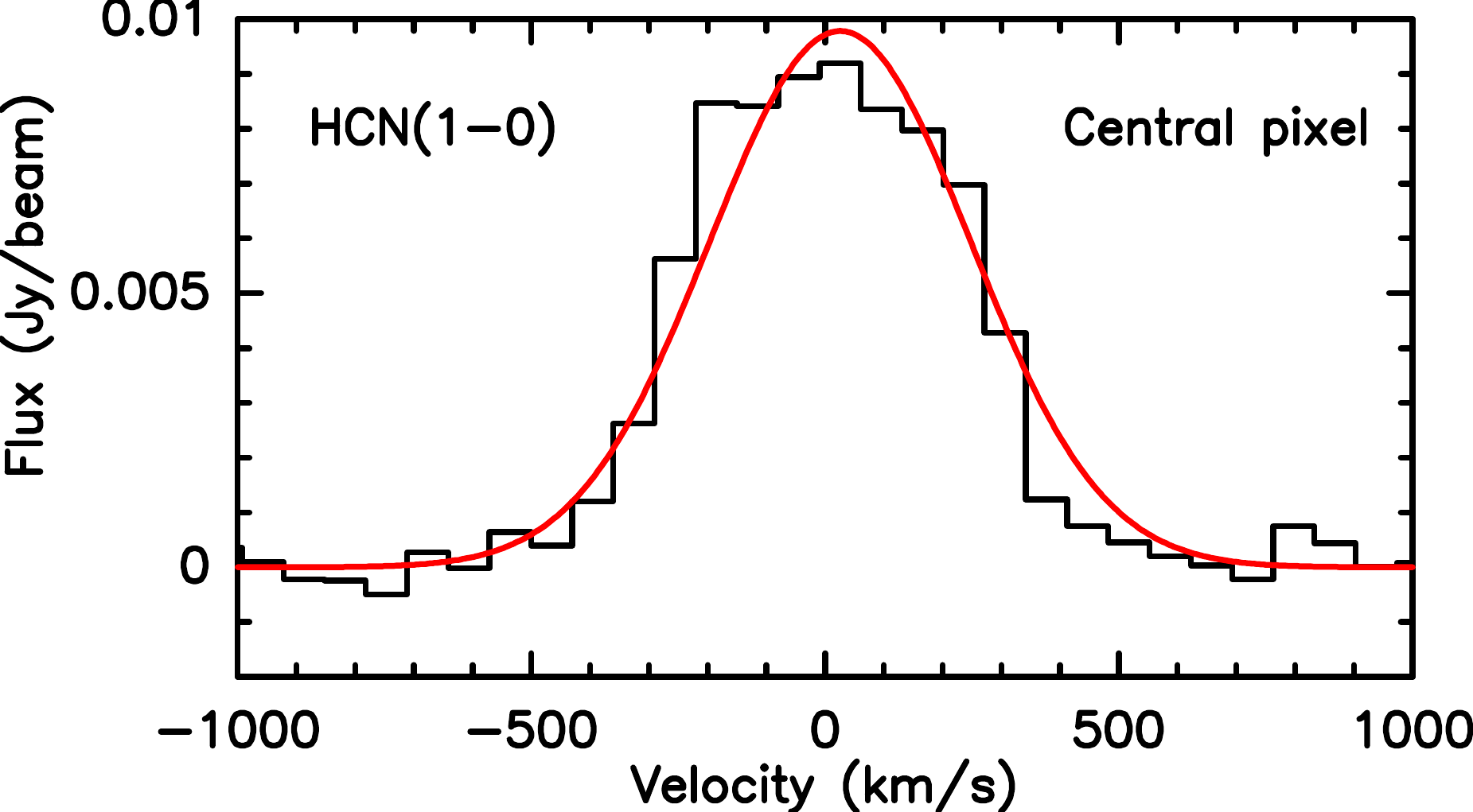}
        \end{subfigure}\hspace*{5mm}
\vspace*{2mm}
        \begin{subfigure}{0.3\textwidth}
                \includegraphics[width=\textwidth]{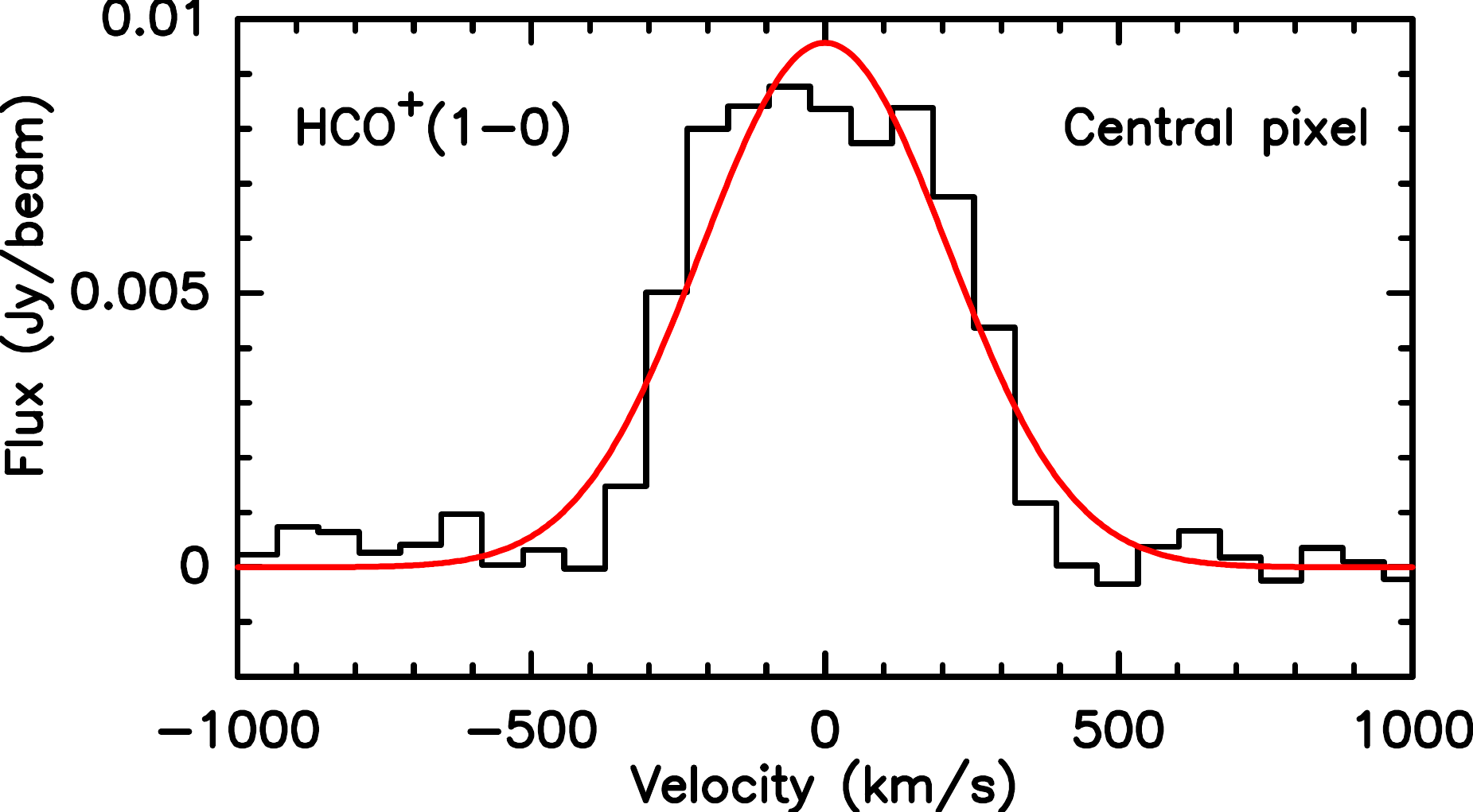}  
        \end{subfigure}\hspace*{\fill}
                \begin{subfigure}{0.3\textwidth}
                        \includegraphics[width=\textwidth]{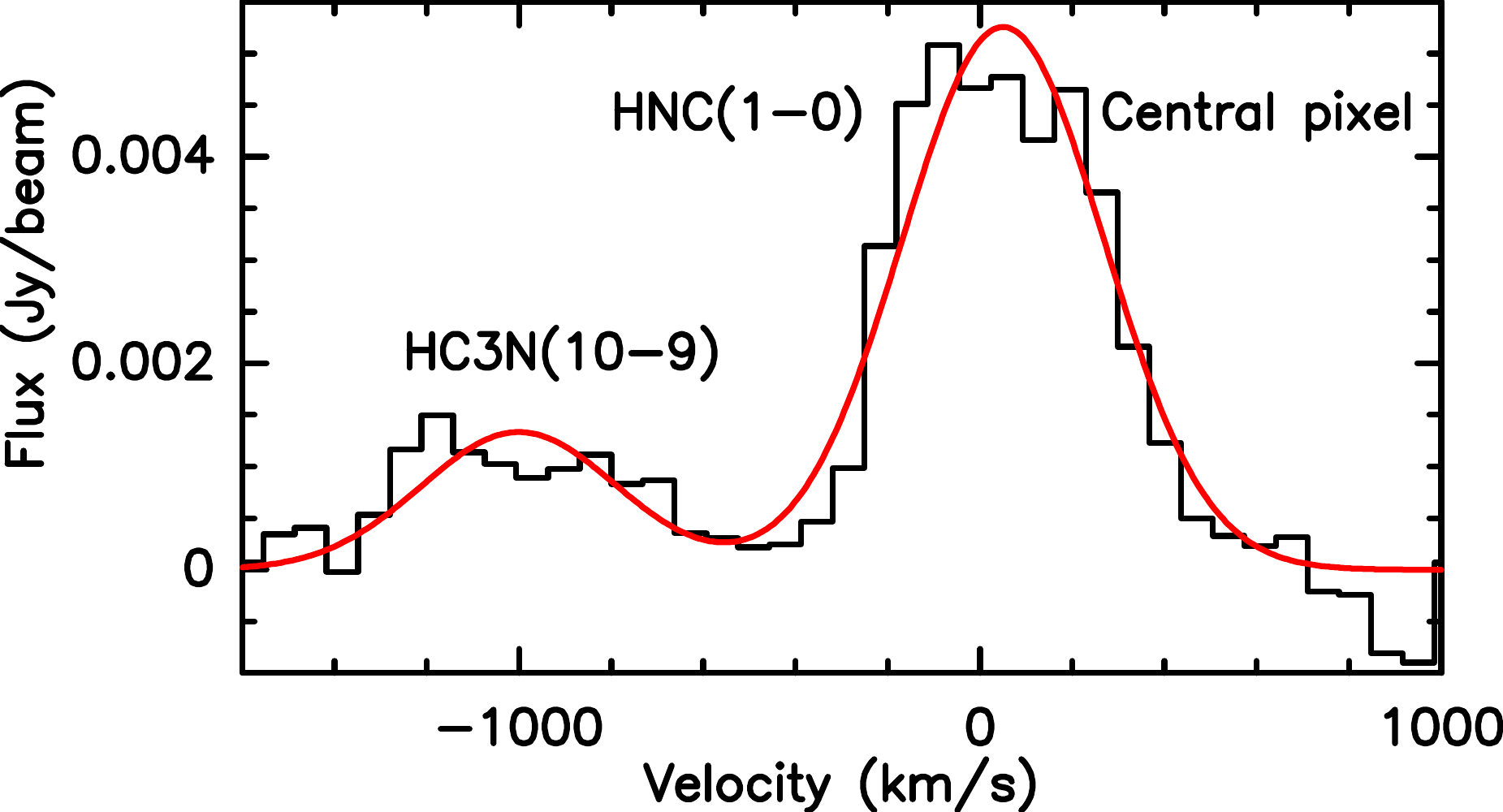}  
                \end{subfigure}\hspace*{\fill}          
                        \vfill{}        
        \begin{subfigure}{0.3\textwidth}
        \includegraphics[width=\textwidth]{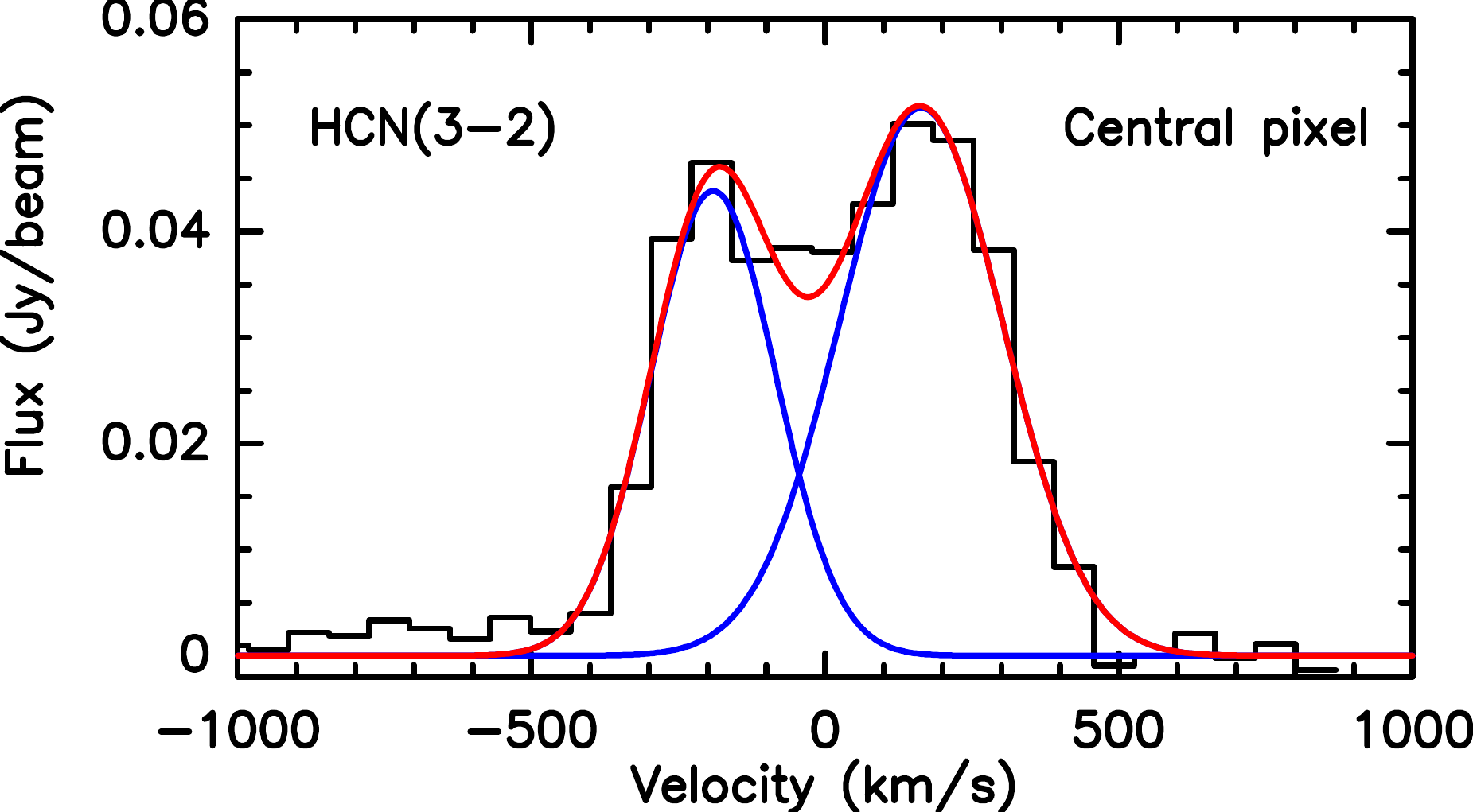}
        \end{subfigure}\hspace*{5mm}
        \begin{subfigure}{0.3\textwidth}
        \includegraphics[width=\textwidth]{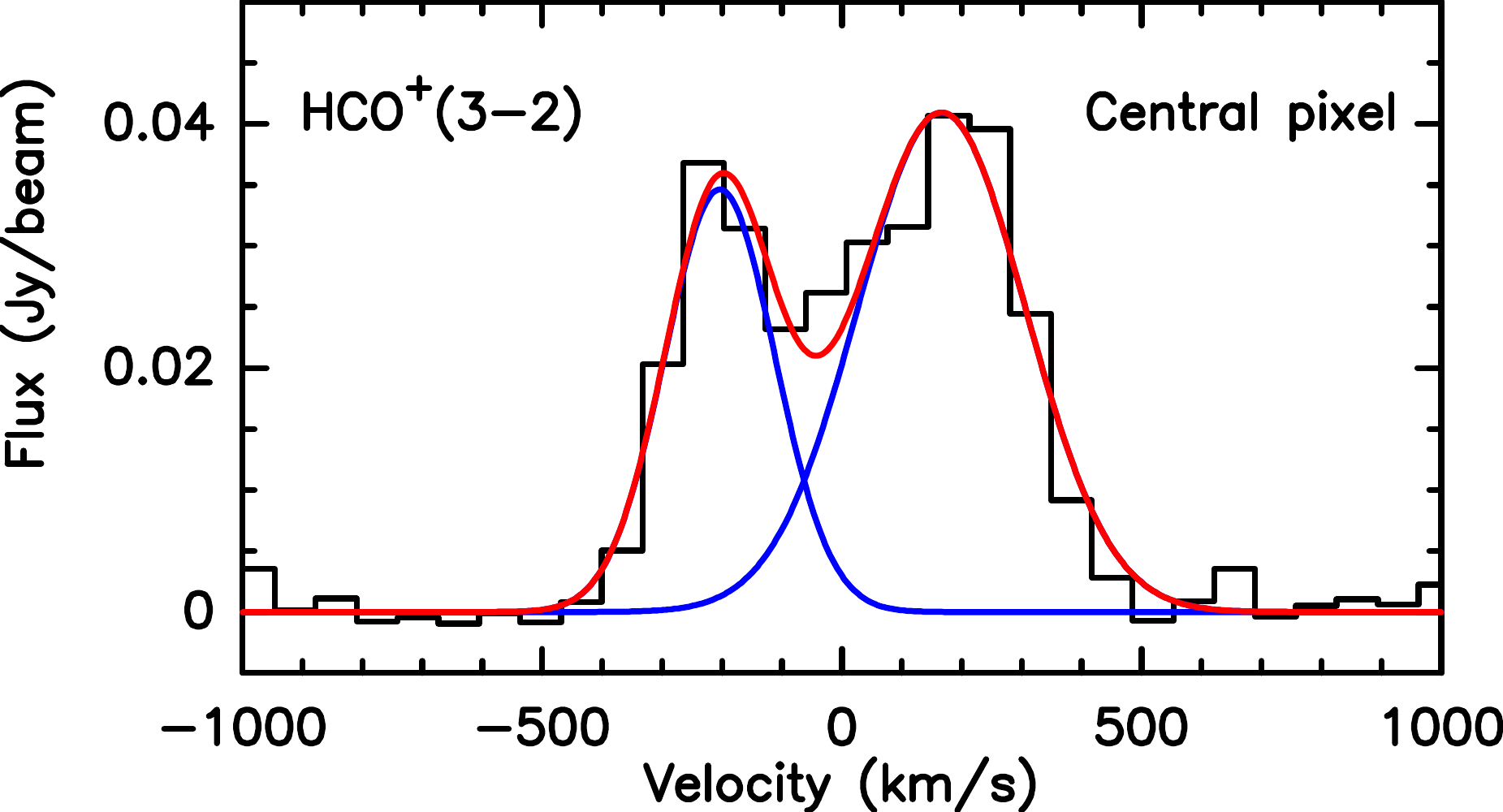}  
        \end{subfigure}\hspace*{\fill}
        \begin{subfigure}{0.3\textwidth}
        \includegraphics[width=\textwidth]{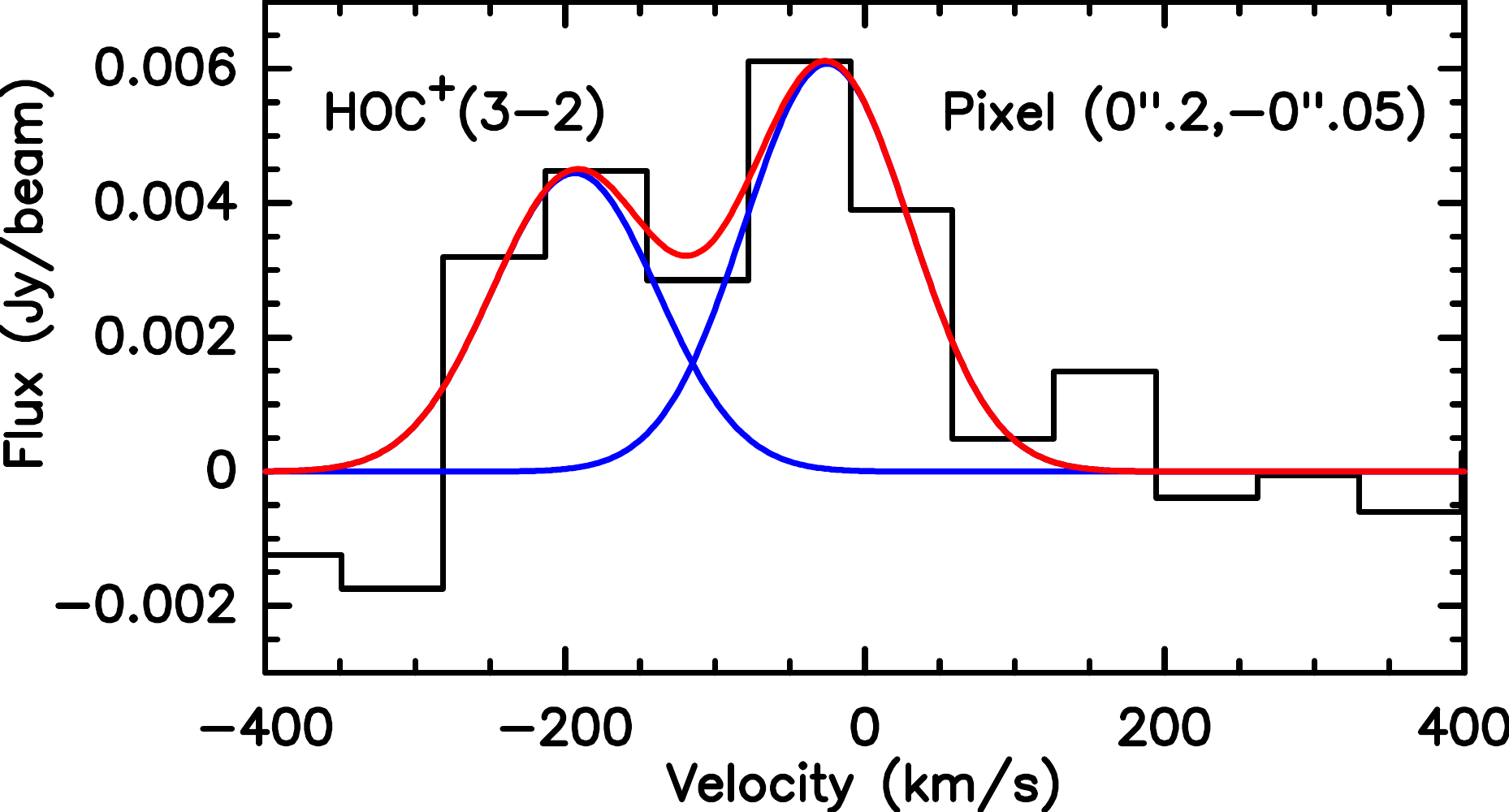}                 
        \end{subfigure}\hspace*{\fill}
        \caption{Emission lines detected with NOEMA (black histograms) and Gaussian fits (total:   red lines; if two components are present, individual components are displayed in blue colour).  The velocity resolution is 68\,km\,s$^{-1}$ in all cases. The labels in the top right corners indicate if the spectra were extracted from  the integrated emission (``Integrated'', top panels),  from the central pixel (bottom panels), or  in the case of HOC$^+$, from the pixel at (0$\ffas$2, $-$0$\ffas$05).}
        \label{fig2}
  \end{figure*}

  \begin{figure*}[ht!]
        \centering
        \begin{subfigure}{0.3\textwidth}
                \includegraphics[width=\textwidth]{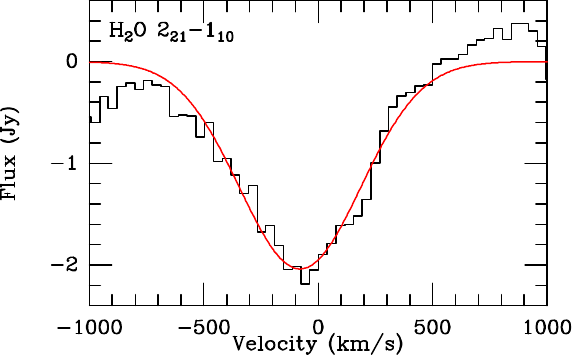}
        \end{subfigure}\hspace*{5mm}
  \vspace*{2mm}
        \begin{subfigure}{0.3\textwidth}
                \includegraphics[width=\textwidth]{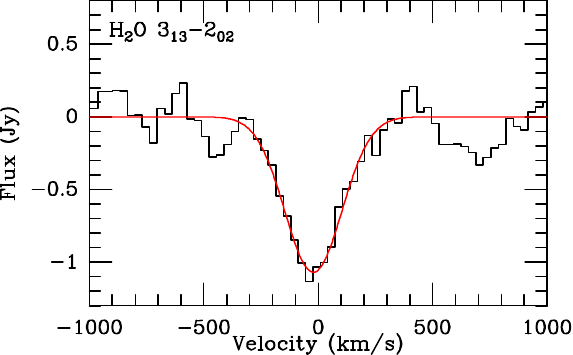}  
        \end{subfigure}\hspace*{\fill}
        \begin{subfigure}{0.3\textwidth}
                \includegraphics[width=\textwidth]{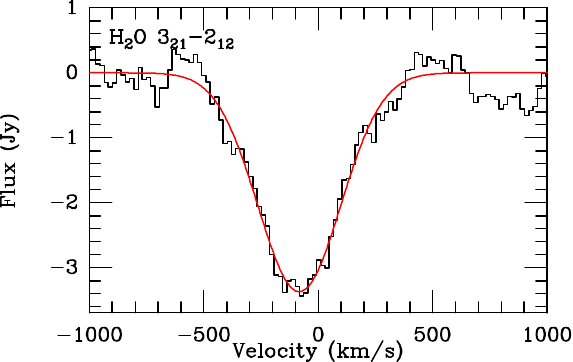}                 
        \end{subfigure}\hspace*{\fill}  
        \vfill{}                        
        \begin{subfigure}{0.3\textwidth}
                \includegraphics[width=\textwidth]{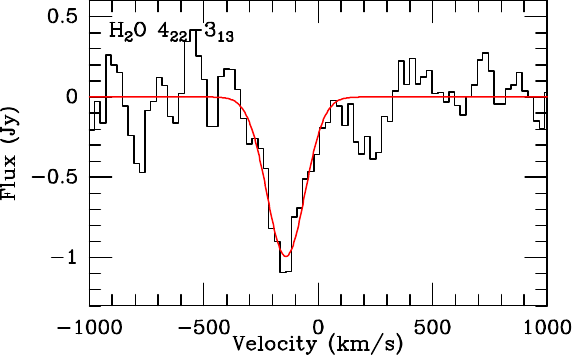}                 
        \end{subfigure}\hspace*{\fill}
        \begin{subfigure}{0.3\textwidth}
        \includegraphics[width=\textwidth]{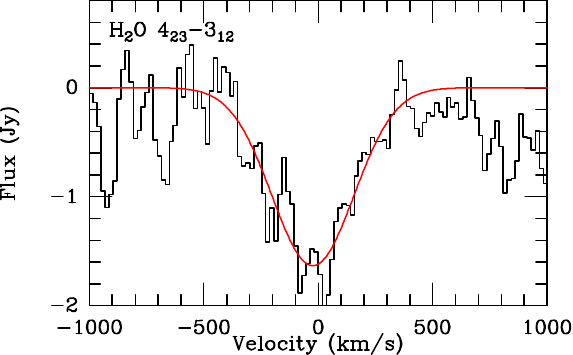}             
  \end{subfigure}\hspace*{\fill}
        \begin{subfigure}{0.3\textwidth}
                \includegraphics[width=\textwidth]{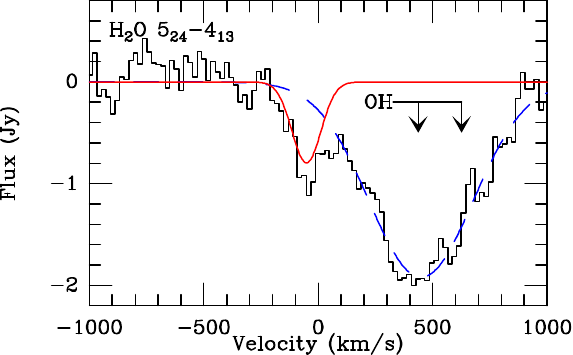}                 
        \end{subfigure}\hspace*{\fill}
        \caption{H$_2$O absorption lines (black histograms) detected with \emph{Herschel}/PACS and Gaussian fits (red lines). The velocity sampling is 20-40\,km\,s$^{-1}$.}
        \label{fig3}
  \end{figure*}

  \begin{figure*}[ht!]
        \centering
        \begin{subfigure}{0.3\textwidth}
                \includegraphics[width=\textwidth]{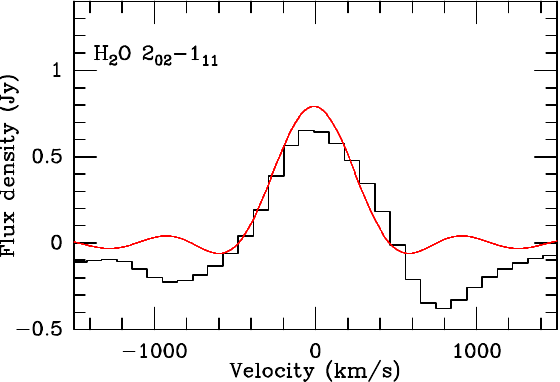}  
        \end{subfigure}\hspace*{\fill}
  \vspace*{2mm}
        \begin{subfigure}{0.3\textwidth}
                \includegraphics[width=\textwidth]{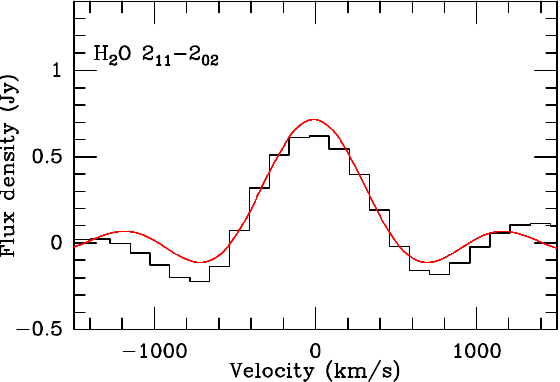}                 
        \end{subfigure}\hspace*{\fill}  
        \begin{subfigure}{0.3\textwidth}
                \includegraphics[width=\textwidth]{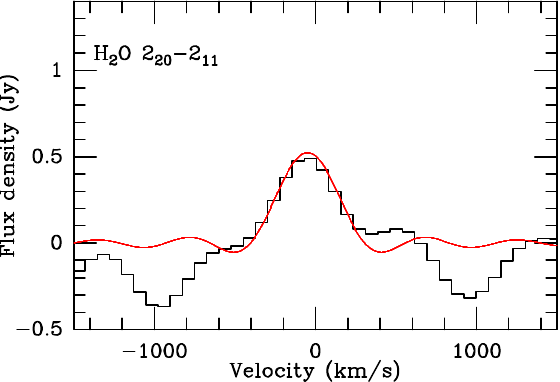}                 
        \end{subfigure}\hspace*{\fill}
  \vfill
        \begin{subfigure}{0.3\textwidth}
                \includegraphics[width=\textwidth]{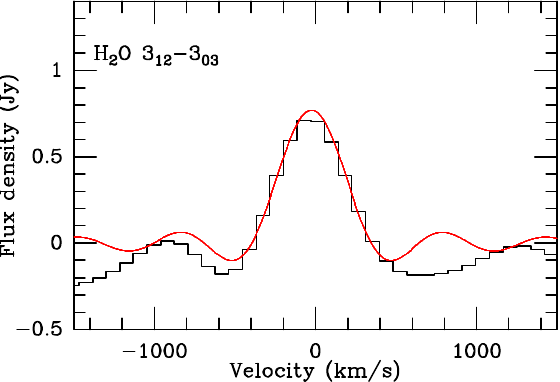}                 
        \end{subfigure}\hspace*{\fill}
        \begin{subfigure}{0.3\textwidth}
                \includegraphics[width=\textwidth]{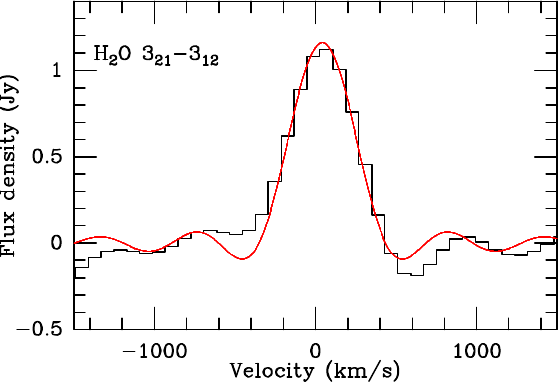}                 
        \end{subfigure}\hspace*{\fill}
        \begin{subfigure}{0.3\textwidth}
        \includegraphics[width=\textwidth]{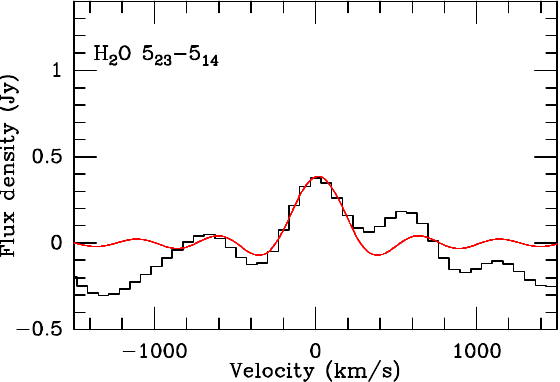}               
\end{subfigure}\hspace*{\fill}
        \caption{H$_2$O  lines (black histograms) detected with \emph{Herschel}/SPIRE, and Gaussians fits (convolved with sinc functions).}
        \label{fig4}
  \end{figure*}

  \begin{table*}
  	\caption{Gaussian fit parameters to the  emission lines detected with NOEMA.}   
  	\centering                        
  	\begin{tabular}{l c c c c c l}       
  		\hline\hline                
  		&Flux& Line peak&       FWHM &Central velocity  & Pixel \\   
  		&[Jy\,km\,s$^{-1}$]\,$^*$ & [mJy]\,$^*$&        [km\,s$^{-1}$] &  [km\,s$^{-1}$]  &\\   
  		\hline                        
  		HCN\,$(1-0)$ & $6.0\pm0.9$  &10.9 &$520\pm85$ &$0\pm0$ &All\\  
  		HCO$^+(1-0)$ & $5.8\pm0.8$  &11.0 &$501\pm78$ &$0\pm0$ &All\\  
  		HNC\,$(1-0)$ & $3.9\pm0.4$  & 7.1&$517\pm59$ &$50\pm0$ &All\\  
  		HC$_3$N\,$(10-9)$ & $1.2\pm0.3$  &2.2 &$500\pm0$ &$0\pm 0$ &All\\         
  		\hline
  		HCN\,$(1-0)$ & $5.4\pm0.8$  &9.8 &$524\pm79$ &$26\pm0$ &(0$''$, 0$''$)\\  
  		HCO$^+(1-0)$ & $5.0\pm0.7$  &9.6 &$495\pm78$ &$0\pm0$ &(0$''$, 0$''$)\\  
  		HNC\,$(1-0)$ & $2.9\pm0.3$  &5.2 &$516\pm51$ &$50\pm0$ &(0$''$, 0$''$)\\  
  		HC$_3$N\,$(10-9)$ & $0.7\pm0.2$  &1.3 &$500\pm0$ &$0\pm 0$ &(0$''$, 0$''$)\\       
  		\hline
  		HCN\,$(3-2)_{\,1}$ & $14.9\pm0.8$  & 63.3&$220\pm179$ &$-198\pm71$ & All\\     
  		HCN\,$(3-2)_{\,2}$& $28.8\pm0.5$  & 76.6& $354\pm170$&$154\pm83$&All\\ 
  		HCO$^+(3-2)_{\,1}$\      &  $13.3\pm0.7$& 58.7& $213\pm136$&$-197\pm66$&All\\ 
  		HCO$^+(3-2)_{\,2}$      &$24.7\pm4.2$&68.7&$338\pm159$ &$140\pm75$ &All\\
  		\hline          
  		HCN\,$(3-2)_{\,1}$& $11.7\pm0.6$  & 43.8&$251\pm130$ &$-190\pm63$ &(0$''$, 0$''$)\\     
  		HCN\,$(3-2)_{\,2}$& $18.0\pm0.1$  &51.7 & $328\pm134$&$164\pm63$&(0$''$, 0$''$)\\ 
  		HCO$^+(3-2)_{\,1}$       &  $8.0\pm1.0$&34.6 & $216\pm0$&$-203\pm92$&(0$''$, 0$''$)\\ 
  		HCO$^+(3-2)_{\,2}$      &$14.4\pm0.6$&40.9&$331\pm199$ &$166\pm89$ &(0$''$, 0$''$)\\               
  		HOC$^+(3-2)_{\,1}$& $0.6\pm0.3$ & 4.4& $130\pm0$&$-197\pm0$&(0.2$''$,-0.05$''$)\\
  		HOC$^+(3-2)_{\,2}$&$1.0\pm0.2$& 7.6&$130\pm0$&$-27\pm0$ &(0.2$''$,-0.05$''$)  \\
  		\hline    
  	\end{tabular}
  	\caption*{Notes: The last column indicates whether the spectra were extracted from only one pixel, or from all pixels showing significant emission.  $^*$: The units of flux and line peak estimated  in a single pixel are  Jy\,km\,s$^{-1}$\,beam$^{-1}$ and mJy\,beam$^{-1}$ , respectively.      For the $(J-J')$=$(3-2)$ lines near $\lambda$=1\,mm, we denote the two Gaussian components with the subscripts ``1'', and ``2''. Parameters with zero errors were fixed.}
  	\label{table1}         
  \end{table*}

  \begin{table*}
  	\caption{Gaussian fit parameters to the water lines detected with \emph{Herschel}. }   
  	\centering                          
  	\begin{tabular}{l c c c c c c c l}       
  		\hline\hline                 
  		&       $\lambda_{\rm rest}$&$E_{\rm low}$ / $E_{\rm upper}${*}&Flux& Line peak&      FWHM &Central velocity   \\   
  		&[$\mu$m]&[K]&[Jy\,km\,s$^{-1}$] & [mJy]&       [km\,s$^{-1}$] &  [km\,s$^{-1}$]  &\\   
  		\hline                       
  		H$_2$O\,$(2_{2,1}-1_{1,0})$  &108.073&61& $-1356\pm87$& $-2.0\pm0.1$& $625\pm48$& $-79\pm21$ & \\  
  		H$_2$O\,$(3_{1,3}-2_{0,2})$  &138.528&101&$-334\pm46$  &$-1.1\pm0.1$ &$294\pm49$ & $-21\pm20$  & \\      
  		H$_2$O\,$(3_{2,1}-2_{1,2})$ & 75.381&114& $-1531\pm60$&$-3.4\pm0.2$ & $427\pm19$&  $-82\pm8$ & \\                   
  		H$_2$O\,$(4_{2,2}-3_{1,3})$& 57.636& 205& $-240\pm62$  & $-1.0\pm0.1$& $215\pm70$& $-140\pm26$& \\    
  		H$_2$O\,$(4_{2,3}-3_{1,2})$&78.742 & 275& $-733\pm138$  &$-1.6\pm0.2$ & $422\pm94$ &  $-23\pm38$ & \\    
  		H$_2$O\,$(5_{2,4}-4_{1,3})$ & 71.067&396& $-127\pm28$  &  $-0.8\pm0.1$&$150\pm35$ &$-51\pm14$ & \\                   
  		\hline          
  		\hline
  		H$_2$O\,$(2_{0,2}-1_{1,1})$  &303.456&101&         $421\pm74$        &$0.8\pm0.1$&$207\pm41$&$-24\pm75$\,$^*$&\\      
  		H$_2$O\,$(2_{1,1}-2_{0,2})$  &398.643&137&         $408\pm42$        &$0.71\pm0.06$&$164\pm40$&$-26\pm32$\,$^*$&\\      
  		H$_2$O\,$(2_{2,0}-2_{1,1})$  &243.974&196&         $220\pm67$  &$0.5\pm0.1$&$148\pm75$&$-63\pm56$\,$^*$&\\      
  		H$_2$O\,$(3_{1,2}-3_{0,3})$  &273.193&249&         $223\pm52$        &$0.8\pm0.1$&$137\pm38$&$-39\pm38$\,$^*$&\\      
  		H$_2$O\,$(3_{2,1}-3_{1,2})$  &257.795&305&         $545\pm62$  &$1.2\pm0.1$&$171\pm3$&$28\pm25$\,$^*$&\\      
  		H$_2$O\,$(5_{2,3}-5_{1,4})$  &212.526&642&         $128\pm43$  &$0.4\pm0.1$&$65\pm51$&$3\pm54$\,$^*$&\\      
  		\hline                            
  	\end{tabular}
  	\caption*{Notes: Some lines are affected by strong noise that increases the errors in the Gaussian fittings. {*} The first six H$_2$O lines were detected in absorption with PACS, and we show the energy of their lower level. The  last six lines were detected with SPIRE in emission, so we show the energy of their upper level. \,$^*$ Due to the low velocity resolution of the SPIRE instrument $\sim$250-360\,km\,s$^{-1}$, these values are very uncertain and we do not consider them in our discussion.}
  	\label{table2}       
  \end{table*}

\subsubsection{Nuclear emission}
\label{linesprofilesNOEMA}
Figure\,\ref{fig2} shows the  spectra of all observed lines with NOEMA extracted from  the whole region (top panels), and  from the central pixel of the observations (bottom panels). The total emission was integrated using  masks in the moment zero maps of the HCO$^+(1-0)$  and $(3-2)$ lines, which show the most extended emission at 1\,mm and 3\,mm (Table\,\ref{table3}). These masks were also used to integrate the emission of the more compact species observed at similar frequencies (i.e. one mask over HCO$^+(1-0)$ for all 3\,mm species, and other mask over HCO$^+(3-2)$ for all 1\,mm species).  
While HOC$^+(1-0)$ is not detected,  the  $(3-2)$ transition is seen arising from a very compact region near the nucleus. When integrating the overall flux of the map, the emission of the HOC$^+(3-2)$ line  drops below the  noise and, therefore, for comparison with the other molecules, we measure its flux in the  pixel where it peaks (see Sect.\ref{momentmaps} for details). There, HOC$^+(3-2)$ is detected with a signal-to-noise  ratio (S/N) of seven. Additionally, its central velocity is blue-shifted  compared to the other lines observed with NOEMA (Table\,\ref{table1}).

 The 3\,mm lines (HCN, HCO$^+$, HNC and HOC$^+(1-0)$, and HC$_3$N$(10-9)$)  have  roughly Gaussian-like profiles, although their peaks are slightly flat-topped.  These lines were fitted with single Gaussian velocity profiles (Fig.\,\ref{fig2}). On the other hand, the 1\,mm lines (HCN, HCO$^+$ and HOC$^+(3-2)$) are double-peaked.  To account for these profiles, we fitted two Gaussians (their parameters are listed in Table~\ref{table1}).   The intensity of the dip between the double peaks is 19\,mJy and 26\,mJy for the HCN and HCO$^+(3-2)$ lines, respectively. The dip appears at slightly blue-shifted velocities, specifically at -30\,km\,s$^{-1}$ for HCO$^+(3-2)$, and -60\,km\,s$^{-1}$ for HCN$(3-2)$. The nature of these double-peaked profiles is further discussed in Sect.\,\ref{asymmetry}.

Figures\,\ref{fig3} and \ref{fig4} show the H$_2$O line profiles observed with \emph{Herschel} PACS and SPIRE and the best Gaussians  fits. All transitions were fitted by single Gaussian profiles (for the SPIRE lines, the Gaussians were convolved with a sinc function, Sect.\,\ref{obs-Herschel}).   H$_2$O\,$(5_{24}-4_{13})$ is partially blended with an OH line at 71$\mu$m.  To disentangle the emission of the two  species, we fitted a double Gaussian profile to the observed spectrum.

It is important to note that all Gaussian fits to the  water lines observed with PACS are  blue-shifted to velocities between -20\,km\,s$^{-1}$ and -140\,km\,s$^{-1}$ (Table\,\ref{table2}). Interestingly, these values are, to within the errors, the same as the velocities of the dips in the profiles of HCO$^+$ and HCN$(3-2)$. The connection between the two is discussed in Sect.\,\ref{asymmetry}.

\begin{figure}
        \centering
        \includegraphics[width=0.5\textwidth]{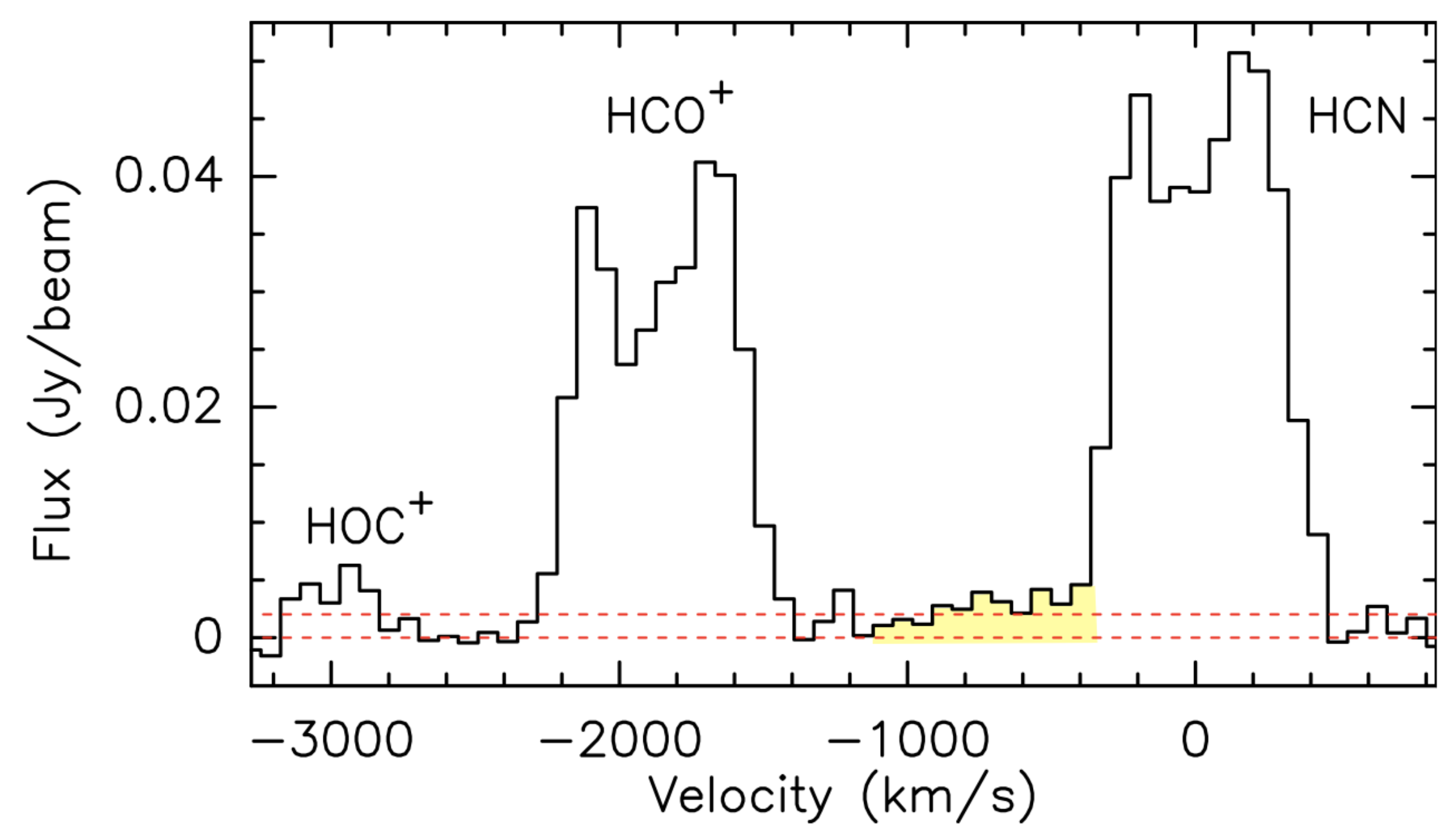}   
        \caption{Spectrum of the central pixel for the $(J-J')$=$(3-2)$ lines of HCN, HCO$^+$ and HOC$^+$. The baseline of order 0  and the $3\times {\rm rms}$  flux (calculated at the final velocity resolution of 68\,km\,s$^{-1}$) are marked with  horizontal dashed lines. The outflow emission at the blue-shifted velocities of HCN is highlighted in yellow.}
        \label{fig5}%
\end{figure}

\subsubsection{High-velocity emission}
\label{high-vel-emission}

None of the 3\,mm and FIR water lines show obvious extended line wings, which would reveal  the Mrk\,273 molecular outflow previously detected with CO, OH, and H$_2$ by \citet{U}, \citet{Veilleux13}, \citet{Cicone14},  and \citet{Gonzalez17}.
For the 1\,mm observations, the spectrum of the overall integrated emission has no evident signatures of line wings either. However, in the central pixel, the  HCN profile exhibits a line wing that extends between -400 and $\sim$1000\,km\,s$^{-1}$, while the red side shows no wing (Fig.\,\ref{fig5}).  This emission is detected with an S/N of $\sim$5.

 We used the JPL catalog \citep{Pickett98} to look for  lines arising from 266.2 to 266.9\,GHz, which correspond to the velocity range  [-350,-1200]\,km\,s$^{-1}$ where the HCN$(3-2)$ wing-like feature is seen. CH$_2$NH\,$(4_{1,3}-3_{1,2})$ is the most likely line arising at these frequencies, with an energy level of $E_{\rm low}$=19\,K. This transition was  detected in the LIRGs  IC\,860 and Zw\,49-57, where its flux density is three to four times fainter than HCN$(3-2)$ \citep{Aalto15}. Assuming a similar ratio and  excitation conditions in Mrk\,273, CH$_2$NH would then have a peak flux between 4.5 and 6\,mJy, that is, brighter than the emission we see. The HCN shoulder does not have a (single or double) Gaussian profile similar to the detected lines, but has the shape of a line wing. For these reasons, it seems unlikely that  the emission comes from the CH$_2$NH line, although a potential contamination cannot be ruled out.

The two nuclei of Arp\,220 have HCN\,$(3-2)$  and $(4-3)$ blue-shifted wings very similar to what we observe here in Mrk\,273 \citep{Martin16}. The fact that these bumps appear in both HCN transitions, while there is no corresponding CH$_2$NH line  close to the HCN$(4-3)$ frequency, strengthens our claim that the line wing  in Mrk\,273  comes from HCN$(3-2)$, and that it is tracing gas outflowing at high velocities. In summary, given the feature intensity, spectral shape, and integrated emission (see following section), we conclude that the HCN$(3-2)$ blue-shifted shoulder comes from the outflowing gas moving at approximate velocities between -400 and -1000\,km\,s$^{-1}$.  In Sects.\,\ref{mom0outflow} and \ref{outflow} we discuss in more detail the properties of this molecular outflow.

\begin{figure*}
        \centering
        \begin{subfigure}{0.3\textwidth}
                \includegraphics[width=\textwidth]{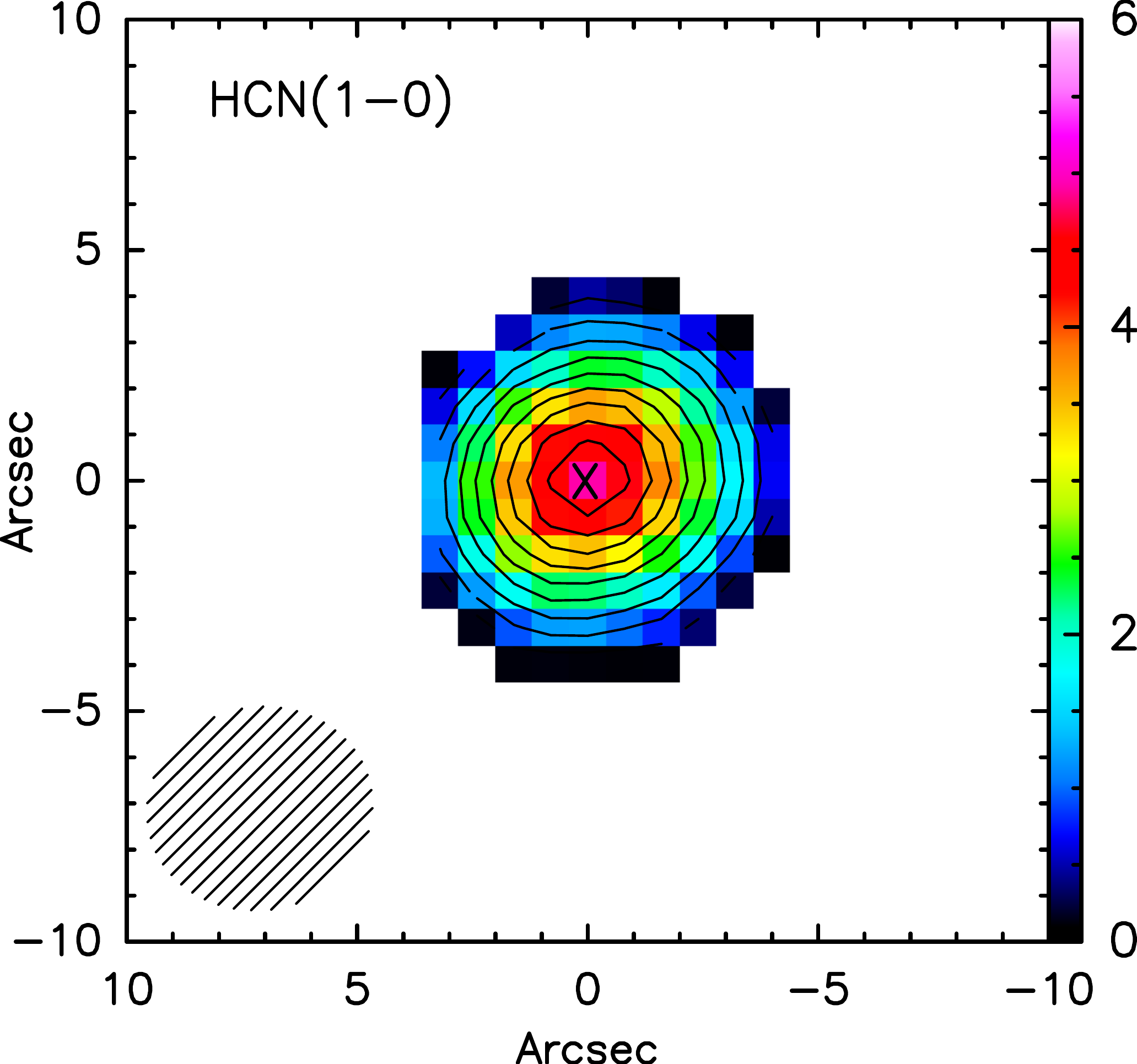}
        \end{subfigure}\hspace*{\fill}
        \vspace{5mm}
        \begin{subfigure}{0.3\textwidth}
                \includegraphics[width=\textwidth]{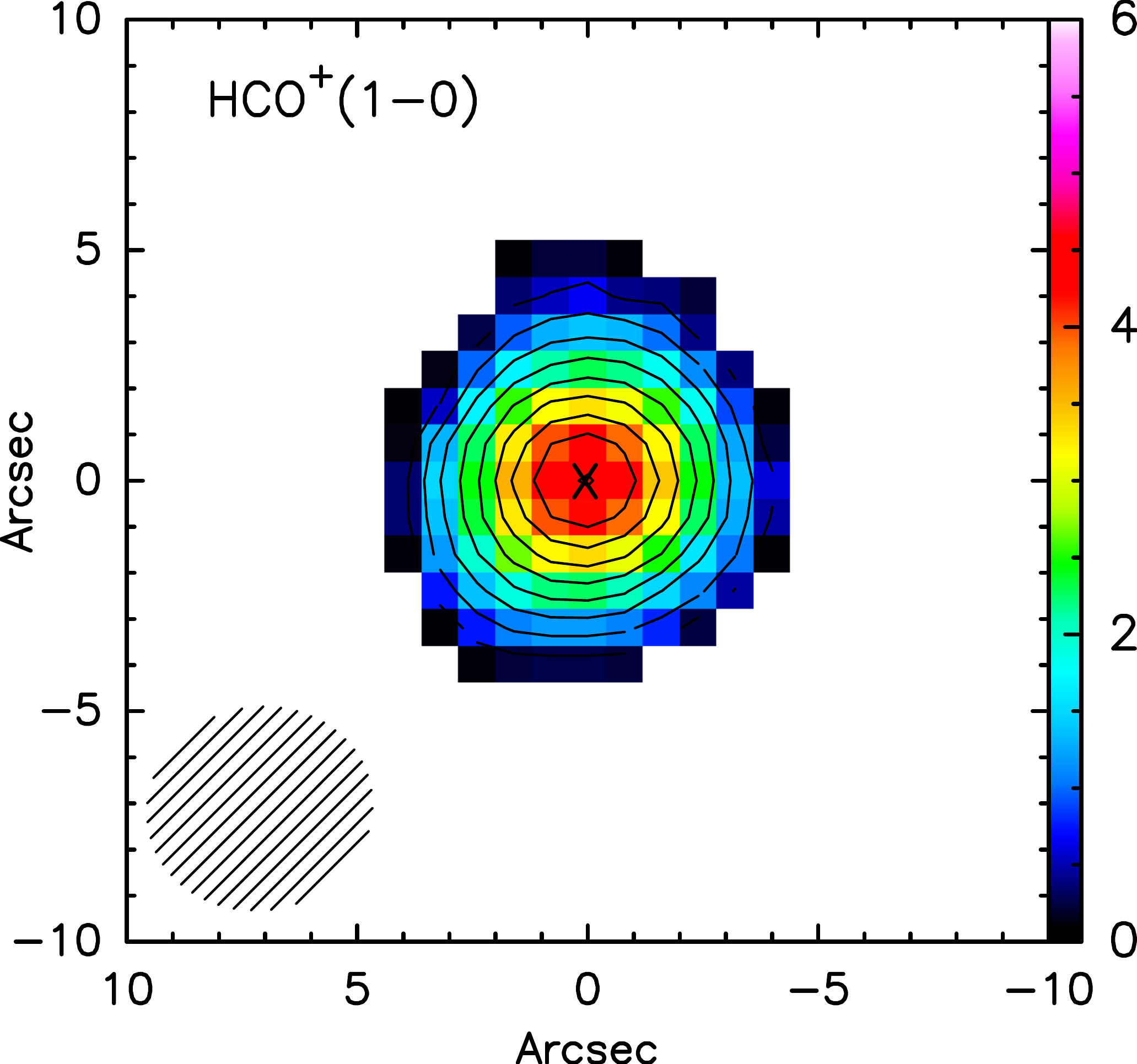}
        \end{subfigure}\hspace*{\fill}
        \begin{subfigure}{0.3\textwidth}
                \includegraphics[width=\textwidth]{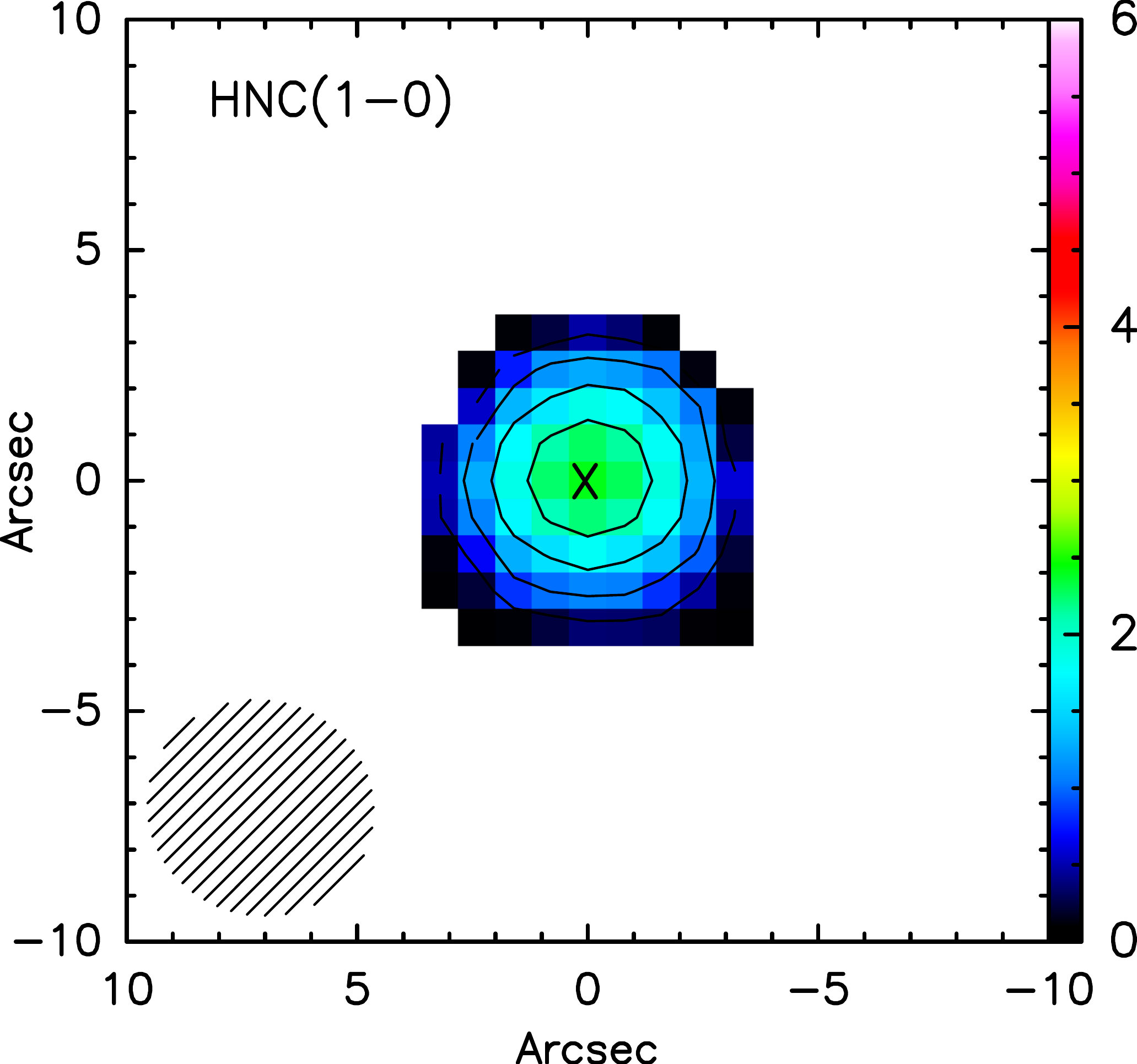}
        \end{subfigure} \hspace*{\fill} 
        \begin{subfigure}{0.3\textwidth} 
                \includegraphics[width=\textwidth]{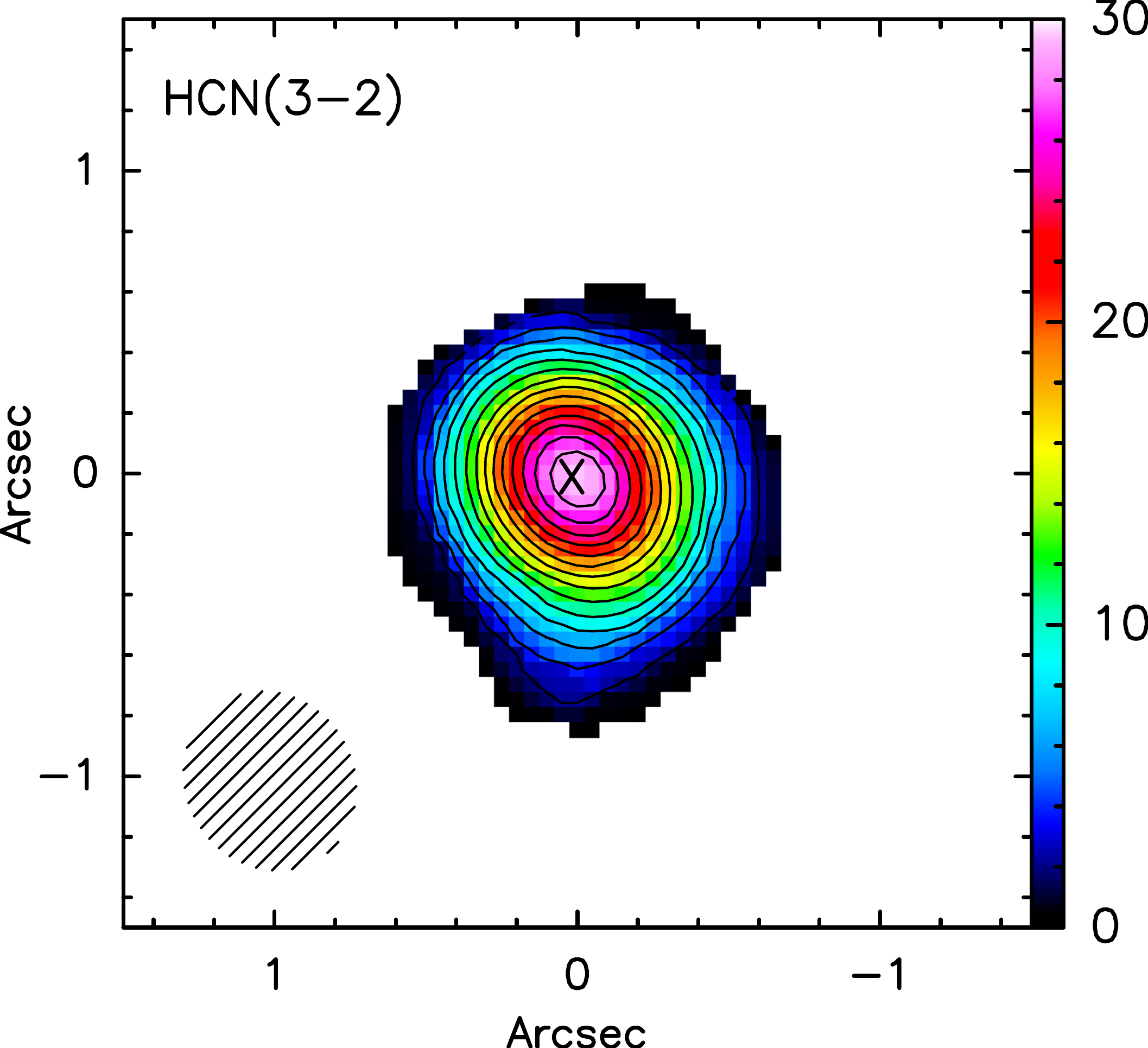}        
        \end{subfigure}\hspace*{\fill}
        \begin{subfigure}{0.3\textwidth}
                \includegraphics[width=\textwidth]{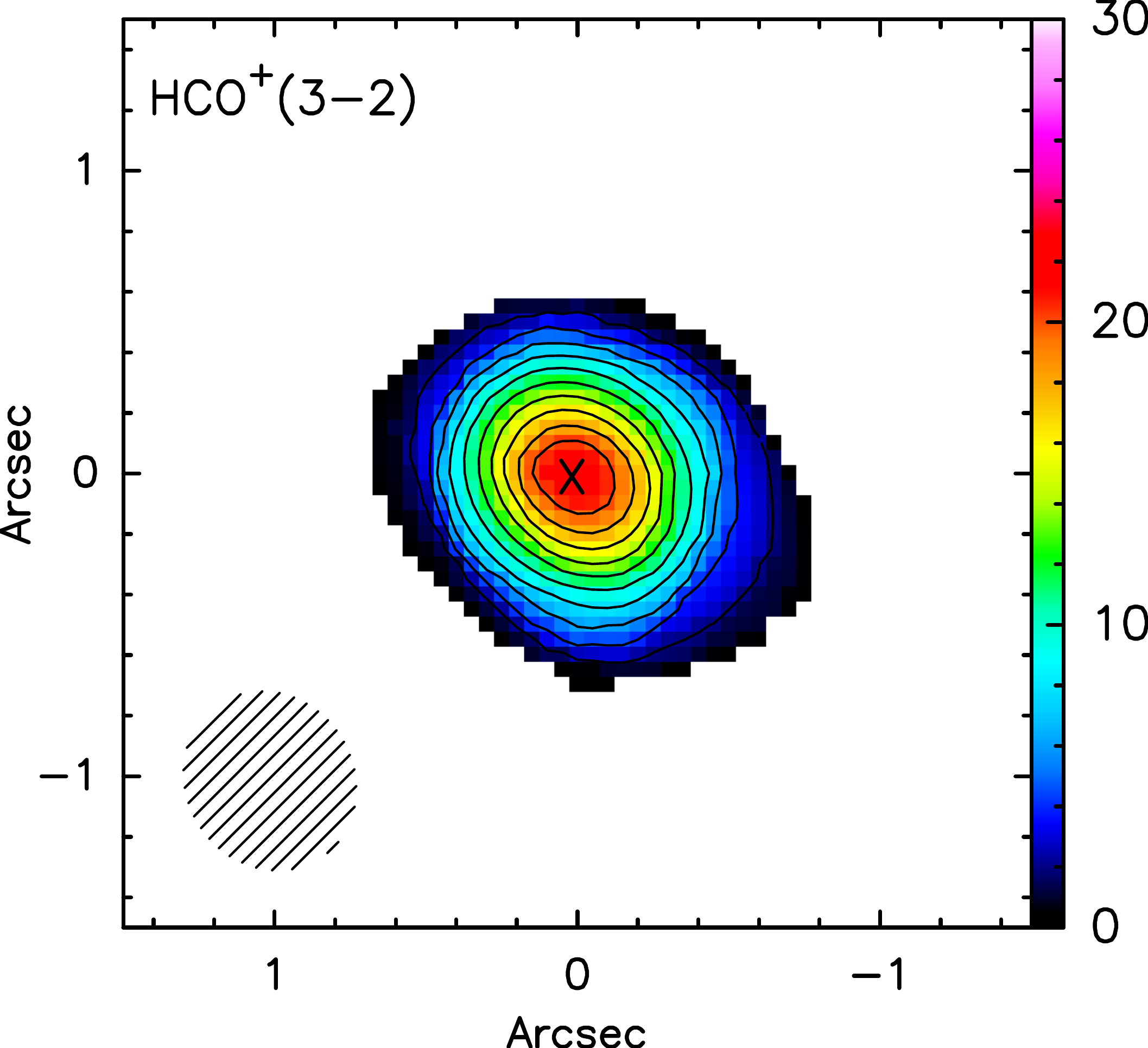}       
        \end{subfigure} \hspace*{\fill} 
        \begin{subfigure}{0.3\textwidth}        
                \includegraphics[width=\textwidth]{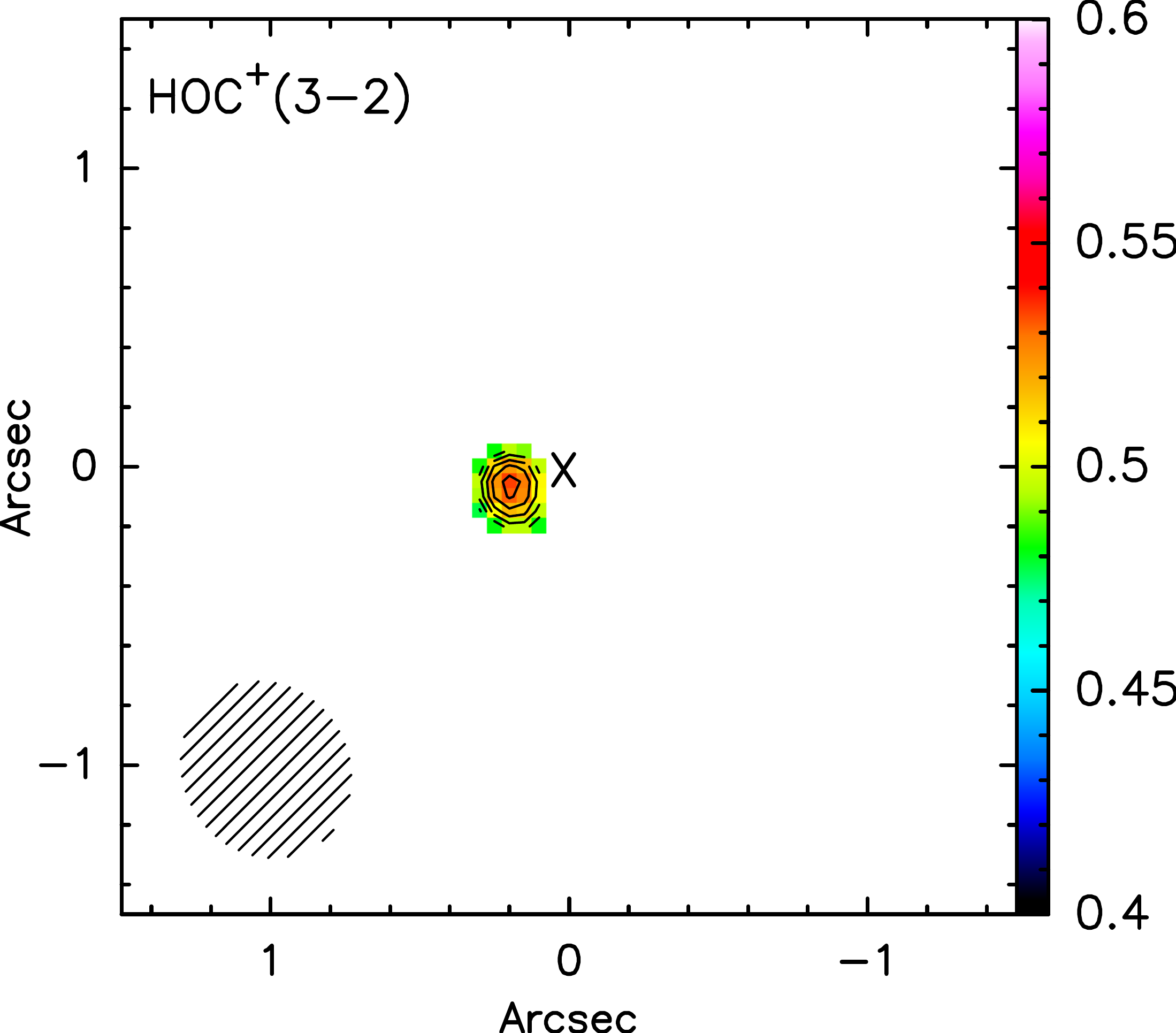}                  
        \end{subfigure}\hspace*{\fill}
        \caption{Integrated intensities (moment zero maps). Contour levels for HCN, HCO$^+$ and HNC\,$(1-0)$ go from 0 to 6\,Jy\,kms$^{-1}$beam$^{-1}$ with a step of 0.5\,Jy\,kms$^{-1}$beam$^{-1}$. For HCN\,$(3-2)$ and HCO\,$^+(3-2)$  the levels range from 2 to 30\,Jy\,kms$^{-1}$beam$^{-1}$ with a step of 2\,Jy\,kms$^{-1}$beam$^{-1}$. For HOC$^+(3-2)$, the contours are from 0.4 to 0.6\,Jy\,kms$^{-1}$beam$^{-1}$  with a step of 0.04\,Jy\,kms$^{-1}$beam$^{-1}$. We highlight the different  scales between the $(1-0)$ and the $(3-2)$ lines. The crosses in the centre of each panel indicate the location of the nuclear source. The synthesised beam is shown in the bottom-left corner of each panel. North is up, and east is to the left.}
        \label{fig6}%
\end{figure*}

\begin{figure}[ht!]
        \centering
\includegraphics[width=0.45\textwidth]{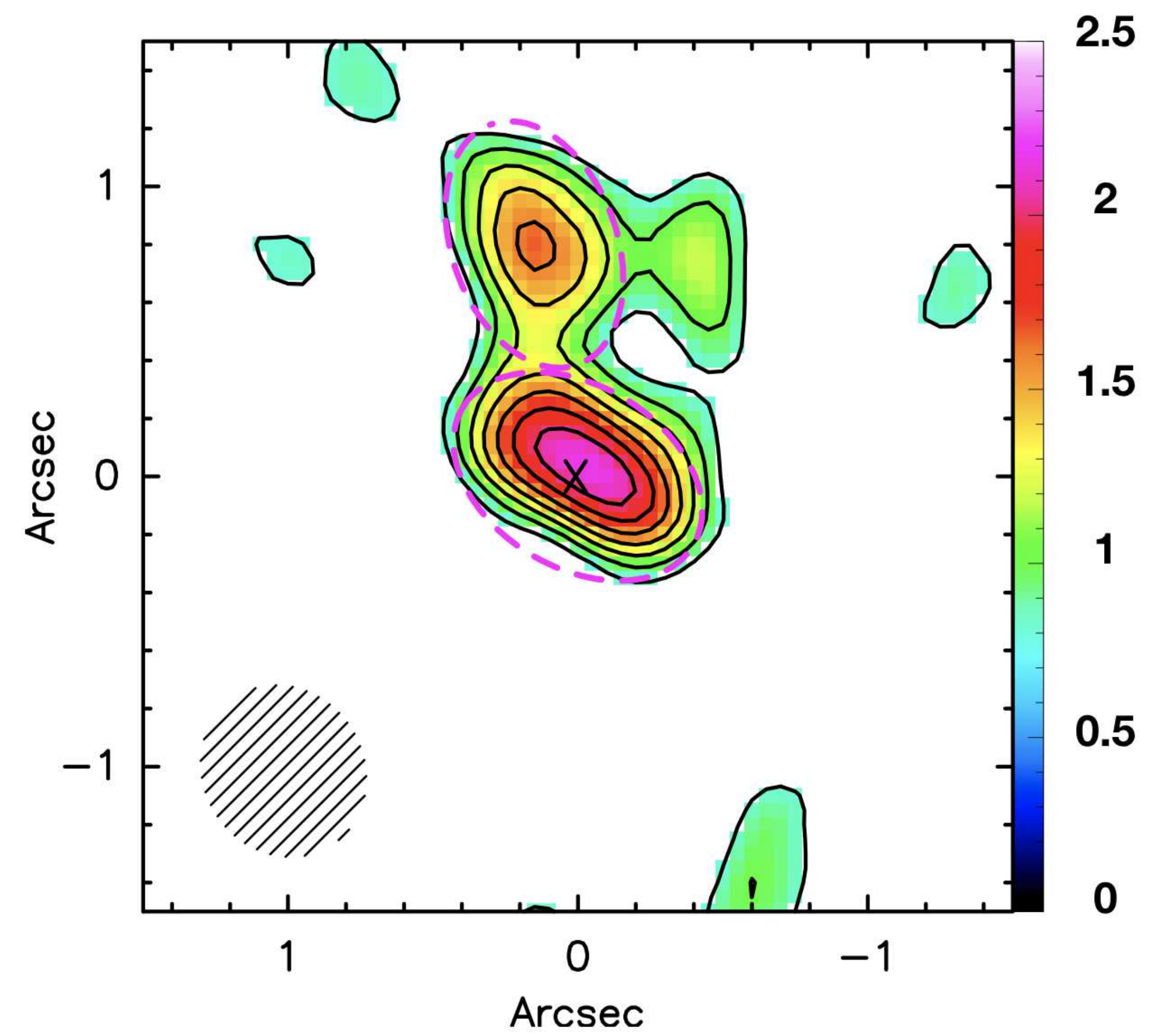}
        \caption{Integrated intensity of the outflow seen in HCN\,$(3-2)$ between -1100\,km\,s$^{-1}$ and -400\,km\,s$^{-1}$ (yellow region marked in Fig.\,\ref{fig5}). The cross in the centre marks the location of the nuclear source. The contours start with a 5$\sigma$ flux and go from 0.8\,mJy\,kms$^{-1}$\,beam$^{-1}$ to 2.5\,Jy\,kms$^{-1}$\,beam$^{-1}$ with steps of 0.2\,mJy\,kms$^{-1}$\,beam$^{-1}$. The magenta dashed ellipses show the regions fitted to the two main components. The synthesised beam is plotted in the bottom-left corner.}
        \label{fig7}%
\end{figure}

\subsection{Moment maps of HCN, HCO$^+$, HNC and HOC$^+$}
\label{momentmaps}
\subsubsection{Position and size of the nuclear emission}
\label{intint}
Figure\,\ref{fig6} and Table\, \ref{table1} show the integrated intensities of the HCN, HCO$^+$, HNC and HOC$^+$ lines. The deconvolved emission sizes and position angles of all lines were measured in the uv plane by fitting elliptical Gaussians with the task \emph{uv\_fit} within MAPPING, and are  listed in Table\,\ref{table3}.
The continuum and the HCN, HCO$^+$,  and HNC lines have their peak intensities at the central pixel, which we take as the location of the northern nuclear source.  On the contrary, the HOC$^+$  maximum is found at an offset  $(0\ffas2,-0\ffas05)$, corresponding to $(152, -38)$\,pc to the south-east. The  position accuracy of our observations can be calculated from

\begin{equation}
\Delta \alpha, \Delta \delta  \simeq 0.6\times(SNR)^{-1} \times \theta_b,
\end{equation}

where $\Delta \alpha, \Delta \delta$ are the errors in RA and DEC, and $\theta_b$ is the synthesised beam size \citep{Condon97,Ivison07}. For HOC$^+(3-2)$,  we have a $\Delta \alpha, \Delta \delta \simeq$ 0$\ffas$1 ($\sim$80\,pc), confirming our claim that its shift is real, at least  in RA (because the HOC$^+$ offset  in declination  is smaller than our position accuracy). In Sect.\,\ref{hoc+}, we  further discuss the origin of  HOC$^+$. 

\citet{DS98} estimated a CO\,$(1-0)$  emission deconvolved size of ($0\ffas9\times0\ffas6$) with a beam of ($1\ffas4\times1\ffas3$). We obtain larger sizes for the HCN and HCO$^+(1-0)$ lines ($\sim(2''\times2'')$ equivalent to $\sim$1.5\,kpc, Table\,\ref{table3}), most probably because we collect more emission within our larger beam  of ($4\ffas9\times4\ffas5$). In the following, we refer to this region as the outer disc.
The HCN and HCO$^+(3-2)$ lines, however, are confined to a much smaller region of  ($0\ffas4\times0\ffas3)$ ($\sim$(300$\times230)$\,pc, Table\,\ref{table3}), implying that the denser, star-forming gas is more concentrated in the nucleus. We refer to this as the inner disc. Despite our high angular resolution at 267\,GHz, the nucleus is still unresolved and does not show any structures in the moment zero maps, implying a very compact core that does not expand more than $<0\ffas$3 (230\,pc). Indeed, from our H$_2$O modelling,  we estimate the size of the core to be $\sim$50\,pc (see Sect.\,\ref{modelling-water} for details).

HC$_3$N\,$(10-9)$ and HOC$^+(3-2)$ are  faint (although we detect the latter with an S/N of $\sim$7  in the  pixel where it peaks) and unresolved at our resolution. Therefore, we cannot measure  their emission sizes.

\begin{figure*}
        \centering
        \begin{subfigure}{0.3\textwidth}
    \includegraphics[width=\textwidth]{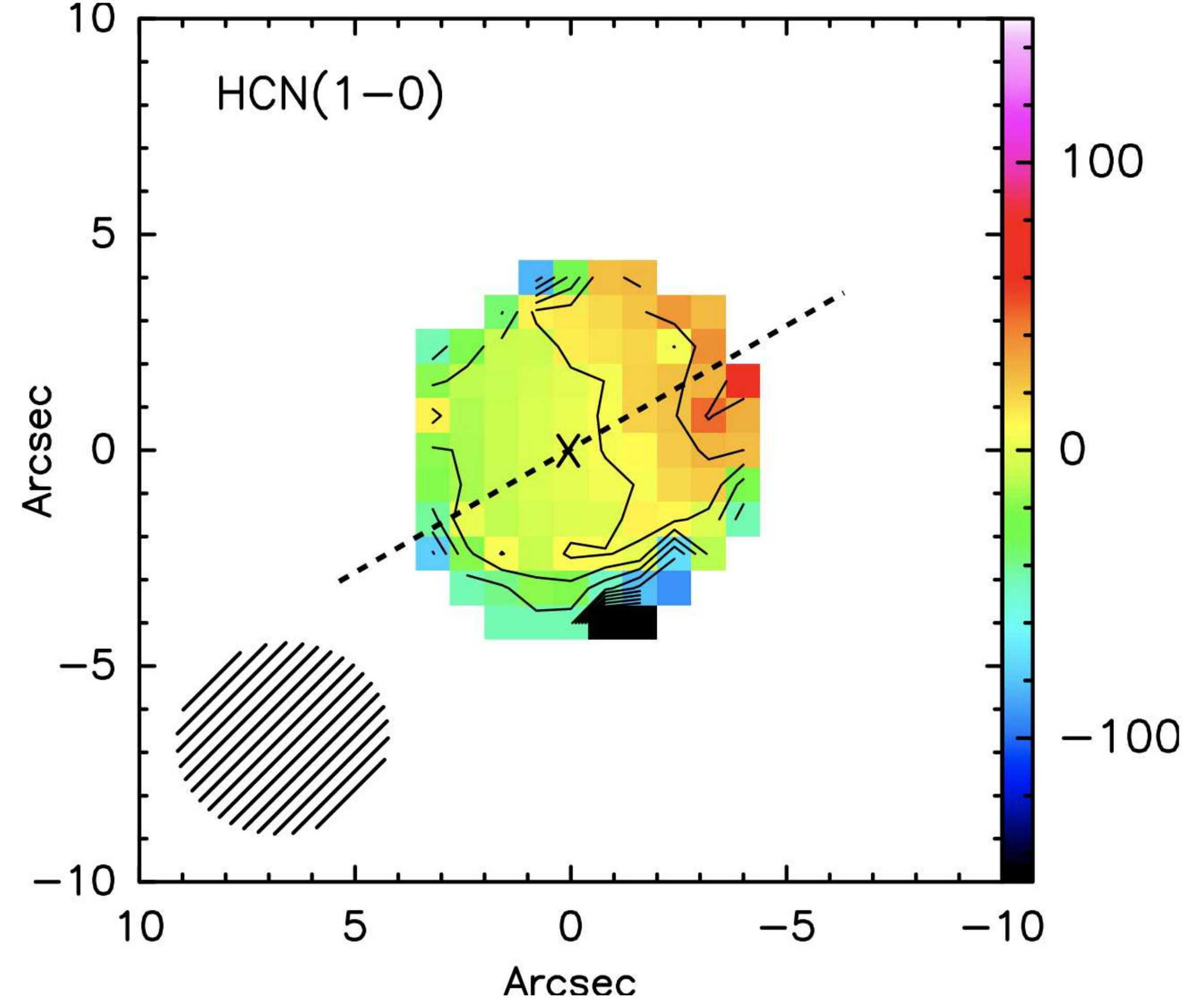}
                \end{subfigure}
        \begin{subfigure}{0.3\textwidth}
        \includegraphics[width=\textwidth]{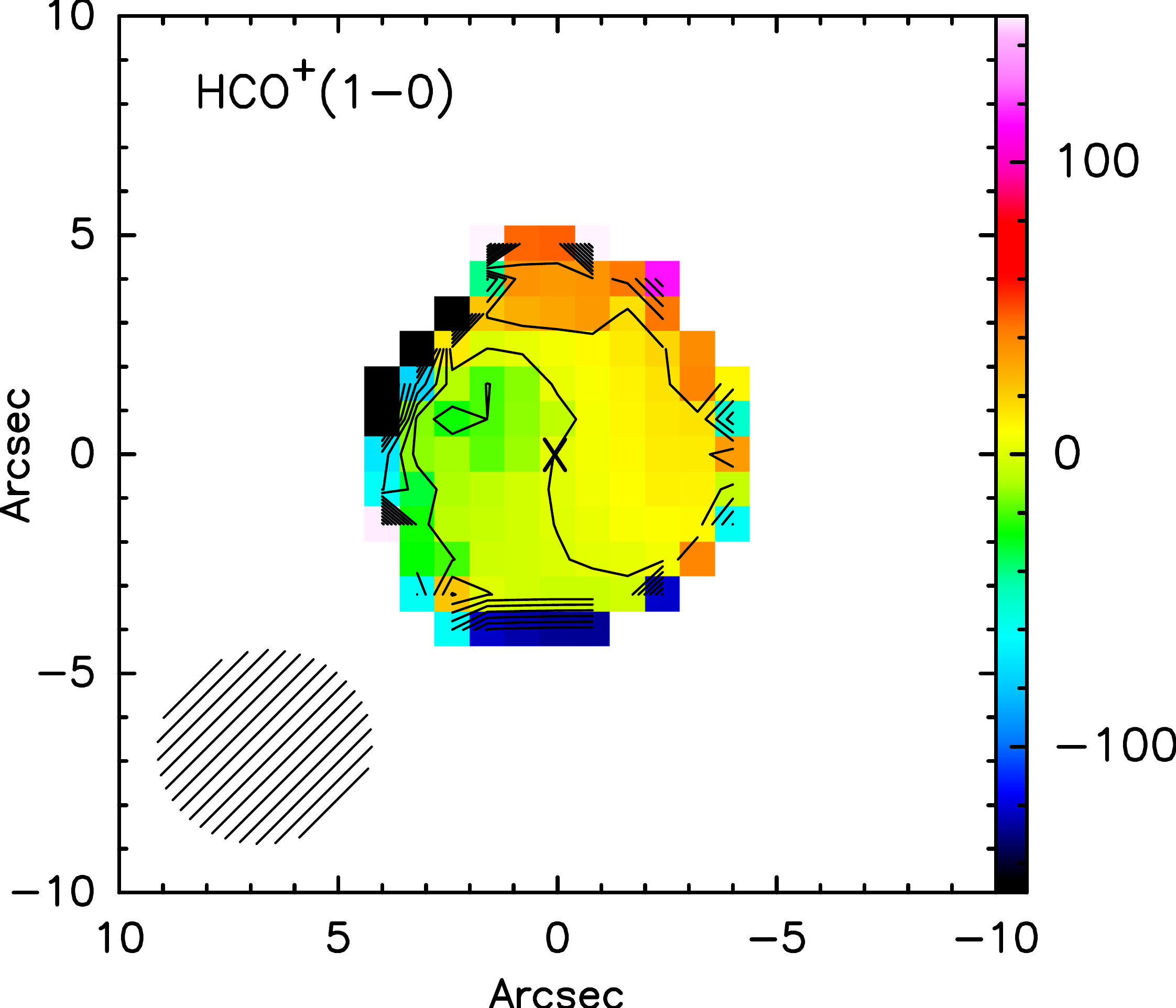}
        \end{subfigure}
        \begin{subfigure}{0.3\textwidth}
        \includegraphics[width=\textwidth]{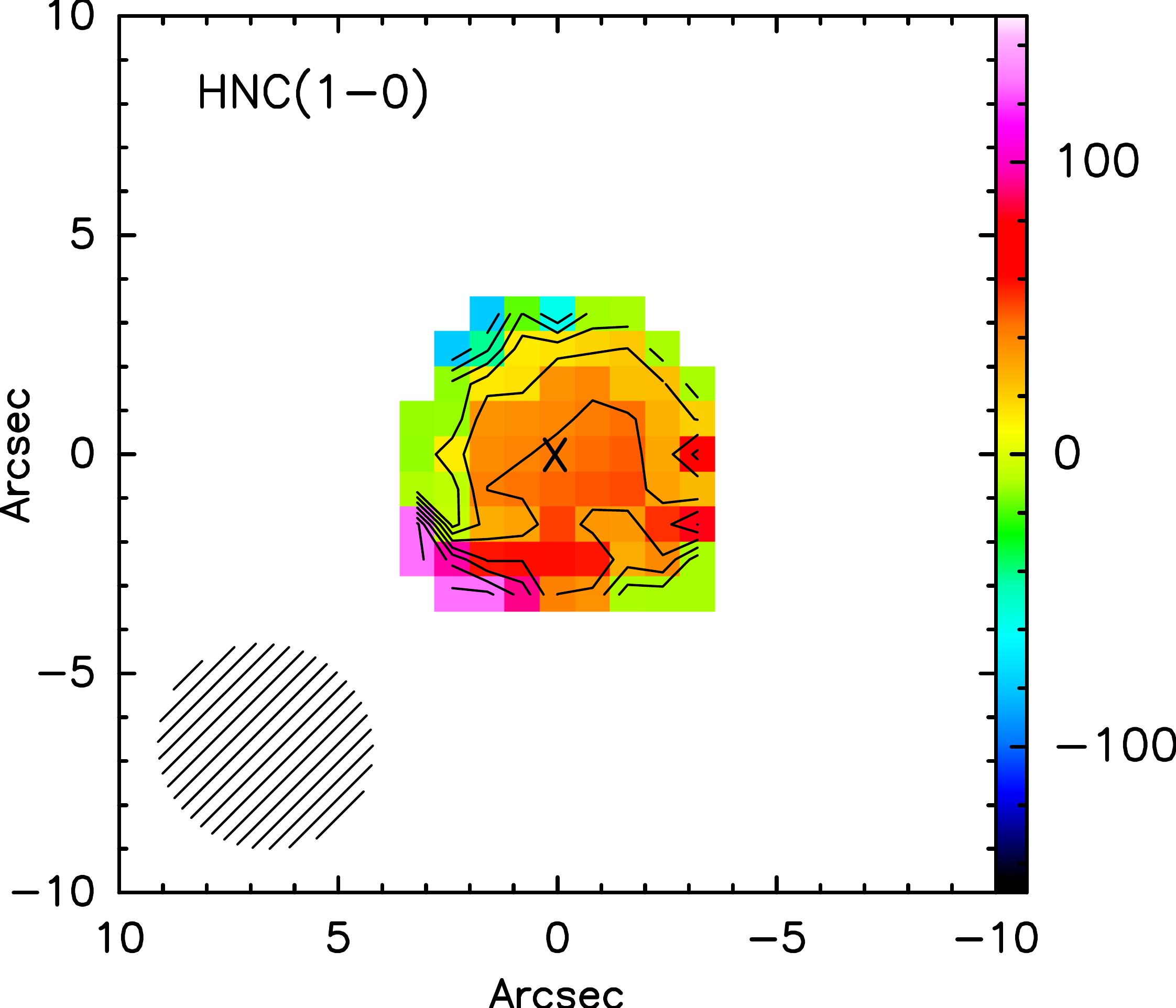}      
        \end{subfigure}
                \vspace{5mm}
        \begin{subfigure}{0.3\textwidth}
     \includegraphics[width=\textwidth]{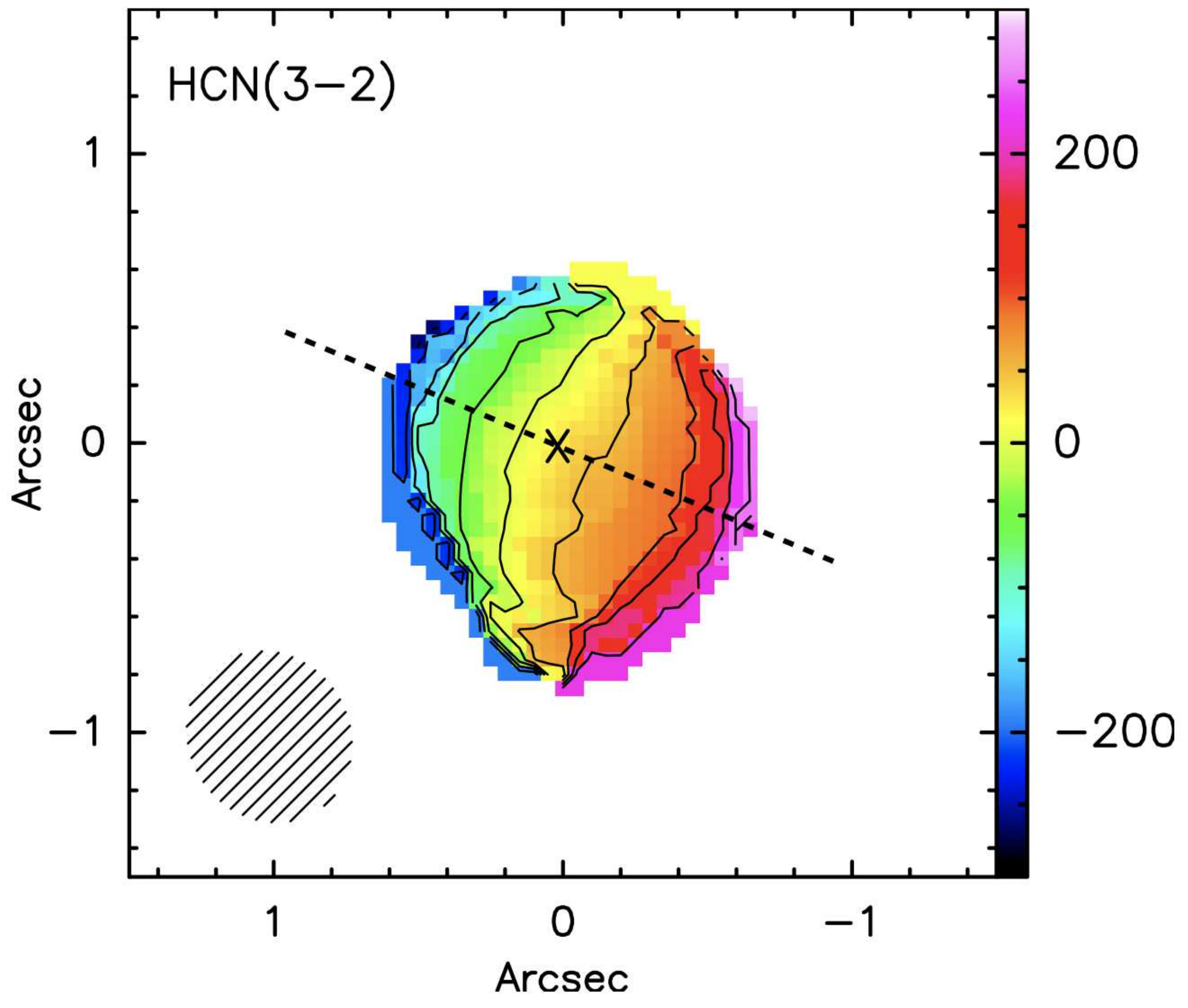}    
                \end{subfigure}
        \begin{subfigure}{0.3\textwidth}
        \includegraphics[width=\textwidth]{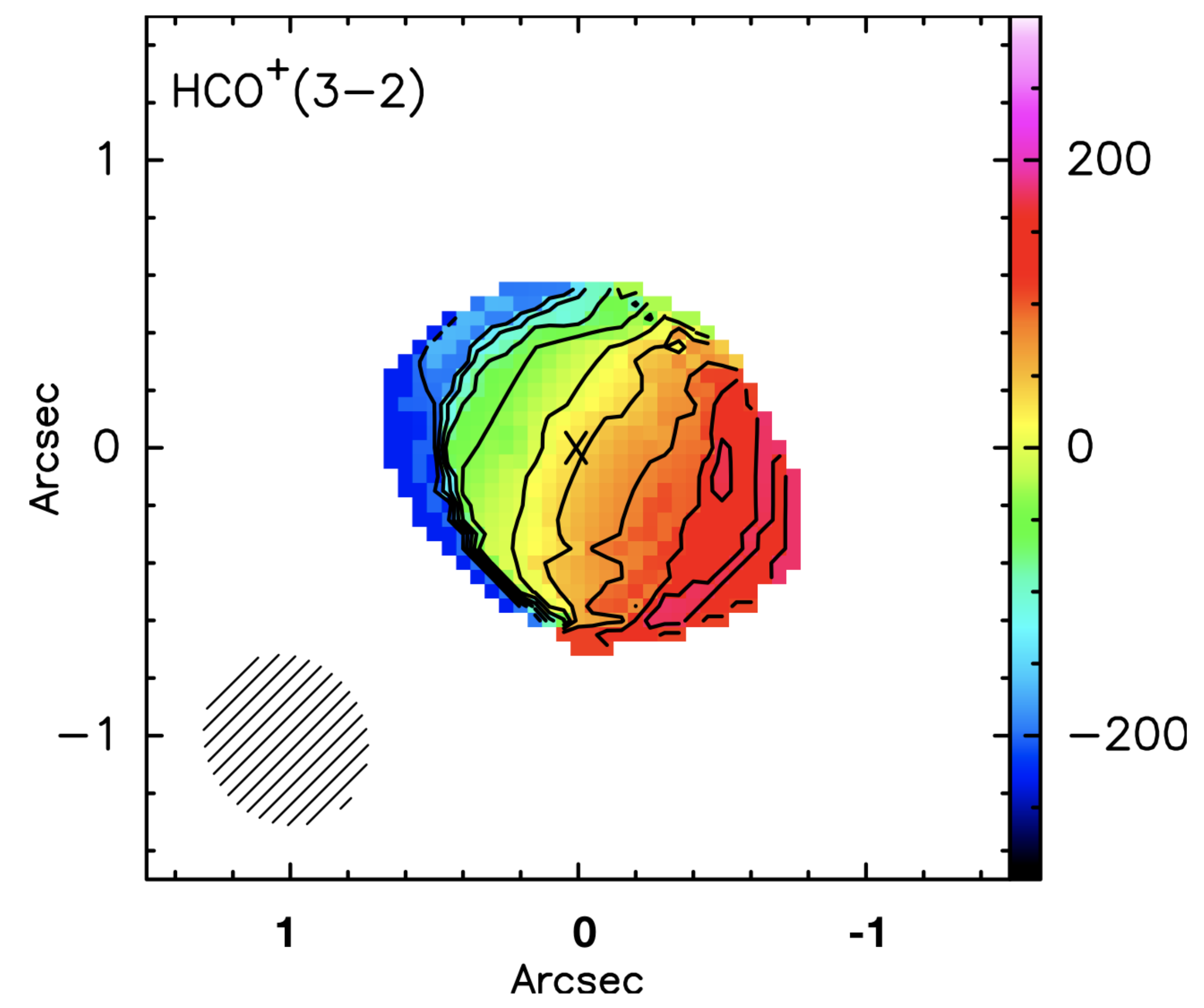}       
                \end{subfigure} 
        \begin{subfigure}{0.3\textwidth}
        \includegraphics[width=\textwidth]{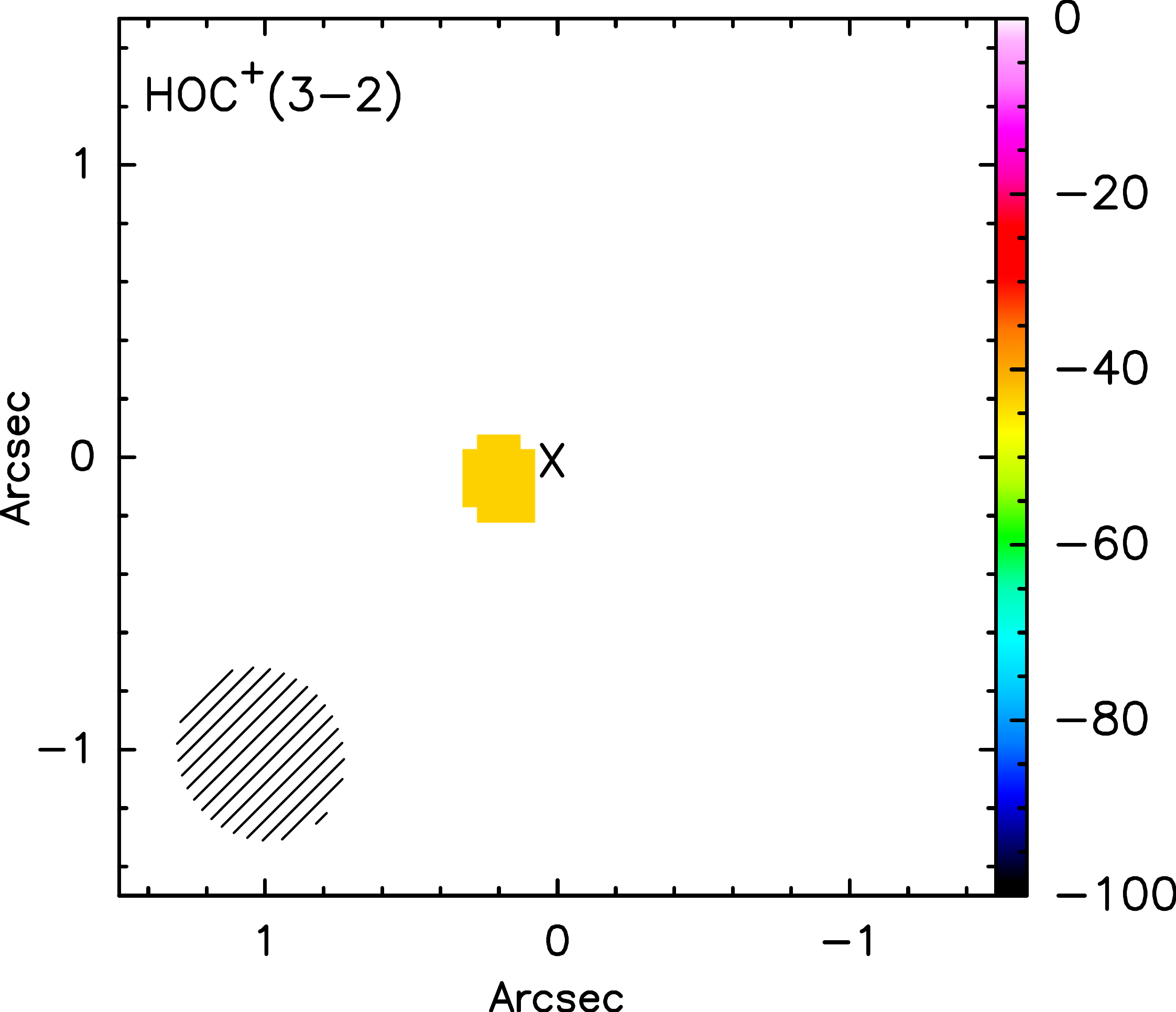}       
                \end{subfigure}                 
        \caption{Velocity fields (moment 1 maps). The coloured velocity scale  (right axis) is in km\,s$^{-1}$. The step in contours is 20\,kms$^{-1}$ for all lines. We note the blue-shifted velocities of HOC$^+(3-2)$.
        The crosses in the centre of each panel indicate the position of the nuclear source. The beam is shown in the lower left  corner of each panel. North is up, and east is to the left. The dashed lines in the HCN plots indicate the cut for the p-v diagrams shown in Fig.\,\ref{fig9}.}                
        \label{fig8}%
\end{figure*}

\begin{figure*}[t!]
        \centering              
                \begin{subfigure}{0.3\textwidth}
      \includegraphics[width=\textwidth]{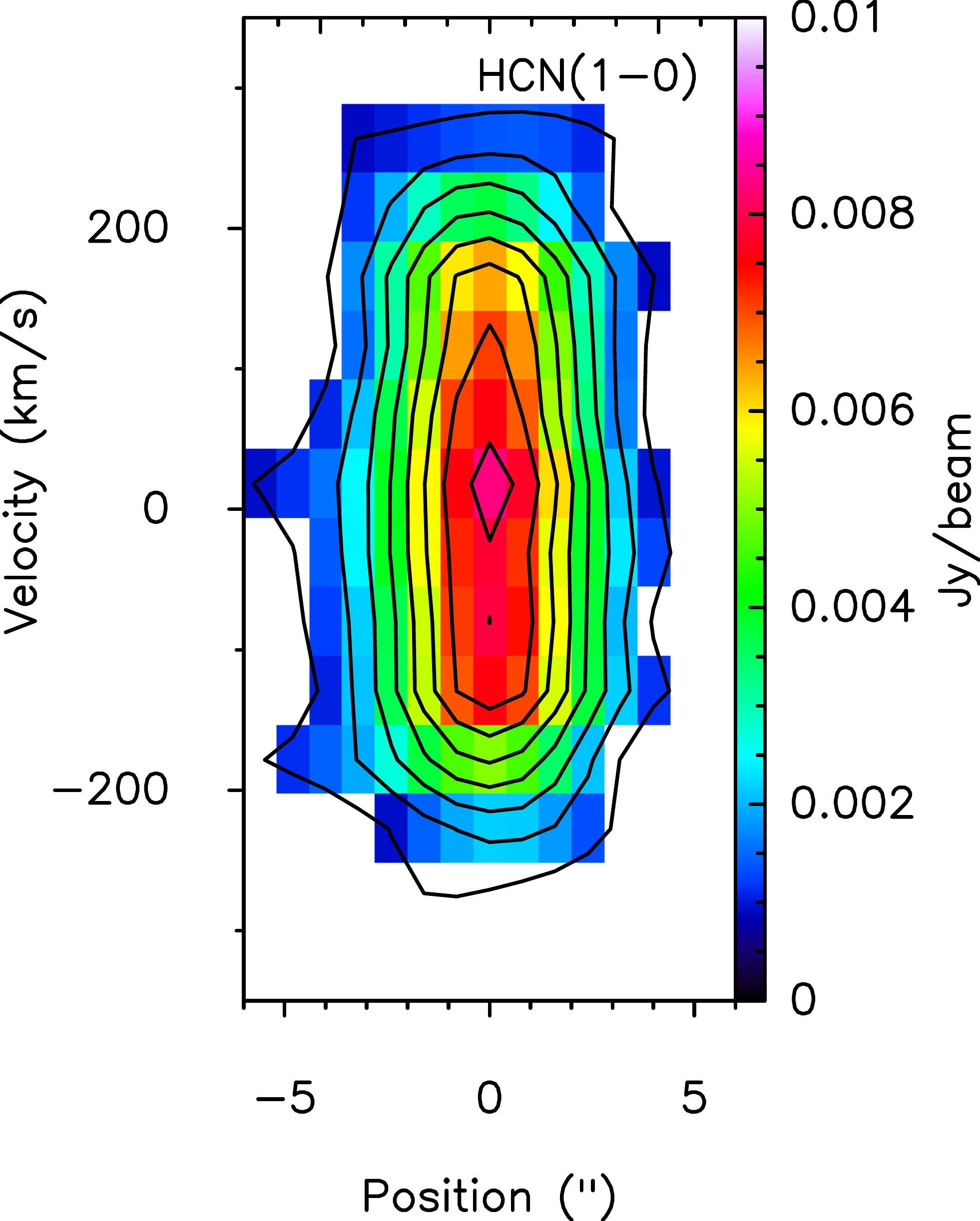}
                \end{subfigure}\hspace*{1.5cm}
                \vspace{7mm}                                    
                \begin{subfigure}{0.3\textwidth}
                \includegraphics[width=\textwidth]{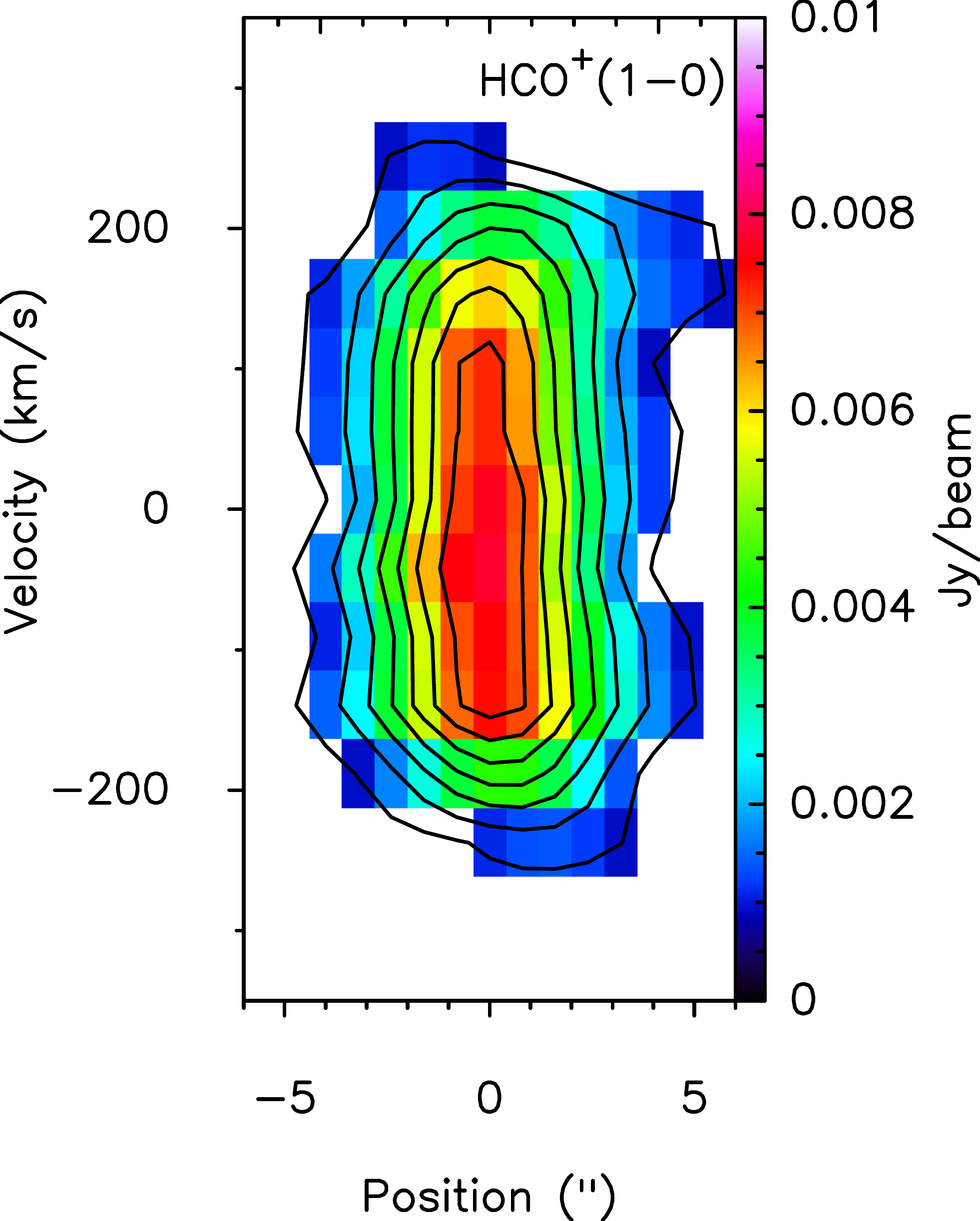}
                \end{subfigure}\hspace*{1.5cm}  

        \begin{subfigure}{0.3\textwidth}
        \includegraphics[width=\textwidth]{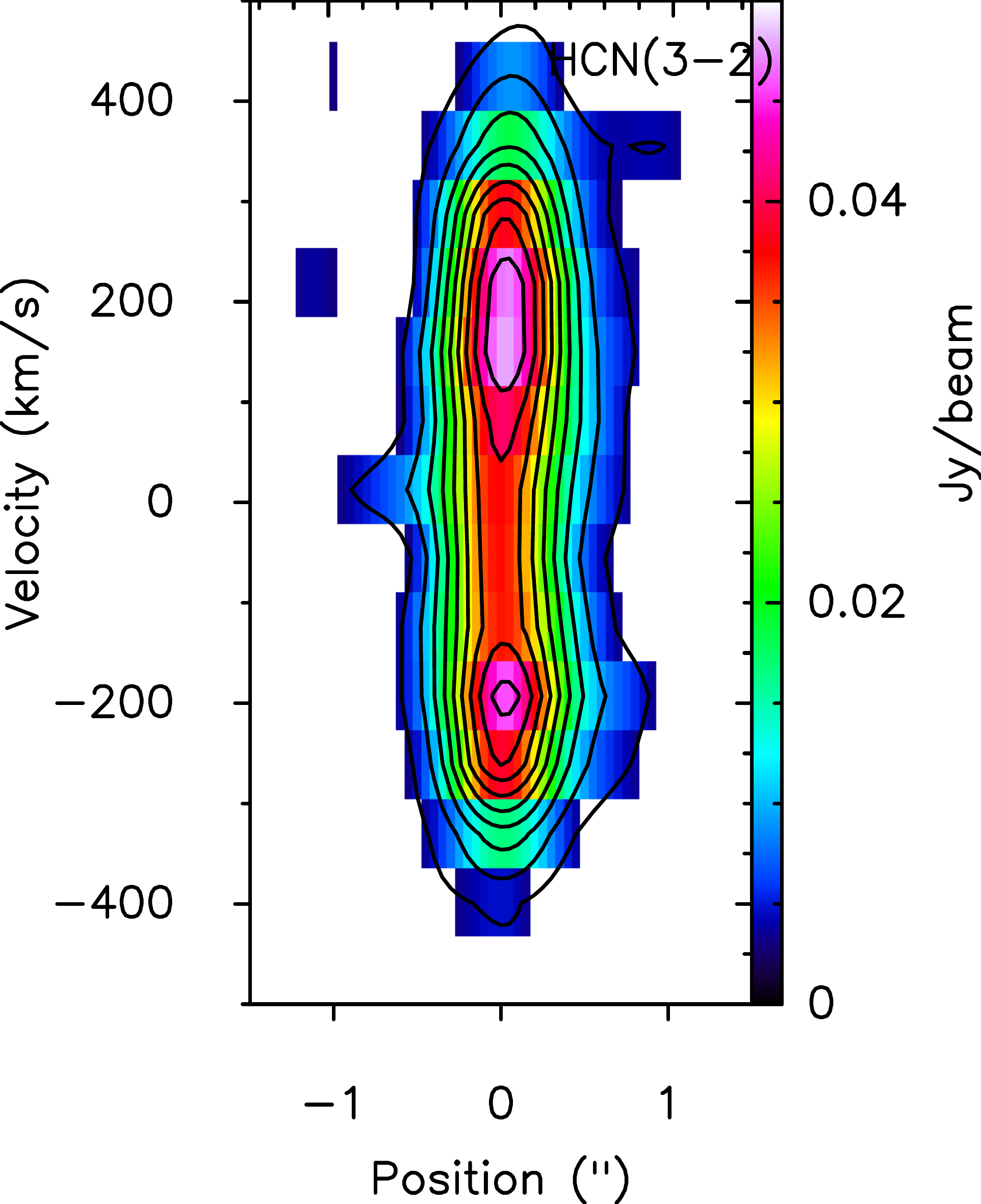}
        \end{subfigure}\hspace*{1.5cm}
        \vspace{7mm}    
        \begin{subfigure}{0.3\textwidth}
        \includegraphics[width=\textwidth]{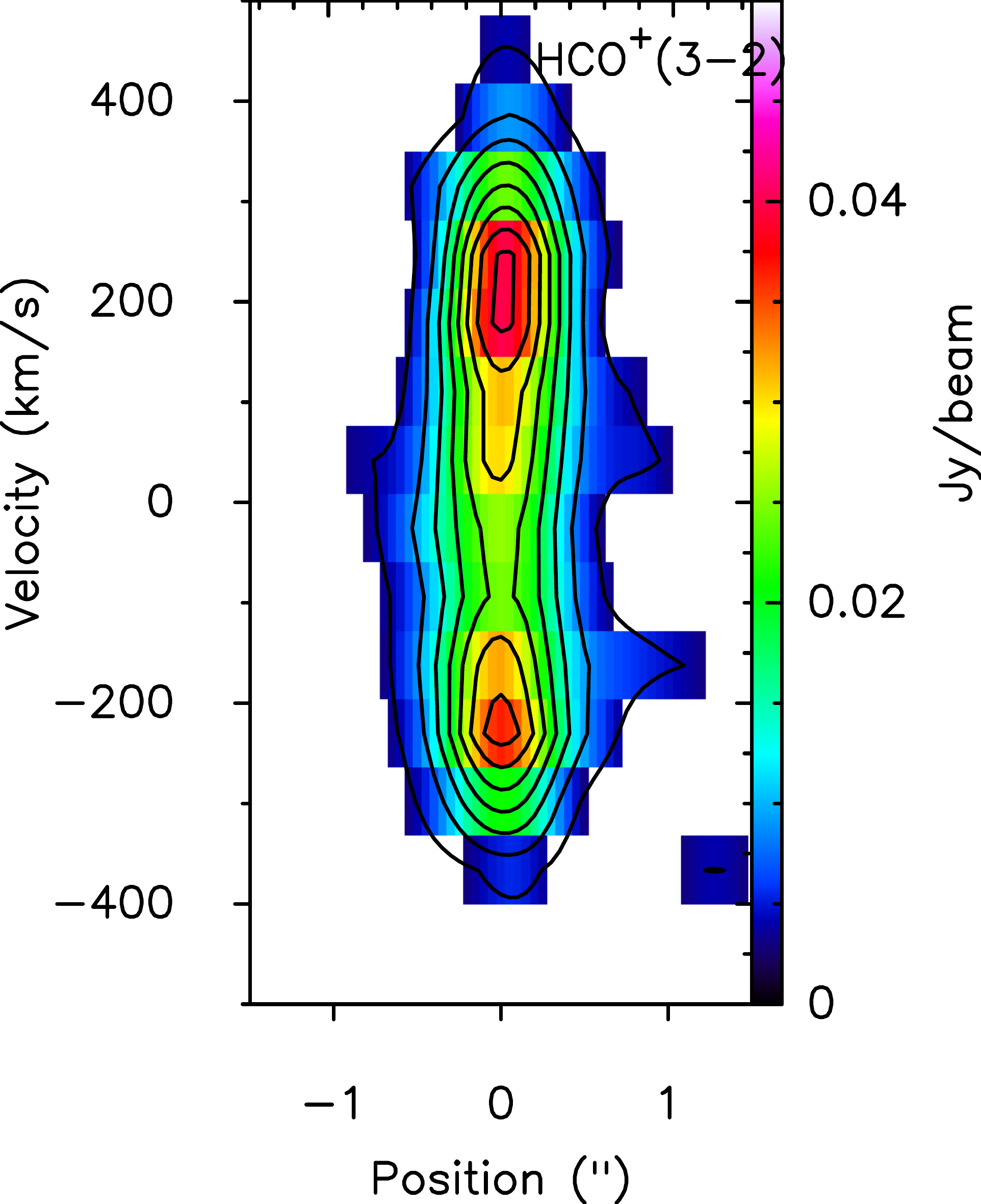}
        \end{subfigure}\hspace*{1.5cm}
        \caption{Position-velocity (p-v) maps of HCN and HCO$^+$. The cuts along the axes of rotation  are shown by the dashed lines in Fig.\,\ref{fig8}. Contour steps of the $(1-0)$ maps go from 0.9\,mJy\,beam$^{-1}$  (3$\sigma$) to 7.9\,mJy\,beam$^{-1}$ with steps of 1\,mJy\,beam$^{-1}$. For the $(3-2)$ maps, contours go from 3.9\,mJy\,beam$^{-1}$ (3$\sigma$) to  59\,mJy\,beam$^{-1}$  with  steps of 5\,mJy\,beam$^{-1}$.}
        \label{fig9}%
\end{figure*}

 \begin{table*}
        \caption{Global properties of the lines detected with NOEMA.}
        \centering                         
        \begin{tabular}{l c c c c c c c c }       
                \hline\hline                 
                & $\nu_{\rm rest}$      &K\,Jy$^{-1}$& Size$^a$ &  P.A.$^a$ &L$'$\\   
                &[MHz]&&[$''$]& [$\degree$]&[K\,km\,s$^{-1}$ pc$^2$]\\
                \hline                      
                HCN\,$(1-0)$ &88631.8&7.1675&(2.0$\pm$0.2)\,$\times$\,(1.6$\pm$0.2)& $-39\pm20$ &$2.5\times10^8$& &  \\
                HCO$^+(1-0)$&89188.6&7.0783&(2.2$\pm$0.2)\,$\times$\,(1.7$\pm$0.2)&$1\pm16$&$2.8\times10^8$     \\
                HNC\,$(1-0)$&90663.6&6.4696&(2.4$\pm$0.4)\,$\times$\,(2.0$\pm$0.3)&$-36\pm35$&$2.2\times10^8$   \\
                HC$_3$N\,$(10-9)$&90979.0&6.4248&$<$1.6&---&---\\
                HCN\,$(3-2)$    & 265886.4&51.4126&(0.35$\pm$0.01)\,$\times$\,(0.31$\pm$0.01)&$68\pm11$& $4.4\times10^8$\\ %
                HCO$^+(3-2)$       &  267557.6&50.7724  &(0.40$\pm$0.01)\,$\times$\,(0.32$\pm$0.01)&$71\pm5$& $2.6\times10^8$\\  
                HOC$^+(3-2)$     & 268451.1& 50.4350& $<$0.3&---&---\\
                \hline                                 
        \end{tabular}
        \caption*{
                Notes: $^a$ Calculated as the FWHM  of the elliptical Gaussian fitted to the line emission in the uv-plane (see Sect.\,\ref{intint}). The line luminosity is $L=\pi R^2 I$, where $R$ is the source size in column four in parsec ($1''=761$\,pc), and $I$ is the integrated intensity (shown in Table\,\ref{table1}) in units of  K\,km\,s$^{-1}$. }  
        \label{table3}   
 \end{table*}

\subsubsection{Integrated intensity of the outflow}
\label{mom0outflow}
The moment-zero map of the outflow traced by HCN$(3-2)$ (highlighted in yellow in Fig.\,\ref{fig5}) is shown in Fig.\,\ref{fig7}. After centering the spectrum at the HCN$(3-2)$ rest frequency, we integrated the emission of all pixels  in the velocity range [-1100, -400]\,km\,s$^{-1}$. Its morphology is composed of two main peaks of emission;  a stronger one with an elliptical shape around the centre, and a fainter and rounder feature at  $\sim$0.8$''$ ($\sim$600\,pc) to the north. In the central pixel, the peak flux  has a S/N of $\sim$5 (measured at the final velocity resolution), while the integrated flux density measured in the moment-zero map  is 2.4\,Jy\,km\,s$^{-1}$.

We measured the size of the outflow by fitting two ellipses to the  main peaks seen in the moment-zero map (see Fig.\,\ref{fig7}). We take the size of the outflow as the  projected distance between the central pixel and the centre of the northern ellipse. We measure a size of 0$\ffas$61\,$\pm$\,0.$''$05, equivalent to 464\,pc, with a position angle of 10\,$\pm$\,3$\degree$. The direction of the flow heading to the north is consistent with the outflowing gas detected with CO\,$(1-0)$ by \citet{Cicone14}.

\subsubsection{Velocity fields  and position-velocity maps}
\label{mom1-pv}
Figures\,\ref{fig8} and \ref{fig9} show the velocity fields and position-velocity (p-v) maps of HCN, HCO$^+$ and HNC. The velocity fields  reveal the rotation of the Mrk\,273 nuclear disc (see \citet{DS98} for a detailed study of the disc properties). There are significant differences in the morphologies of the iso-velocity contours of the various observed  lines. HCN and HCO$^+(1-0)$ trace rotating gas in the outer  disc showing a south-east to north-west  direction. However, the less extended gas traced by HNC\,$(1-0)$ shows a  north-east to south-west  rotation, similar to the gas in the inner disc traced by the HCN\,$(3-2)$ and HCO$^+(3-2)$ lines. The rotation in the central $<0\ffas5$ follows the velocity fields traced by the warm and  compact gas traced by H$_2$, Br$\gamma$ and [FeII] \citep{Medling14,U}. This is consistent with the two kinematic systems of the disc already discovered by \citet{DS98} using the CO\,$(1-0)$ and CO\,$(2-1)$ lines with beam sizes $(1\ffas4,1\ffas3)$ and $(0\ffas6,0\ffas6)$. The agreement between the \citet{DS98} velocity fields and ours, despite the difference in angular resolution, suggests that this effect is not due to the larger beam  size of our 3\,mm data.

Figure\,\ref{fig9} shows the p-v diagrams of the HCN and HCO$^+$ lines along cuts through the centre and  perpendicular to the axes of  rotation as plotted with dashed lines in Fig.\,\ref{fig8}. The outer disc traced by the $(1-0)$ lines shows maximum velocities $\pm$\,300km\,s$^{-1}$, though most of the gas exhibits velocities within the smaller range of $\pm$\,150km\,s$^{-1}$. The p-v diagrams of HCN and HCO$^+(3-2)$ show that the gas in the inner disc rotates faster, reaching maximum velocities of $\pm$\,400\,km\,s$^{-1}$, although the average speed is $\pm$\,200\,km\,s$^{-1}$ at the edges of the disc, which are separated by 0$\ffas$2 ($\sim$150\,pc). 
  
The  dynamical mass of the inner, starbursting disc, measured from the kinematics of the $(3-2)$ lines, is $(4-5)\times10^9M_\odot$, as calculated from $RV_{rot}^2/G$. Here,  $V_{\rm rot}$ is the average rotational velocity corrected for the 45 degree inclination of the disc  \citep{DS98}, $R$ is the size of the emission in pc (Table\,\ref{table3}), and $G$ is the gravitational constant.

 \subsubsection{Velocity dispersions}

\begin{figure*}
        \centering
        \begin{subfigure}{0.3\textwidth}
                \includegraphics[width=\textwidth]{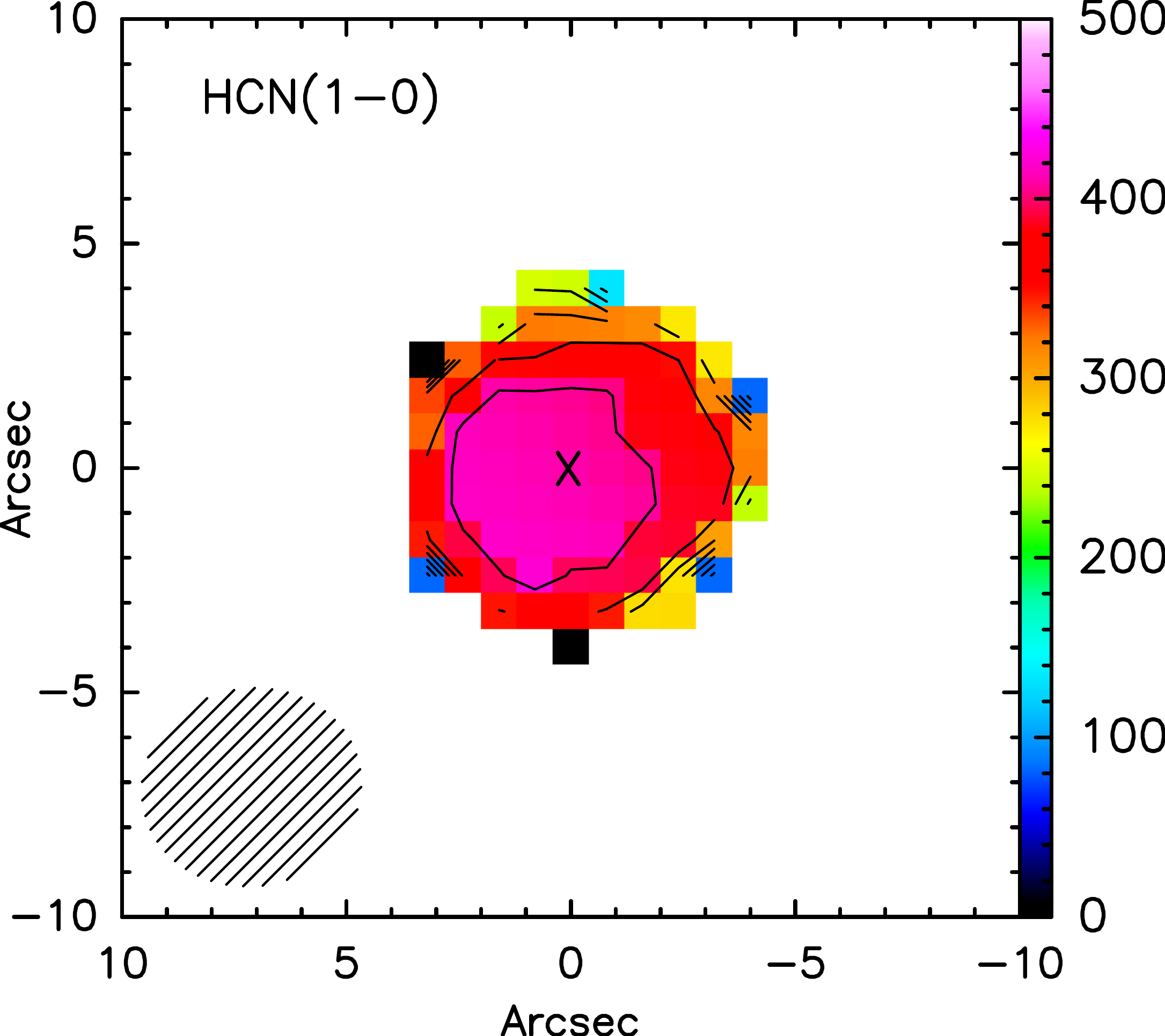}
        \end{subfigure}
\hspace{4mm}
        \vspace{5mm}
        \begin{subfigure}{0.3\textwidth}
                \includegraphics[width=\textwidth]{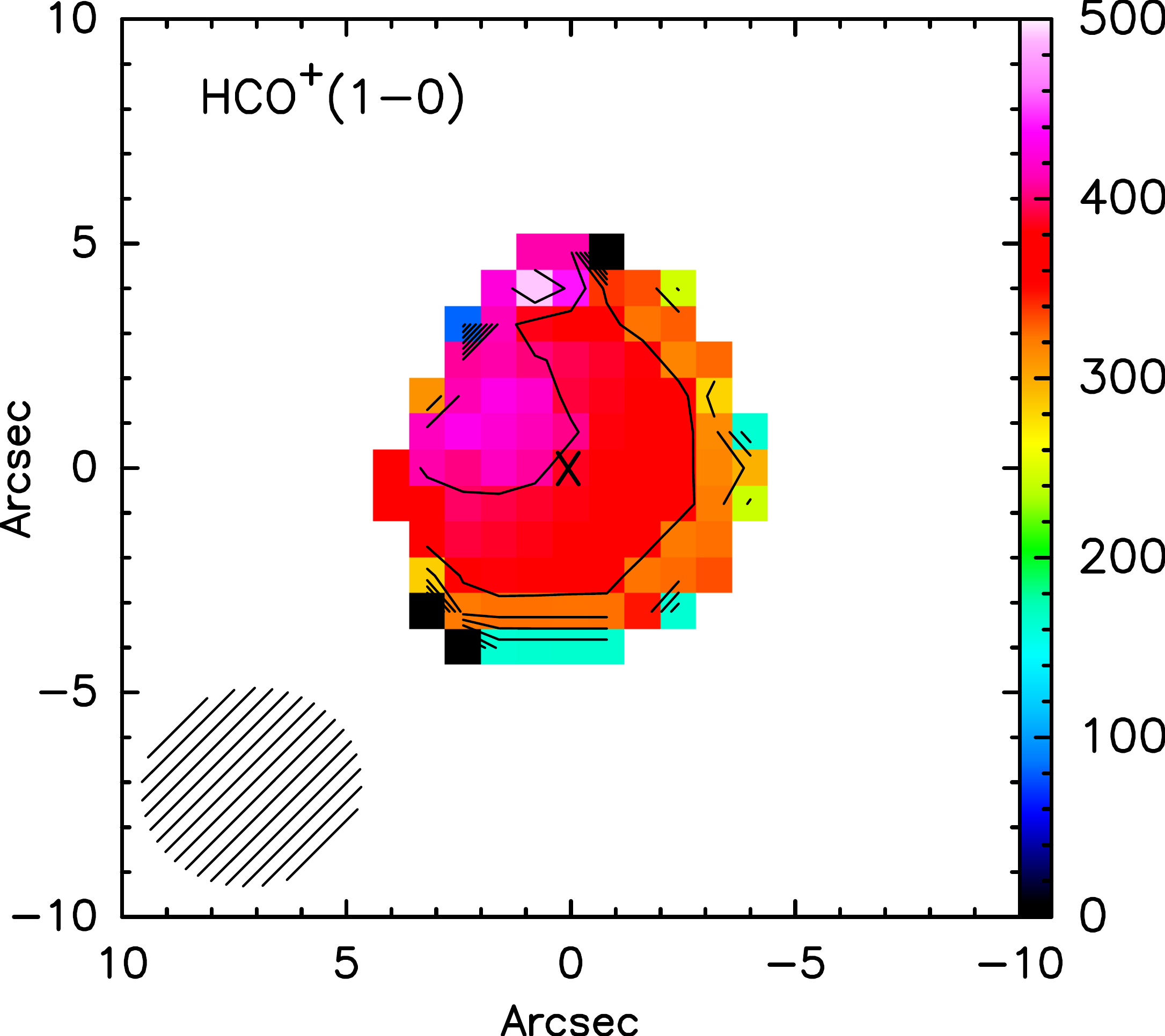}
        \end{subfigure}
\hspace{4mm}
        \begin{subfigure}{0.3\textwidth}
                \includegraphics[width=\textwidth]{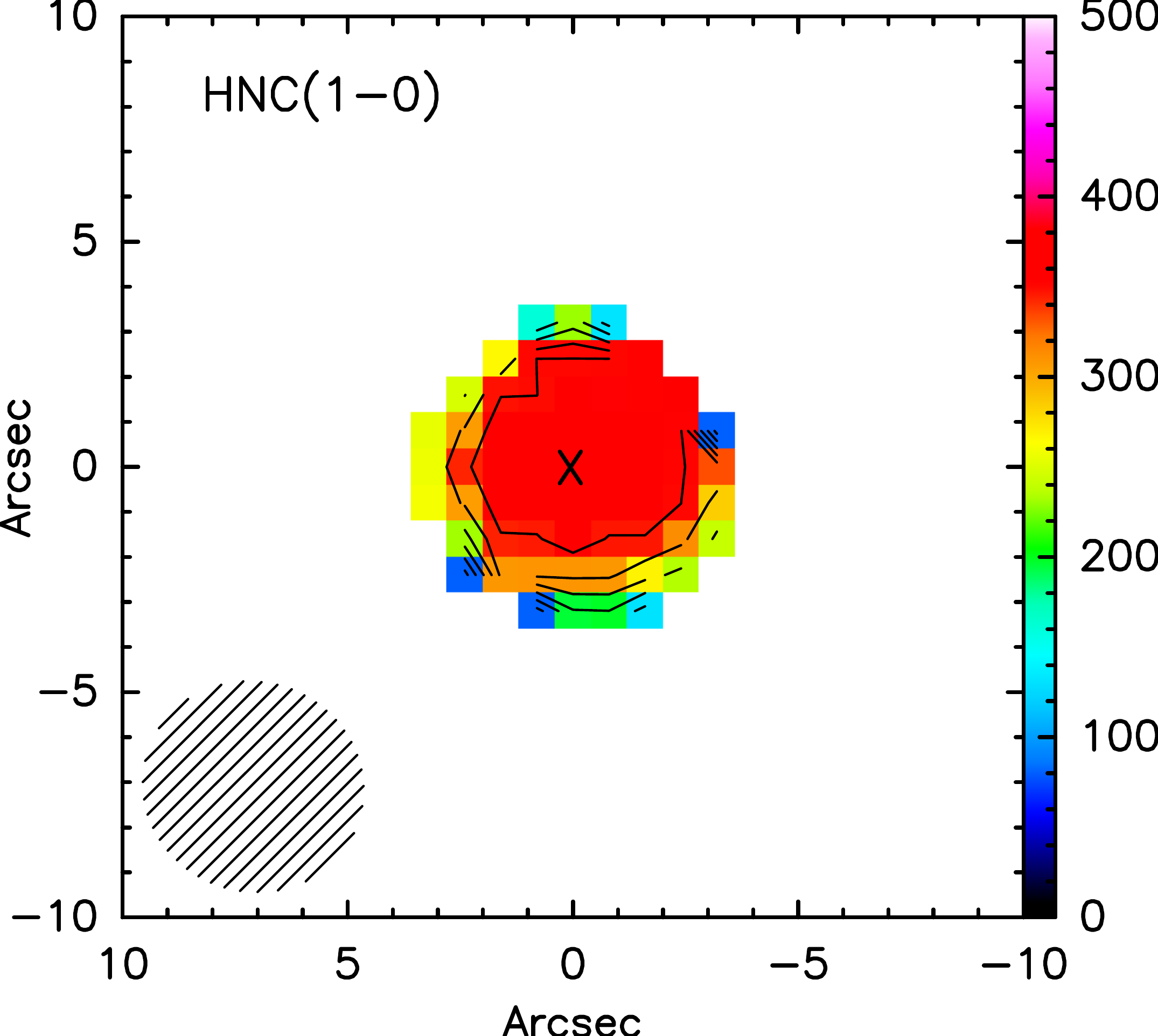}
        \end{subfigure}
        \begin{subfigure}{0.3\textwidth}
                \includegraphics[width=\textwidth]{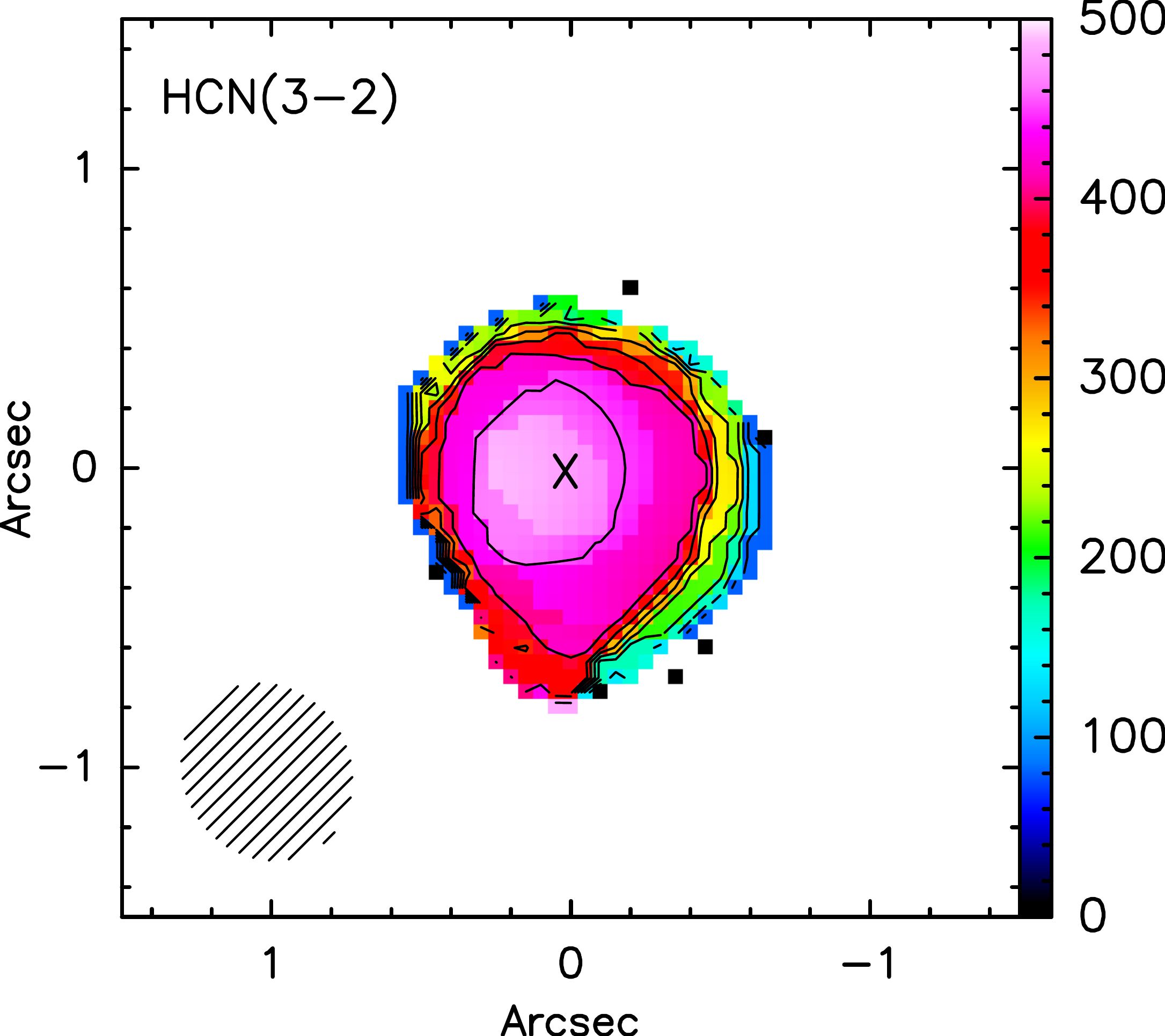}        
        \end{subfigure}
\hspace{4mm}
        \begin{subfigure}{0.3\textwidth}
                \includegraphics[width=\textwidth]{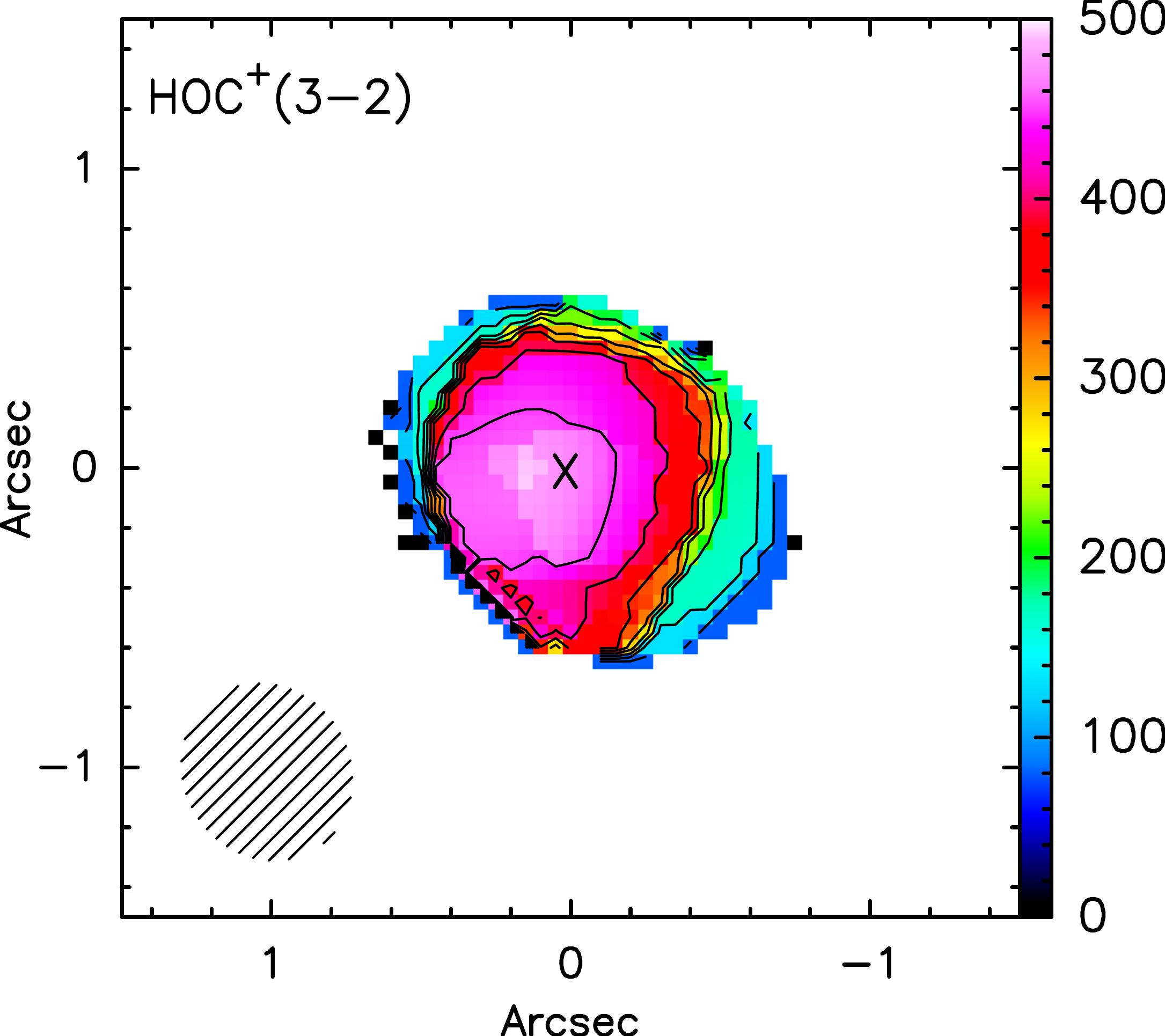}               
        \end{subfigure}                                                                 
        \caption{Velocity dispersions. Contours  go from 0 to  500\,kms$^{-1}$  with steps of 50\,kms$^{-1}$ for all lines. The crosses in the centre of each panel indicate the position of the nuclear source. The synthesised beam is shown in the bottom-left corner of each panel. North is up and East is to the left.}
        \label{fig10}%
\end{figure*}

The velocity dispersions of the NOEMA data were calculated as 
        \begin{equation}
        \sigma_v=\frac{FWHM}{2\times \sqrt{2\times ln(2)}}
        .\end{equation}
        
The moment-2 maps of HCN, HCO$^+$ and HNC are shown in Fig.\,\ref{fig10}.       All lines  reach similar maximum dispersions of  500\,km\,s$^{-1}$. We explored the \citet{Toomre64} stability criterion, $Q\ge1$,  for the inner, starbursting, gaseous disc to check its stability against gravitational perturbations, as

        \begin{equation}
Q=\frac {\sigma_v \times \kappa}{\pi \times G \times \Sigma}
,\end{equation}

where $\kappa$ is the epicyclic frequency, $G$ is the gravitational constant, and $\Sigma$ is the surface density of the gas. The surface density in the inner 380\,pc of Mrk273 is 37500 $M_\odot/pc^2$  \citep{Yun95}.  We note that this value was calculated in a region slightly larger than the radius of the inner disc (of $\sim$300\,pc), but it  is still a good approximation if compared to the size of the outer disc (1.5\,kpc). 
For the epicyclic frequency, we assumed a Keplerian disc. Therefore, $\kappa=\omega$,  $\omega$ being the angular velocity. For consistency, we calculated $\omega$ in the same radius of 380 pc, and used the average rotational velocities from our HCN and HCO$^+(3-2)$ velocity maps (200 km\,s$^{-1}$, Figs.\,\ref{fig8} and \ref{fig9}). We obtain $Q=0.5$, which  indicates that the inner disc is unstable and clumpy/turbulent enough to  form further self-gravitating condensations of gas. \\

\subsubsection{HCN and HCO$^+$ channel-velocity maps}
\begin{figure*}[h!]
        \centering
        \includegraphics[width=0.9\textwidth]{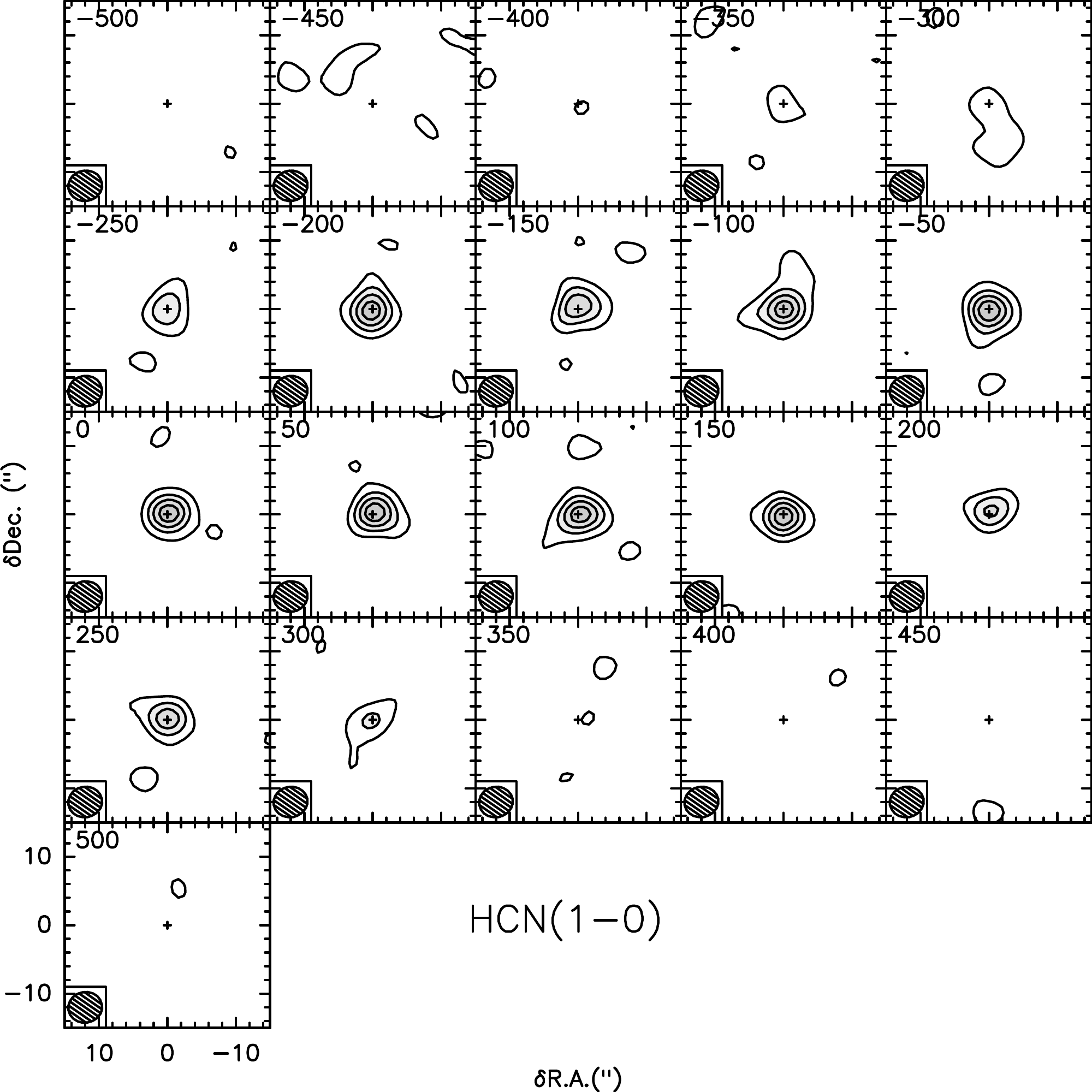}         
        \caption{Channel-velocity maps of HCN\,$(1-0)$ in the velocity range [-500, 500]\,km\,s$^{-1}$ with steps of 50\,km\,s$^{-1}$ . Contours go from 1.5\,mJy\,beam$^{-1}$ (5$\sigma$) to 9.5\,mJy\,beam$^{-1}$ with  a spacing of 2\,mJy\,beam$^{-1}$. The synthesised beam is plotted in the bottom-left corner. North is up and east is to the left.}
        \label{fig11}%
\end{figure*}

\begin{figure*}[h!]
        \centering
        \includegraphics[width=0.9\textwidth]{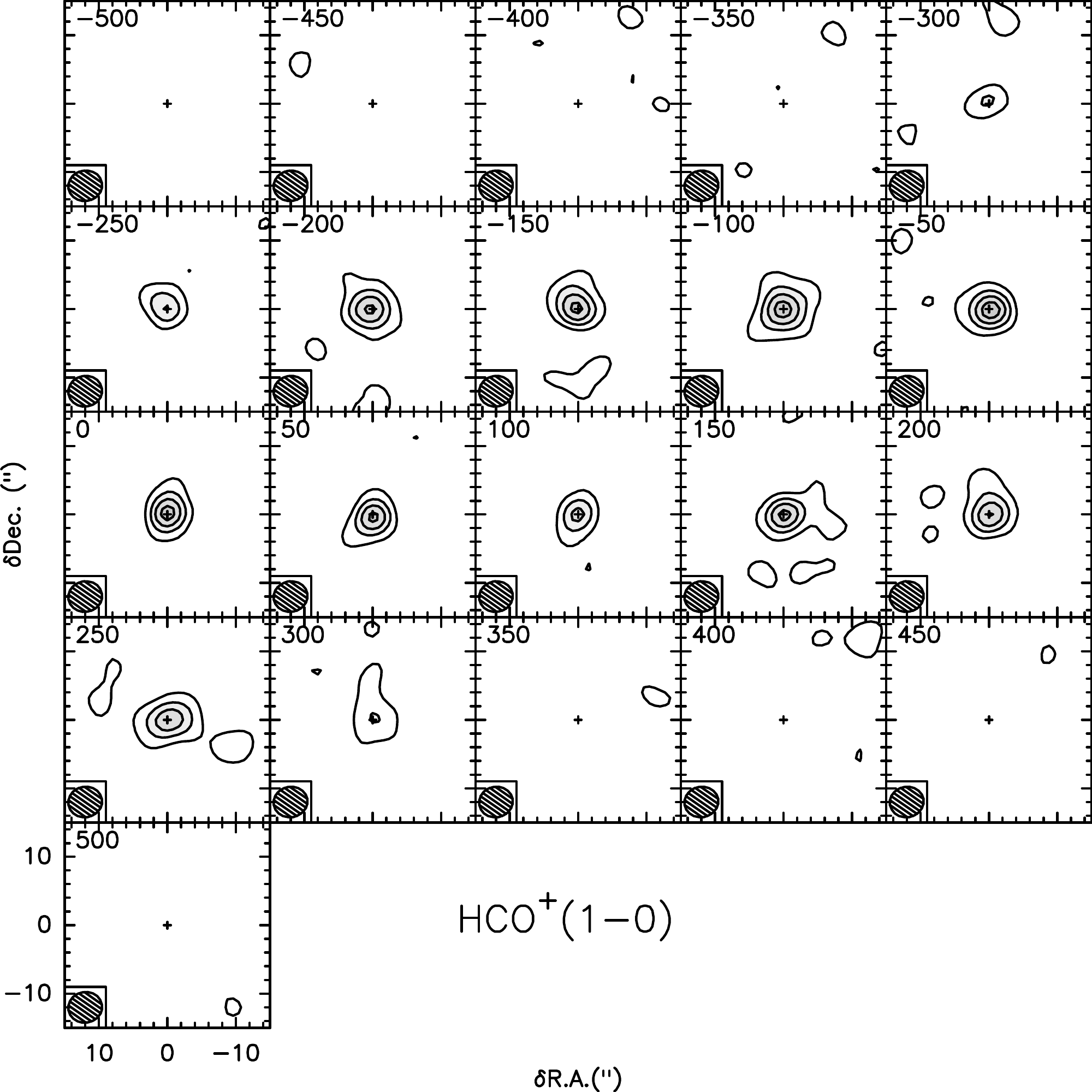} 
        \caption{Channel-velocity maps of  HCO$^+(1-0)$ in the velocity range [-500, 500]\,km\,s$^{-1}$ with steps of 50\,km\,s$^{-1}$ . Contours go from 1.5\,mJy\,beam$^{-1}$ (5$\sigma$) to 9.5\,mJy\,beam$^{-1}$ with  a spacing of 2\,mJy\,beam$^{-1}$. The synthesised beam is plotted in the bottom-left corner. North is up and east is to the left.}
        \label{fig12}%
\end{figure*}

\begin{figure*}[h!]
        \centering
        \includegraphics[width=0.9\textwidth]{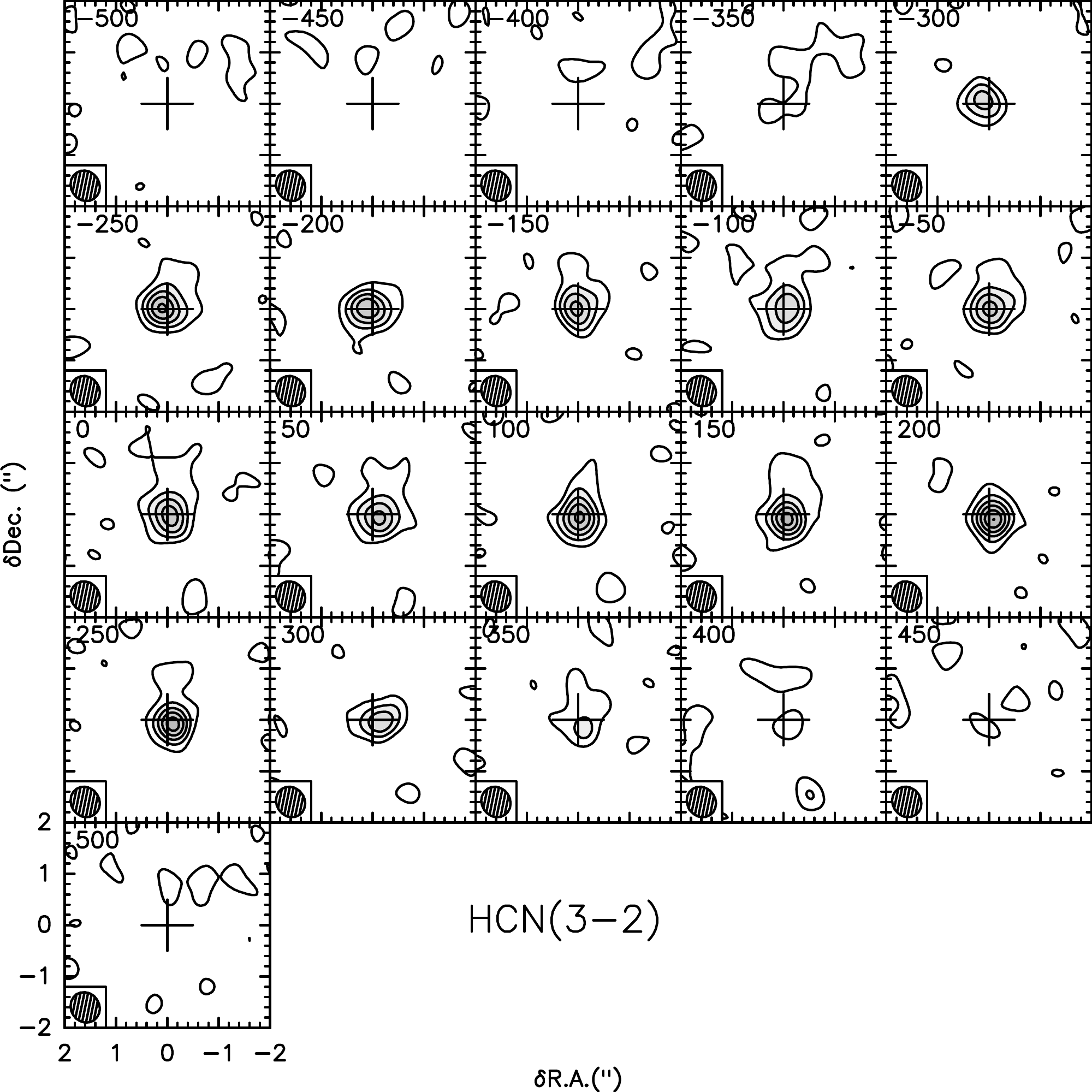}         
        \caption{Channel-velocity maps of HCN\,$(3-2)$ in the velocity range [-500, 500]\,km\,s$^{-1}$ with steps of 50\,km\,s$^{-1}$. Contours go from 6.5\,mJy\,beam$^{-1}$ (5$\sigma$)  to 56\,mJy\,beam$^{-1}$ with  a spacing of 10\,mJy\,beam$^{-1}$. We highlight the significantly smaller spatial scale relative to the channel map of the $(1-0)$ line shown in Fig.\,\ref{fig11}. The synthesised beam is plotted in the bottom-left corner. North is up and east is to the left.}
        \label{fig13}
\end{figure*}

\begin{figure*}[h!]
        \centering      
        \includegraphics[width=0.9\textwidth]{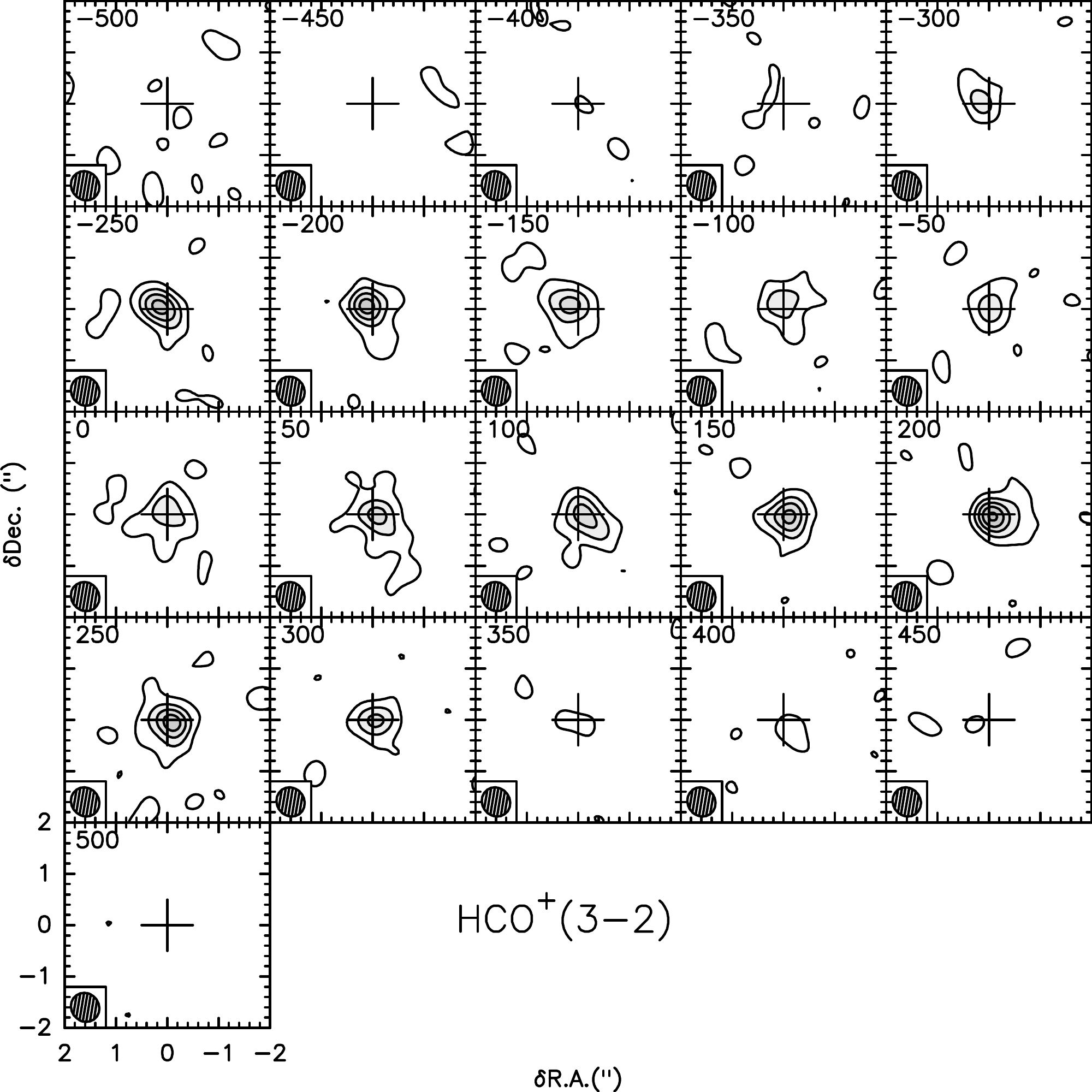}         
        \caption{Channel-velocity maps of  HCO$^+(3-2)$ in the velocity range [-500, 500]\,km\,s$^{-1}$ with steps of 50\,km\,s$^{-1}$. Contours go from 6.5\,mJy\,beam$^{-1}$ (5$\sigma$)  to 56\,mJy\,beam$^{-1}$ with  a spacing of 10\,mJy\,beam$^{-1}$. We highlight the significantly smaller spatial scale relative to the channel maps of the $(1-0)$ line shown in Fig.\,\ref{fig12}. The synthesised beam is plotted in the bottom-left corner. North is up and east is to the left.}
        \label{fig14}
\end{figure*}

Figures\,\ref{fig11} to \ref{fig14} show the channel maps of HCN and HCO$^+(1-0)$ and $(3-2)$ between -500 and +500\,km\,s$^{-1}$ in steps of 50\,km\,s$^{-1}$, with a beginning intensity contour level of 5$\sigma$. 
The $(1-0)$ lines show  emission away from the nucleus in all directions. In particular, the HCN$(1-0)$ channel map reveals gas in the northern direction as far as 10$''$  from the centre at negative velocities, as well as elongations towards the south at  $\pm$300\,km\,s$^{-1}$. Some extensions to the east and south-east  are also seen in some channels. The HCO$^+(1-0)$ emission is similarly extended, but the structure towards the south is perhaps the most distinct (e.g. channels $\pm$\,150\,km\,s$^{-1}$). Nevertheless,  extensions to the north ($\pm$\,200\,km\,s$^{-1}$ and $\pm$\,300\,km\,s$^{-1}$), east ($\pm$\,250\,km\,s$^{-1}$), and west (+150\,km\,s$^{-1}$ and +250\,km\,s$^{-1}$) are also seen.

The HCN and HCO$^+(3-2)$ channel maps trace  denser gas in the inner parts of the galactic disc ($\le300$\,pc), although some emission can be seen in the central $\pm$1$''$ ($\pm$800\,pc), and their elongations are even more obvious than in the $(1-0)$ channel maps. HCN  shows clear signs of  emission towards the north in most channels (best seen between  -250\,km\,s$^{-1}$ and +250\,km\,s$^{-1}$). On the other hand,  the gas traced by HCO$^+$  is more extended towards the south (e.g. -200\,km\,s$^{-1}$ and +50\,km\,s$^{-1}$). 

The most significant extensions to the north (from the HCN maps) and to the south (from the HCO$^+$ maps) are  signatures of outflowing gas, and are further discussed in Sect.\,\ref{outflow}.

\begin{figure*}[h!]
        \centering
        \includegraphics[angle=0,width=\textwidth]{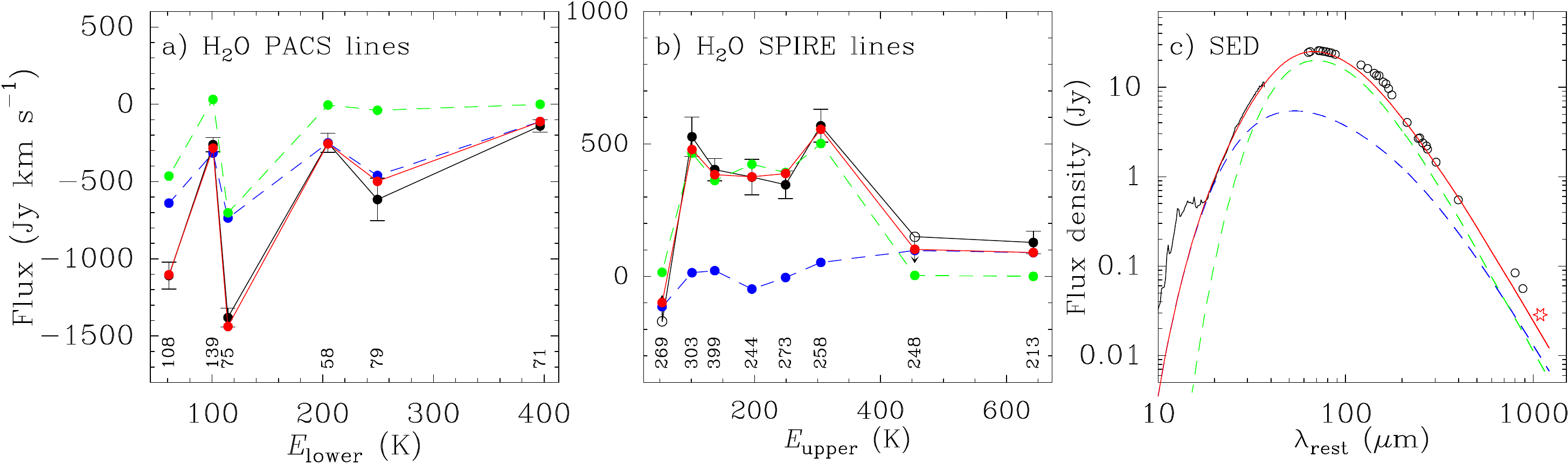}         
        \caption{Comparison between the observed (black symbols) and modelled
                (coloured symbols and lines) H$_2$O fluxes in Mrk~273 
                within a) PACS and b) SPIRE. The model includes two components: the core (in
                blue), which accounts for most absorption lines observed within PACS, and the
                inner disc (in green), which dominates the emission of the sub-mm lines
                with $E_{\mathrm{upper}}<400$ K observed with SPIRE (see Table\,\ref{tab:h2o}). Red      colours and symbols indicate total  modelled fluxes. The numbers at the bottom of panels (a) and (b) indicate rounded-up transition wavelengths     in $\mu$m. c) The SED of Mrk~273, including the {\it Spitzer}/IRS spectrum, {\it Herschel}/PACS and SPIRE continuum data from observations of both H$_2$O and OH lines, sub-mm data at 800 and 880 $\mu$m \citep{rig96,wil08}, and our measured flux density at 1mm (starred-red symbol), is compared with the   prediction of our composite model.}             
        \label{fig15}
\end{figure*}

\begin{center}
        \begin{table*}
                \caption{Results for the two-component modelling of the H$_2$O lines in
                        Mrk~273.}
                \label{tab:h2o}
                \begin{tabular}{lcccccc}
                        \hline
                        \noalign{\smallskip}
                        Component & $T_{\mathrm{dust}}$ & $\tau_{\mathrm{100\mu m}}$ & $R$ &
                        $L_{\mathrm{IR}}$ & $N(\mathrm{H_2O})$ & $X(\mathrm{H_2O})$ \\
                        & [K] & & [pc] & [$10^{11}$\,L$_{\odot}$] & [$10^{16}$\,cm$^{-2}$] & 
                        [$10^{-6}$] \\
                        \noalign{\smallskip}
                        \hline
                        \noalign{\smallskip}
                        Core  & $95$ ($\gtrsim80$) & $5.0$ ($\gtrsim4$)   & $51$ ($40-60$) 
                        & $3.8$ ($2.3-5.0$) & $800$ ($\gtrsim400$) & $1.2$ ($0.6-3$) \\
                        Inner disc  & $55$ ($45-65$)     & $0.5$ ($0.2-1$)  & $284$ ($180-360$) 
                        & $8.0$ ($6-12$)  & $7.8$ ($5-20$) & $0.12$ ($0.08-0.5$) \\
                        \noalign{\smallskip}
                        \hline
                \end{tabular}
                \caption*{Notes: Numbers in parenthesis indicate the most plausible
                        ranges, as inferred from all combinations with $\chi^2$ not exceeding
                        $1.7\times$ the minimum value. }
        \end{table*}
\end{center}

\subsection{Modelling of  H$_2$O}
\label{modelling-water}

We have used the library of H$_2$O models generated by \citet{Gonzalez14}
to fit the H$_2$O emission and absorption observed in Mrk~273. 
The models assume spherical symmetry, are non-local, and include excitation by
both the FIR field emitted by warm dust (which is mixed with the
H$_2$O molecules) and collisions with H$_2$. The collisional rates were taken
from \citet{Dubernet09} and \citet{Daniel11}, and a gas-to-dust ratio of 100 was
adopted \citep{wil08}. The models assume uniform physical properties ($T_{\mathrm{dust}}$, $T_{\mathrm{gas}}$, gas and dust densities, H$_2$O abundance). The source is divided into a set of spherical shells where the 
statistical equilibrium level populations are calculated. We assume
a H$_2$O ortho-to-para ratio of three. Line broadening is simulated by including a micro-turbulent velocity ($V_{\mathrm{turb}}$), with no systemic motions. 

The modelled line fluxes and continuum flux densities scale as
$(R/D_{\mathrm{L}})^2$ where $R$ is the source radius and $D_L$ is the
luminosity distance, so that they are easily scalable to the properties of any
source. Following \cite{fal17} and \cite{Gonzalez17}, we have fit the line fluxes
with a combination of $N_C$ model components by minimising $\chi^2$, with the
radius $R$ of each component being the only free parameter that is varied. We
required $N_C=2$ components to properly fit  the PACS and SPIRE H$_2$O fluxes
simultaneously. Since the models also make specific predictions for the
spectral energy distribution (SED) of each component, and since H$_2$O probes the galaxy FIR emission responsible for its excitation and more specifically the transition from the mid- to the far-IR \citep{Gonzalez10,Gonzalez14}, 
we also included in the fit the continuum flux densities at 30 and
60\,$\mu$m. 

Figure\,\ref{fig15} compares the observed H$_2$O fluxes and SED with the results of our
best model fit, and Table\,\ref{tab:h2o} lists the properties of the two model 
components (shown with blue and green colours in Fig.\,\ref{fig15}). The two components
show very different behaviours relative to the H$_2$O absorption and emission. 
We require a very compact (effective radius $R\sim50$ pc), very warm 
($T_{\mathrm{dust}}\sim95$\,K), and very optically thick 
($\tau_{\mathrm{100\mu m}}\gtrsim4$) component (referred to as the ``core'')
to account for the observed PACS absorption in several lines
(primarily the $3_{13}-2_{02}$ line at 138\,$\mu$m, the $4_{22}-3_{13}$ line at
58\,$\mu$m, the $4_{23}-3_{12}$ line at 79\,$\mu$m, and the $5_{24}-4_{13}$
line at 71\,$\mu$m) and also the SPIRE emission in the $5_{23}-5_{14}$ line at
212\,$\mu$m (although this line is only marginally detected at the $3\sigma$
level). However, the core component predicts negligible emission in most SPIRE
lines (and even absorption in the $2_{20}-2_{11}$ line), which indicates the
presence of a more extended component ($R\sim280$ pc), moderately warm 
($T_{\mathrm{dust}}\sim55$\,K), and with lower column 
($\tau_{\mathrm{100\mu m}}\sim0.5$). This extended
component, mostly responsible for the H$_2$O emission observed with SPIRE, is naturally identified 
with the inner disc traced by the $J=3-2$ lines of HCO$^+$ and HCN ($\sim$300\,pc). 

The two water components together provide a good fit to the FIR emission, though
the combined model underestimates, to some extent, the flux densities between
100 and 200\,$\mu$m. This probably indicates a range in $T_{\mathrm{dust}}$
for the disc component, rather than a single value. It is also worth noting
that the source luminosity is dominated by the disc (i.e. the starburst), with
the core component accounting for
$L_{\mathrm{IR}}\sim4\times10^{11}$\,L$_{\odot}$. The latter corresponds to
$\sim30$\% of the total IR luminosity, which is close to the estimated AGN
contribution based on the $15-$to$-30$\,$\mu$m diagnostic 
\citep[e.g.][]{vei09}. The blueshift of the absorption lines, which are
dominated by the core component, is similar to that seen in the excited OH
lines \citep{Gonzalez17}, suggesting that the core is expanding (more details in Sect.\,\ref{expansion}). Our model for the core also predicts significant emission at 265 GHz, $\sim10$\,mJy, though 
this value is relatively uncertain because of its dependence on the actual
continuum optical depth and the mass absorption coefficient of dust
($\kappa_{\lambda}$) at millimetre wavelengths\footnote{For the core component, we modified the $k_{\lambda}$ curve in Fig.~2 of \citet{Gonzalez14} in such a way that $\kappa_{1.3\mathrm{mm}}=0.9$\,cm$^2$\,g$^{-1}$ of dust, more similar to the value used by \citet{DS98}.}.

\section{Discusion}
\label{discusion}

\begin{table*}
                \caption{Brightness temperature ratios in Mrk\,273  (row/column), evaluated over the entire emission. }
        \centering                         
        \begin{tabular}{l c c c c c c c l}        
                \hline\hline               
                & HCN$(1-0)$  & HCO$^+(1-0)$&HNC$(1-0)$  & HC$_3$N$(10-9)$& HCN$(3-2)$ &HCO$^+(3-2)$ & HOC$^+(3-2)^\dagger$   \\    
                \hline                        
                 HCN$(1-0)$&1.00$\pm$0.07&1.00$\pm$0.07&1.8$\pm$0.2&6$\pm$1&---&---&---\\
                 HCO$^+(1-0)$ &1.00$\pm$0.07&1.00$\pm$0.07&1.8$\pm$0.2&6$\pm$1&---&---&---\\
                 HNC$(1-0)$ &0.55$\pm$0.05&0.55$\pm$0.05&1.0$\pm$0.1&3.2$\pm$0.8&---&---&---\\
                 HC$_3$N$(10-9)$&0.17$\pm$0.04&0.17$\pm$0.04&0.31$\pm$0.08&1.0$\pm$0.3&---&---&---\\
                 HCN$(3-2)$&---&---&---&---&1.00$\pm$0.06&1.13$\pm$0.06&10$\pm$5 \\
                 HCO$^+(3-2)$&---&---&---&---&0.88$\pm$0.05&1.00$\pm$0.06&9$\pm$4\\
                 HOC$^+(3-2)^\dagger$&---&---&---&---&0.10$\pm$0.05&0.11$\pm$0.06&1.0$\pm$0.7\\
                \hline                                 
        \end{tabular}
        \caption*{Notes: Due to differences in the observed areas, we do not list the ratios between the 3\,mm and 1\,mm lines (see Sect.\,\ref{lineratios}). 
        $^\dagger$ For HOC$^+(3-2),$ we used the temperature measured in the  pixel where it peaks (see Sect.\,\ref{linesprofilesNOEMA}). }
                \label{table4}       
\end{table*}

\begin{table*}
        \caption{Nuclear brightness temperature ratios in Mrk\,273 (row/column), evaluated in the central pixel.}
        \centering                         
        \begin{tabular}{l c c c c c c c l}       
                \hline\hline                
                & HCN$(1-0)$  & HCO$^+(1-0)$&HNC$(1-0)$  & HC$_3$N$(10-9)$& HCN$(3-2)$ &HCO$^+(3-2)$ & HOC$^+(3-2)^\dagger$   \\    
                \hline                        
                HCN$(1-0)$&1.00$\pm$0.07&1.03$\pm$0.07&2.2$\pm$0.2&9$\pm$2&---&---&---\\
                HCO$^+(1-0)$ &0.97$\pm$0.06&1.00$\pm$0.07&2.1$\pm$0.2&9$\pm$2&---&---&---\\
                HNC$(1-0)$ &0.45$\pm$0.04&0.46$\pm$0.04&1.0$\pm$0.1&4$\pm$1&---&---&---\\
                HC$_3$N$(10-9)$&0.11$\pm$0.02&0.12$\pm$0.03&0.25$\pm$0.06&1.0$\pm$0.3&---&---&---\\
                HCN$(3-2)$&---&---&---&---&1.00$\pm$0.06&1.28$\pm$0.08&7$\pm$3 \\
                HCO$^+(3-2)$&---&---&---&---&0.78$\pm$0.05&1.00$\pm$0.07&5$\pm$3\\
                HOC$^+(3-2)^\dagger$&---&---&---&---&0.14$\pm$0.07&0.18$\pm$0.09&1.0$\pm$0.7\\
                \hline                                
        \end{tabular}
        \caption*{Notes:   Due to differences in the observed areas, we do not list the ratios between the 3\,mm and 1\,mm lines (see Sect.\,\ref{lineratios}). $^\dagger$ For HOC$^+(3-2)$ we used the temperature measured in the  pixel where it peaks (see Sect.\,\ref{linesprofilesNOEMA}).}
        \label{table5}       
\end{table*}

\subsection{A rotating disc with continuum absorption}
\label{asymmetry}

The integrated intensities, velocity fields, and p-v diagrams of HCN and HCO$^+$ show the typical pattern of a rotating disc  (Figs.\,\ref{fig6}, \ref{fig8} \& \ref{fig9}). 
Such a rotating body should in principle  be reflected in the spectra as  double-peaked lines. 
However, due to our relatively large beam size used to observe the $(1-0)$ lines ($4\ffas9\times4\ffas5$), most of the gas is concentrated in the central pixel and the velocity gradients within the beam are not well reproduced. Therefore, our HCN and HCO$^+(1-0)$ lines show a single Gaussian-like profile. On the other hand, the $(3-2)$ transitions were observed with a resolution significantly higher ($0\ffas61\times0\ffas55$), so the velocity gradients  are well traced and the spectral lines reveal the expected pattern of the inner rotating disc (see Sect.\,\ref{intint}).

In addition,  continuum absorption is also apparently contributing to the shape of the HCN and HCO$^+(3-2)$ profiles. Firstly, we note that the $\sim$20-25\,mJy drop flux of the lines is very similar to the continuum flux density at these frequencies, of 29\,mJy. Secondly, the velocity of the minimum flux,  around -50\,km\,s$^{-1}$, roughly coincides with the peak absorption velocities of the H$_2$O lines observed with PACS, and of the OH\,84\,$\mu$m and OH\,65\,$\mu$m lines presented in \citet{Gonzalez17}. The foreground warm gas traced by H$_2$O and OH is absorbing the continuum (e.g. \citealt{Gonzalez17}), and the coincidence in velocities and continuum values of  the dip in HCN and HCO$^+$ indicates that the dense gas might also be absorbing the background dust emission.

In this context,  the two peaks of the HCN and HCO$^+(3-2)$ lines are  probing the edges of the inner rotating disc, while the absorbed flux indicates the positions of the maximum column densities of the gas. The channel maps shown in Figs.\,\ref{fig13} and  \ref{fig14} also show a minimum emission  around -50\,km\,s$^{-1}$ and -100\,km\,s$^{-1}$  (better seen in the HCO$^+(3-2)$ map), velocities at which there is a maximum absorption of the continuum.

Self-absorption might also be an extra factor affecting the line shapes if there is cooler foreground gas with high-enough column densities. This, however, is difficult to  disentangle from the continuum absorption in our data.  In addition, it is possible that the opacity of the gas in the centre of Mrk\,273 is high enough to result in flat-topped profiles such as some of those observed in the HCN and HCO$^+(1-0)$ transitions, even though these species are less abundant than CO. A hint of such a profile might be seen in the HCN, HCO$^+(1-0)$ transitions. We note that any kind of absorption  implies that our estimations of line fluxes, luminosities, and molecules gas mass are  lower limits to the actual values.

 \subsection{The Mrk\,273 molecular  outflow}
 \label{outflow}

 \begin{figure*}[t!]
        \includegraphics[width=0.5\textwidth]{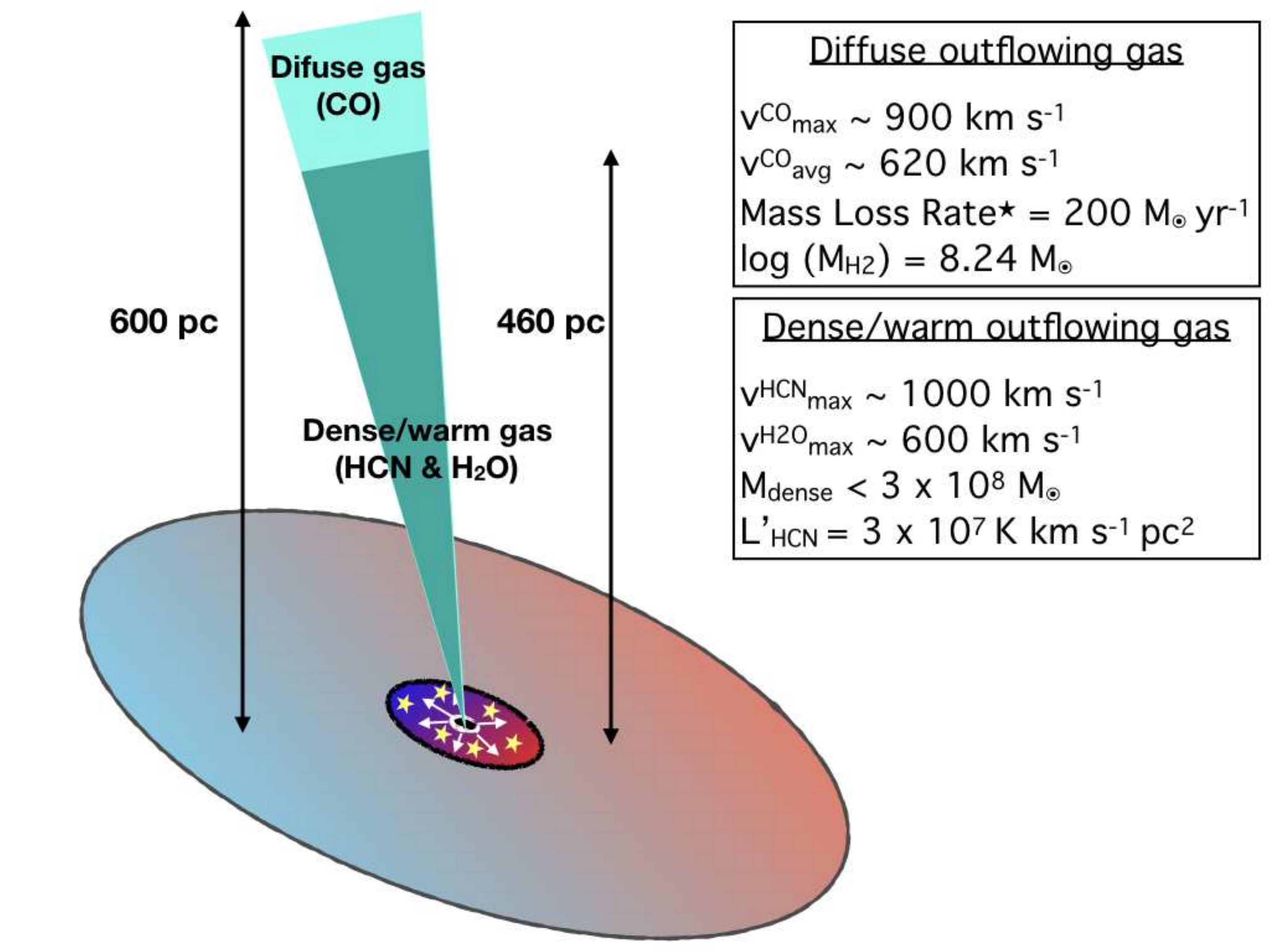}         
        \includegraphics[width=0.5\textwidth]{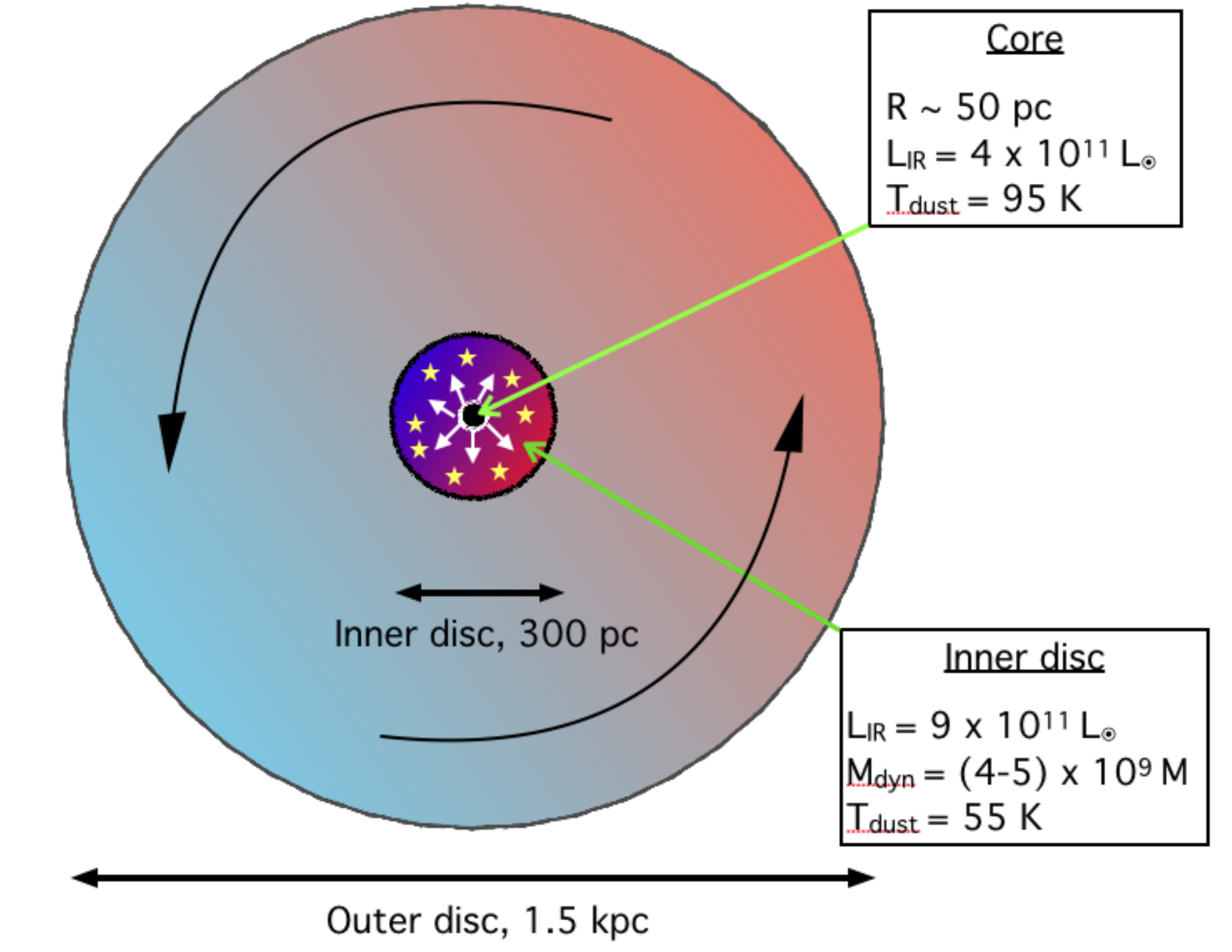}         
        \caption{Left: Sketch of the blue-shifted molecular outflow in the northern nucleus of Mrk\,273. The dense and warm outflowing gas found in our HCN and H$_2$O data, and its properties, is compared to the diffuse outflow found by \citet{Cicone14} by observing CO\,$(1-0)$. $^\star$We note that the mass loss rate is calculated as \.{M}=$v\,(M_{\rm OF}/R_{\rm OF})$      (see Sect.\,5.5 in \citealt{Gonzalez17}), which is a factor of three lower than the value given by \citet{Cicone14}. The position angles of the outflow (10$\degree$) and the inner disc (70$\degree$) are represented.
        Right:  A face-on sketch  of the three disc components identified in our data (outer disc, inner disc, and core), which are plotted in scaled sizes. The decoupled kinematics of the outer and inner discs are represented by the blue and red colours, which depict the orientation of the blue-shifted  and  red-shifted rotating gas. The rotation direction is indicated by the curved arrows. The intensity of the colours illustrate the velocities; the inner disc rotates at a higher speed than the outer disc. The stars in the inner disc aim to show the region where most of the starburst is located. The radial arrows represent the low-velocity expansion of the core. 
        }
        \label{fig16}%
 \end{figure*}

 As commented in Sect.\,\ref{linesprofilesNOEMA}, all the water lines observed with {\emph{Herschel}/PACS are consistently blue-shifted with respect to the systemic velocity of Mrk\,273. The absorption in the various lines peaks in the range $[-20, -140]$\,km\,s$^{-1}$  and extends as far as $-$600\,km\,s$^{-1}$. Velocity shifts are also observed in OH\,65\,$\mu$m and OH\,84\,$\mu$m \citep{Gonzalez17}, and trace the low-velocity gas of the approaching component of the outflow \citep{Gonzalez17}. 
 
 The channel maps of HCN and HCO$^+$ (Figs.\,\ref{fig11} to \ref{fig14}) show clear emission elongated to the north and south. This seems to correspond to the red-shifted low-velocity component (|v-v$_{\rm sys}$|$<$400\,km\,s$^{-1}$) of the wind heading to the north observed with CO  \citep{Cicone14}. In addition,  we also see  emission in the north-south direction at higher velocities (up to $\sim$400\,km\,s$^{-1}$). In our data, we  distinguish two velocity components of the outflow: one with relatively low velocities (|v-v$_{\rm sys}$|$<$400\,km\,s$^{-1}$) that is seen only in the channel maps (because its emission is blended with that of the disc in the spectra); and a  high-velocity component  (|v-v$_{\rm sys}$|$>$400\,km\,s$^{-1}$)  that is  seen in the spectrum of the central channel as a blue-shifted shoulder of the HCN\,$(3-2)$ line (Fig.\,\ref{fig5}).   The HCN\,$(3-2)$ spectral bump spans  approximately from $-$400\,km\,s$^{-1}$ to  $-$1000\,km\,s$^{-1}$ (when centring the  line at  the rest frequency of HCN$(3-2)$), which is consistent with the outflow velocities measured with CO \citep{Cicone14}.

 As explained in Sect.\,\ref{mom0outflow}, the size of the fast wind traced by HCN$(3-2)$ is  $0.''61\pm0.''05$ ($\sim$460\,pc) and the elongated shape towards the north can be  seen clearly (Fig.\,\ref{fig7}). The CO\,$(1-0)$ outflowing gas extends up to 550-600\,pc in Mrk\,273 \citep{Cicone14}. This difference suggests that the expelled moderate-density gas ($n_{\rm H_2}\le10^3$cm$^{-3}$) travels further than the dense gas ($n_{\rm H_2}\ge10^4$cm$^{-3}$). In the right panel of Fig.\,\ref{fig16}, we show a sketch of the dense and warm outflow properties  derived from our observations of Mrk\,273, as well as the  comparison with the diffuse phase of the outflow  detected in CO by \citet{Cicone14}.
 
 The HCN fast outflow luminosity  is $L^{\rm outflow}_{\rm HCN(3-2)}=\pi R^2 I=8\times10^7$K\,km\,s$^{-1}$\,pc$^2$.  To estimate the dense gas mass contained in the outflow, we used the relation $M_{\rm dense}=10\times L'[HCN (3-2)]$ \citep{GaoSolomon04}. We note that this formula might overestimate the actual value, because it assumes that all gas is virialised, which might not be the case in the outflow. Additionally, the HCN-to-H$_2$ conversion factor, which is not well constrained in ULIRGs, could also lead to an overestimation of the gas mass. A detailed analysis of  the factors  affecting the gas mass estimation can be found in \citet{GaoSolomon04}. We obtain $M^{\rm outflow}_{\rm dense}\le$8$\times10^8M_\odot$, which is consistent with the value obtained from OH observations ($\sim$1.6$\times10^8M_\odot$, \citealp{Gonzalez17}), and  is also  similar to the dense gas mass in the outflow of Mrk\,231 ($\sim$4$\times10^8M_\odot$, \citealp{Aalto15a}). We note that our result refers only  to the high-velocity gas (|v-v$_{\rm sys}$|$>$400\,km\,s$^{-1}$), as it is not possible to separate the slow component of the outflow from the disc emission in the HCN data.
 
 In cases where the outflow and the disc kinematics cannot be distinguished well in the velocity maps (as in our Fig.\,\ref{fig8}), a difference in their position angles  can help to probe the kinematic decoupling between the two, and can also yield information about the nuclear powering source. Gas in starburst-powered  outflows is always expelled perpendicular to the plane of the galaxy (thus showing a change of 90$\degree$ with respect to the P.A. of the starbursting disc), while AGN-powered outflows might have virtually any position angle, because the dusty torus and the accretion disc can be tilted with respect to the disc \citep{Burillo15}. In Mrk\,273, we measure a  difference between the P.A. of the inner disc (of 71$\degree\pm5\degree$, from HCO$^+(3-2)$, Table\,\ref{table3}) and the P.A. of the outflow ($10\degree \pm 3\degree$), of 60$\degree \pm$8$\degree$. This  seems to indicate that the  outflow is powered by the central AGN. Indeed, \citet{Cicone14}  found that the high outflow mass-loss rate of Mrk\,273  is consistent with the linear correlation they observed between the bolometric AGN luminosity and the mass outflow rate in local AGN-host galaxies. 
 
 A couple of interesting issues arising from our data are 1)  the red-shifted component of the outflow traced by HCN\,$(3-2)$ is seen only at low velocities in the channel maps (|v-v$_{\rm sys}$|$<400$\,km\,s$^{-1}$). Indeed, a fast red-shifted wind, which should be visible in the spectra (with a similar intensity as the blue-shifted one), does not appear; and 2) the apparent absence of high-velocity winds in HCO$^+$ (i.e. HCO$^+$ line wings at |v-v$_{\rm sys}$|$>400$\,km\,s$^{-1}$). Regarding the first issue, deeper observations are required to improve the S/N of the outflow, to make sure that the red-shifted component is not present. If that were the case, then it would imply that either there is a density or temperature gradient between the receding and the approaching gas (with the red-shifted gas being less dense and/or warm in general), or there is a chemical differentiation between the two. These are not uncommon characteristics in galactic winds, such as  in Mrk\,231 \citep{Aalto12,Aalto15a}. The chemical differentiation could also explain why there is not enough high-velocity gas traced by HCO$^+$ in the outflow (see \citet{Lindberg16} for details about this phenomena in the wind of Mrk\,231).

\subsection{Expansion of the core}
\label{expansion}
In Mrk\,273, the velocity peaks of the OH\,65\,$\mu$m and OI\,63\,$\mu$m lines are
 shifted in comparison to the emission peak of [CII] at 158\,$\mu$m \citep{Gonzalez17}. The [CII] maximum is found at zero velocities (with respect to the systemic velocity of the galaxy), and  arises from the bulk of the warm gas in the nucleus. The OH\,65\,$\mu$m and OI\,63\,$\mu$m lines, however, have their maximum (absorption/emission) peaks shifted by $\pm$50 km/s, and trace the blue component of the outflowing gas moving at low velocities. Apart from being a sign of a superwind, \citet{Gonzalez17} argue that this change in the redshift also indicates that the large columns of gas close to the central engine are expanding at low velocities. This effect has been found in a number of other ULIRGs where low-velocity outflows are also present \citep{Gonzalez17}.

As discussed in Sect.\,\ref{asymmetry}, the shift of approximately 50\,km\,s$^{-1}$ seen in OH\,65\,$\mu$m and OI\,63\,$\mu$m also appears in our H$_2$O data. In particular, the H$_2$O  lines observed with PACS, which according to our models trace the compact core with a radius $\sim$50\,pc, are also blue-shifted with respect to [CII] by $[-20,-140]$\,km\,s$^{-1}$ (Table\,\ref{table2}). This  indicates that, apart from rotating, the gas in the core is expanding at low velocities, pushing outwards the boundaries with the inner disc  traced by the HCN and HCO$^+(3-2)$ lines. Indeed, the expansion is also reflected in our HCO$^+$ and HCN$(3-2)$ data.  While HCO$^+$ and HCN$(3-2)$ double-peaks probe the outer edges of the inner disc  (at a radius of $\sim$300\,pc),  their peaks of absorption probe the kinematics of the bulk of the disc, which is also shifted by  50-100\,km\,s$^{-1}$. We interpret this as the gas in the inner disc being pushed by the expanding gas of the core. We illustrate this in the sketch shown in the right panel of Figure\,\ref{fig16}.

\subsection{Non-detection  of vibrational emission}
\label{vib}

\begin{table*}
        \caption{Main parameters of the undetected vibrational transitions of HCN and HC$_3$N.}   
        \centering                          
        \begin{tabular}{l c c c c c c c c c}       
                \hline\hline
                Line & Freq & $E_{\rm low}$ &$A_{ul}$ &$n_{\rm crit}$ & Flux & Line Peak & L$'$ \\  
                                & [GHz] & [K] & [$\times10^{-3}$s$^{-1}$]  & [cm$^{-3}$] & [K\,km\,s$^{-1}$] &[mJy] & [K\,km\,s$^{-1}$\,pc$^2$]\\
                \hline
                HC$_3$N\,$(10-9)$ v$_6$=1 &91.202&512&0.15&$>10^{11}$&$\le$1.1&$\le$0.3 & $\le$6.4$\times10^6$\\
                HCN\,$(3-2)$ v$_2$=1 &267.199&720&0.73&$>10^{10}$&$\le$3.7&$\le$1.1 & $\le$7.3$\times10^5$\\             
                \hline                                    
        \end{tabular}   
        \label{table6}          
\end{table*}

Rotational transitions within vibrationally excited levels of HCN and HC$_3$N have been observed in the central $<$100\,pc of several (U)LIRGs, probing regions of high  temperatures of $\ge$100\,K \citep{Sakamoto10,Costagliola15,Aalto15, Martin11,Martin16}, and high column  densities. Their excitation cannot be explained by collisional effects alone, and mid-IR pumping is necessary in order to populate the upper energy levels and fit their observed luminosities \citep{Aalto15}. In that case, the vibrational lines may be more suitable to study the optically thick dust cores of galaxies than the rotational transitions (see \citet{Aalto15} for a detailed discussion). 

In our NOEMA observations of Mrk\,273, we do not detect the vibrationally excited line HCN$(3-2)$ v$_2$=1. The line is split into two components with energy levels of similar intensity, v2=1e and v2=1f, at frequencies 265.8 and 267.2\,GHz.   The first one is completely blended with the HCN\,$(3-2)$ rotational line due to the galactic broad line widths, and it is not possible to estimate its peak temperature.  However, the v$_2$=1f transition would appear as a bump only  partially blended with the red side of the HCO$^+(3-2)$ transition. We calculate a limit to its flux density in the central pixel of 1.1\,mJy, which is slightly below the $1\sigma$  rms  level (of 1.2\,mJy, calculated at the final velocity resolution).  Assuming that the vibrational line has the same FWHM as the HCN rotational transition ($\sim$600\,km\,s$^{-1}$), the limit to the integrated intensity was estimated as $I \le 3\times rms \times \sqrt{FWHM \times Dv}$\,=4\,K\,km\,s$^{-1}$  ($Dv$, the final spectral resolution, is 68\,Km\,s$^{-1}$).  Under the conservative assumption that both rotational and vibrational lines have the same source size, the upper limit to the HCN-vib luminosity is $L'=\pi R^2 I \le7.3\times 10^5$\,K\,km\,s$^{-1}$pc$^2$ ($R$ is the source size of the HCN\,$(3-2)$ emission specified in Table\,\ref{table3}). This is at least a factor of two fainter than the $L'_{\rm HCN_{vib}(3-2)}$ in Mrk\,231 (1.7$\times 10^7$\,K\,km\,s$^{-1}$pc$^2$, \citealp{Aalto15a}). This is interesting, because \citet{Lahuis07} detect the 14$\mu$m HCN-vib lines  in absorption at the same level in both Mrk\,273 and Mrk\,231.

 \citet{Aalto15} found a tentative correlation between the outflow velocity and the intensity of HCN vibrationally excited lines in a moderate sample of ten (U)LIRGs.  Galaxies with fast outflows (i.e. when the outflow velocity  exceeds the escape velocity of the nuclear region) tend to have fainter vibrational lines. With its non detection of HCN$(3-2)$ v$_2$=1, and its fast molecular outflow, Mrk\,273 also follows this relationship. The reason for this correlation, if true, requires further investigation.

Similarly, we do not detect the vibrationally excited line HC$_3$N\,$(10-9)$ v$_6$=1, even though the energy of its lower level is below that of the HCN-vib lines (Table\,\ref{table6}). We obtain an upper limit to the peak flux density of 0.3 mJy. This line was first tentatively detected in the LIRG NGC\,4418 with the IRAM\,30\,m telescope \citep{Aalto07,Costagliola10} and later confirmed by ALMA observations \citep{Costagliola15} with a rotational-to-vibrational intensity ratio of $\sim$11 . If we extrapolate that ratio to Mrk\,273, then, based on our results, one would expect a flux density of $\sim$0.2\,mJy. Under the   assumption that, in Mrk\,273, HC$_3$N$(10-9)$ v$_6$=1 likely  arises from a region smaller than HCN$(1-0)$ (i.e. $\theta<2''.0\times1''.6$), the limit to its luminosity is $6\times10^6$\,K\,km\,s$^{-1}$\,pc$^2$.

Apart from sensitivity issues, two possibilities can potentially explain why we do not detect  vibrationally excited emission of HCN and HC$_3$N in Mrk\,273. One is that the  mid-IR radiation density  in the core of Mrk\,273 is not as strong as in other galaxies hosting compact obscured nuclei, such as Mrk\,231, Arp\,220 or NGC\,4418. The other is that, although the  mid-IR radiation is strong,  the obscuration in the centre of Mrk\,273 is so high that even the emission of vibrationally excited lines is  extinguished.

\subsection{Molecular line ratios }
\label{lineratios}
Tables\,\ref{table4} and \ref{table5} list the brightness temperature ratios of the species detected with NOEMA. The  ratios were obtained by comparing the temperatures of the lines calculated over the entire emission (Table\,\ref{table4}) and in the central pixel (Table\,\ref{table5}). We note that  the 3\,mm lines (HCN, HCO$^+$, HNC(1-0) and HC$_3$N(10-9)) and the 1\,mm lines (HCN, HCO$^+$, HOC$+(3-2)$) were observed with angular resolutions that differ in  the covering area by a factor $>$60 of covering area. In order to avoid confusion and unphysical results, we only discuss ratios in which both lines  were  observed with the same resolution. In the following we discuss the most relevant ratios.

\subsubsection{HCN/HCO$^+$}
\label{hcnhcop}

The integrated intensity ratio between HCN and HCO$^+$  has been widely used to discriminate between starburst-powered galaxies and AGNs  \citep{Kohno01,Krips08,Imanishi09,Izumi13,Izumi16,Martin15}. An enhancement of the HCN abundance  with respect to HCO$^+$ is observed in a fairly large sample of Seyferts, which show a HCN/HCO$^+$ ratio $\ge 1$. Starbursts, on the other hand, are found to have values $\le 1$. The coexistence of both phenomena in the central few hundred parsecs can however lead to outlier values of the ratio if they are not resolved \citep{Aladro15,Privon15}.

The reason why the HCN/HCO$^+$ ratio is a  diagnostic of the galactic central activity is still debated. For more than a decade, the HCN enhancement was claimed to be due to X-ray irradiation in the vicinity of the SMBHs (e.g. \citealp{Kohno01}). However, recent studies of NGC\,1097 point to high temperatures and/or mechanical heating as a more plausible explanation \citep{Martin15,Izumi16}. Systematic high-angular-resolution observations of a larger sample of  galaxies/AGNs is still needed to clarify this issue.

In the case of HCO$^+$, both observations and  models point to an enhancement in starburst galaxies, and in particular in photon-dominated regions (PDRs),  as a result of  strong UV irradiation. Unfortunately, the scenario could be more complex, since chemical models suggest that HCO$^+$ can also be enhanced in regions heavily pervaded by  cosmic ray/X-ray fields, such as in AGNs (Meijerink et al 2006, Aladro et al. 2013). Yet this scenario has not been confirmed with observations, to our knowledge.

Bearing in mind that the reasons for the enhancements of HCN and HCO$^+$ are difficult to assess, the fact is that this ratio diagnoses the power sources of many well-known active galaxies. Nevertheless,  there are some sources which do not seem to be consistent with this scenario. For example, some apparent starbursts  have HCN-to-HCO$^+$ ratios as high as AGNs, indicating that the HCN enhancement cannot be exclusively attributed to AGNs \citep{Privon15}.

In Mrk\,273, we measure the global brightness intensity ratios HCN$(1-0)$/HCO$^+(1-0)$= 1.0$\pm$0.2 and  HCN$(3-2)$/HCO$^+(3-2)$= 1.1$\pm$0.3 (Table\,\ref{table4}). Evaluating the nuclear ratios in the central pixel, gives very similar values of 1.0$\pm$0.2 and 1.3$\pm$0.3 for the $(1-0)$ and $(3-2)$ transitions respectively (Table\,\ref{table5}).
These ratios place the galaxy in an ambiguous region of the diagnostic diagram, where  a starburst, an AGN, or a combination of the two are possible. Depending on the diagnostic used in the literature,  either an AGN or a starburst are claimed to power the northern nucleus of Mrk\,273. Perhaps a mixture of both activities are present in the central hundred parsecs. Unfortunately, our HCN/HCO$^+$ ratio does not help to elucidate  the nature of the Mrk\,273 nuclear source. Nevertheless, we note that both lines are significantly affected by opacity (Sect. \ref{asymmetry}), and that a more thorough  analysis of this ratio including obscuration effects could further elucidate our results.

Using the Nobeyama Millimeter Array  and RAINBOW interferometers at the Nobeyama Radio Observatory \footnote{https://www.nro.nao.ac.jp/en/astronomer/}, \citet{Imanishi06} obtained a flux ratio HCN$(1-0)$/HCO$^+(1-0)>$1.8 in Mrk\,273, within a beam of 1$\ffas$9$\times$1$\ffas$5. This was interpreted as a signature of an AGN-dominated nucleus. This lower limit is much higher than our values, but we note that the latter authors did not detect HCO$^+(1-0)$, and that their HCN$(1-0)$ and continuum fluxes are 70-80\% lower than our values.

\subsubsection{HCN/HNC and HCO$^+$/HNC}

Some (U)LIRGs, such as Arp\,220 and Mrk\,231, are extremely bright in  HNC, with line intensities almost equal to or even surpassing those of HCN  \citep{Huettemeister95,Aalto02,Aladro15}. Several scenarios can produce this unusually high line ratio, such as chemical reactions at moderate densities and temperatures (n$\sim$10$^4-10^5$cm$^{-3}$, T$_{\rm kin}\sim$50\,K), high opacities of HCN with respect to HNC and, very likely, infrared pumping affecting  HNC more than HCN \citep{Aalto02}. The HCN/HNC intensity ratio can reflect the evolutionary stage of starburst regions in galactic centres, with faint HNC  emission (i.e. HCN/HNC $\gg$ 1) being associated with shock-dominated regions, which are common in early starbursts \citep{Aladro15}.

The brightness temperature ratio between HCN$(1-0)$ and HNC$(1-0)$ in Mrk\,273 is 1.7 (2.1 in the central pixel). Previous observations of these lines with the OSO and SEST single-dish telescopes yielded an integrated intensity ratio $\ge$4 \citep{Aalto02}, although that value is based on an upper limit to the HNC line. Taking our result of 1.7 as the minimum value, it is safe to say that this conservative ratio is moderately high, and that Mrk\,273 cannot be classified as an HNC-luminous galaxy.

As discussed above (Sect.\,\ref{hcnhcop}), the HCN/HCO$^+$ brightness temperature ratio we obtain for Mrk\,273 does not allow us to favour a starburst over an AGN-dominated nucleus.  The HNC and HCO$^+$ intensities appear to be anti-correlated in (U)LIRGs, as observed by \citet{Costagliola11} in a fairly large sample of galaxies. From our data, we obtain an HCO$^+$/HNC\,=\,$1.8\pm0.2$ (Table\,\ref{table5}). If the northern nucleus of Mrk\,273 is dominated by a starburst (as claimed  by \citealp{Condon, Majewski93, DS98}),  the low HNC abundance would indicate that the gas comes from warm and dense phases in an early stage, and would explain why HCO$^+$ is relatively abundant. However,  models  by \citet{Rodriguez09} indicate that most of the stellar population in Mrk\,273 has an age of  0.7-10\,Gyr (although there might be a significant fraction of stars younger than 50\,Myr), which challenges an early starburst scenario. In the case of an AGN (as claimed by \citealp{U,Rodriguez, Iwasawa17}), HNC does not necessarily need to be faint, but HCN could be boosted. However, in the latter scenario one would expect a higher HCN/HCO$^+$ ratio.

\subsubsection{An extremely low HCO$^+$/HOC$^+$ ratio}
\label{hcophocp}
HOC$^+$, the metastable isomer of HCO$^+$, is efficiently formed via the following  ion-molecule reactions; 

\begin{equation}
{\rm{CO^+ + H_2 \rightarrow HOC^+ + H}},
\end{equation}
\begin{equation}
{\rm{C^+ + H_2O \rightarrow HOC^+ + H}},
\end{equation}
\begin{equation}
{\rm{H_3^+ + CO \rightarrow HOC^+ + H_2}},
\end{equation}

and is mainly destroyed  by reactions with $\rm H_2$ \citep{Jarrold86,Smith02,Fuente03}:

\begin{equation}
{\rm{HOC^+ + H_2 \rightarrow HCO^+ + H_2}}.
\end{equation}

While typical values of the HCO$^+$/HOC$^+$ ratio in Galactic dense molecular clouds range between 300 and 6000 (Apponi et al. 1997),  values as low as 50-150 are found in several Galactic and extragalactic PDRs and XDRs (X-ray dominated regions), likely as a consequence of high ionisation rates  created by ionisation fields (UV, cosmic rays, and/or X-ray radiation) (\citealp{Fuente03,Usero04}). From our observations, we derive a global brightness temperature ratio HCO$^+(3-2)$/HOC$^+(3-2)$ = $9\pm4$, and a nuclear ratio of $5\pm3$. Such  low values have only been found in other ULIRGs hosting extremely compact obscured nuclei, namely IC\,860, Zw049.057, and Mrk\,231 \citep{Aalto15a,Aalto15}. The physical and chemical reasons for these low ratios are still not known and merit further study, but high opacities of HCO$^+$ could be responsible.

\subsection{A different origin of HOC$^+$ emission}
\label{hoc+}

\begin{figure}[t!]
        \includegraphics[width=0.4\textwidth]{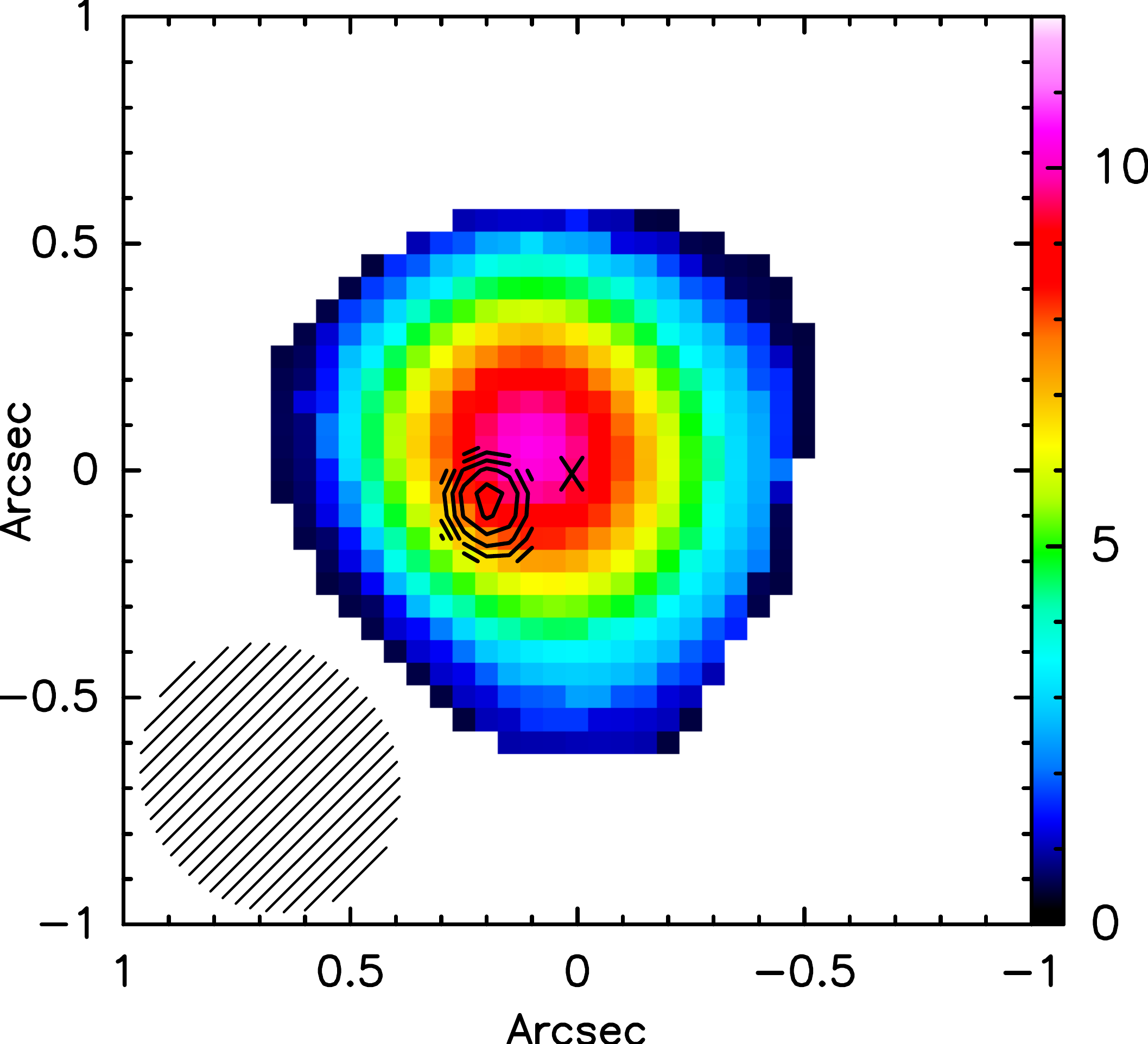}    
        \caption{HCO$^+(3-2)$ blue-shifted  intensity colours (in units of Jy\,km\,s$^{-1}$\,beam$^{-1}$) integrated from -300\,km\,s$^{-1}$ to 0\,km\,s$^{-1}$ with HOC$^+(3-2)$ contours (starting from 3$\sigma$=3.3\,mJy\,beam$^{-1}$) over-plotted with steps of 10\,mJy\,beam$^{-1}$. The cross at the centre marks the continuum peak. The synthesised beam is shown in the bottom-left corner.}
        \label{fig17}%
\end{figure}

As mentioned in Sect.\,\ref{momentmaps}, HOC$^+(3-2)$  peaks at RA\,(J2000)\,=\,13$^h$:44$^m$:42$\ffs$15, DEC\,(J2000)\,=\,55$^\circ$:53$'$:13$\ffas$45, which corresponds to an offset  $(0\ffas2,-0\ffas05)$ south-east of the central pixel. This indicates that its emission has a different origin from the rest of the dense gas tracers. Furthermore, the two HOC$^+$ components are  blue-shifted in velocity with respect to  HCN and HCO$^+(3-2)$ (Table\,\ref{table1}). Figure\,\ref{fig17} shows the HOC$^+$ emission  in contours plotted over the blue-shifted HCO$^+$ emission integrated between -300 and 0\,km\,s$^{-1}$. This plot shows that HOC$^+$  does not  peak at the same  position as the blue-shifted dense gas, and that there is not even detected emission of the species in the central pixel.

Why is the emission of this species shifted with respect to the others? One possibility is that the central pixels are heavily obscured and  HOC$^+$ is completely absorbed there. However, in that case one would expect some kind of symmetric emission around the nucleus, with a ring-like shape, or at least other peaks of emission around the nucleus, which are not seen. This strongly suggests that  HOC$^+$ is not peaking at the very centre.

We checked the literature looking for OH masers, supernovae, radio continuum sources, or any  source emitting at the HOC$^+$ peak coordinates. High-angular resolution NIR observations conducted by \citet{U} resolve the northern nucleus of Mrk\,273 into three components, called N1, N2, and N3. Component N1 is the brightest of the three and is associated with the true nucleus (our ($0'',0''$) position). N3 is found  $\sim$0$\ffas$15 to the south-east of N1, very near the HOC$^+$ maximum flux density. The peak of the HOC$^+$ emission might correspond to N3, but better astrometry is needed in order to strengthen this association. The nature of N3 is, in any case, not clear, but it does have stronger emission from [FeII] than N1. It could be a separated clump, a compact star cluster, or a supernova remnant. N3 does not appear in radio continuum maps \citep{Carilli00,Bondi05}, leaving a supernova remnant as a less likely option. To further address the origin of this species,  deeper high-resolution  observations of HOC$^+$  would be needed to determine its extent and to more accurately measure its position.

\section{Conclusions}
\label{conclusions}

We have used the NOEMA interferometer to observe several spectral lines of HCN, HCO$^+$,  HOC$^+$, HNC, and HC$_3$N with  angular resolutions of ($4\ffas9\times4\ffas5$) and ($0\ffas61\times0\ffas55$) (corresponding to spatial scales of $\sim$($3.7\times3.4$)\,kpc and $\sim$($460\times420$)\,pc). We also included multiple lines of H$_2$O observed with the \emph{Herschel} SPIRE and PACS instruments. Our observations,  extending from the mm to the FIR regime, allowed us to study the properties of the gaseous disc in the northern nucleus of Mrk\,273, as well as the connection between its cold and warm phases. We summarise the main results as follows.

\paragraph{{\bfseries Morphology and kinematics of the disc.}}
The cold and dense gas in the nuclear disc (traced by HCN and HCO$^+$) has two  components with decoupled  kinematics. The low-excitation gas in the outer parts of the disc extends up to a radius of $\sim$1.5\,kpc and rotates from south-east to north-west (with a P.A. of approximately $-40\degree$), while  the more excited dense gas arising from the central star forming region ($<$300\,pc) is characterised by  a north-east to south-west rotation (P.A.$\sim$70$\degree$). This inner disc contains a dynamical mass of $3\times10^9M_\odot$, and a luminosity of $L'_{\rm HCN}=3\times10^8$K\,km\,s$^{-2}$\,pc$^2$. The warm gas, traced by the FIR H$_2$O lines, can  also be separated into two components: a warm and very compact core with a radius  of $\sim$50\,pc and a temperature of 95\,K, and a more extended and relatively cooler component, with R $<$300\,pc and T\,=\,55\,K.

The extended component of the warm gas and the compact component of the cold gas are  co-spatial in the inner $\sim$300\,pc. The H$_2$ column densities and dust properties obtained from our water modelling, as well as the line profiles of the dense gas tracers, show that this  region is significantly affected by dust obscuration. The blue-shifted emission of the bulk of gas (consistently seen in all observed lines) also indicates that the core is  expanding outwards at low velocities (v-v$_{\rm sys}\sim$\,50-100\,km\,s$^{-1}$), likely affected by the outflow.

\paragraph*{{\bfseries Outflow properties.}}
We detected the cold (sub-mm) and warm  (FIR) phases of the Mrk\,273 molecular outflow.  It is a compact outflow, being expelled to distances of $\sim$460\,pc mostly towards the northern direction, but it reaches  high velocities of $\sim$1000\,km\,s$^{-1}$. This fast outflow (|v-v$_{\rm sys}$|$>$400\,km\,s$^{-1}$) has a luminosity of $8\times10^7$\,K\,km\,s$^{-1}$\,pc$^2$, and a mass of dense gas  $M^{\rm outflow}_{\rm dense}\le$8$\times10^8M_\odot$. The difference in P.A. between the major kinematic axis of the inner disc ($71\degree\pm5\degree$) and that of the outflow ($10\degree\pm3 \degree$) suggests that the latter is probably driven by the AGN.

\paragraph*{{\bfseries Chemistry.}}
We explored the chemistry of Mrk\,273 by means of molecular line ratios. The most outstanding ratio is that of HCO$^+$/HOC$^+$. We estimated it to be $<$10,   one of the lowest values ever measured in any galactic or extragalactic source. The reason for this value, however, is still not clear and should be further studied in detail with the help of chemical models. It is worth noting, however, that the origin of HOC$^+$ is  different from the rest of the detected molecular species, since its emission is spatially shifted from the centre.

Regarding the outflow, our non-detection of the high-velocity wind in HCO$^+$, together with the non-detection of the red-shifted outflowing gas either in HCN and HCO$^+$, suggests the possibility of  chemical differentiation. However, we note that, despite our high sensitivity, the fast outflow of Mrk\,273 is very faint, meaning that deeper  observations would be necessary to better probe its chemistry.

\begin{acknowledgements}
 This work is based on observations carried out under project numbers W14DD and E16AK  with the IRAM NOEMA Interferometer. IRAM is supported by INSU/CNRS (France), MPG (Germany) and IGN (Spain). The research leading to these results has received funding from the European Union's Horizon 2020 research and innovation program under grant agreement No 730562 [RadioNet]. This research has made use of NASA's Astrophysics Data System, and the NASA/IPAC Extragalactic Database (NED), which is operated by the Jet Propulsion Laboratory, California Institute of Technology, under contract with the National Aeronautics and Space Administration. We are grateful to the referee for the fast and constructive report, as well as to the  IRAM/NOEMA staff for their help during the observations and data reduction. RA would like to thank Leslie Hunt and Loreto Barcos-Mu\~noz for the useful discussions about Mrk\,273.
\end{acknowledgements}

%
%

\end{document}